\documentclass{article}

\usepackage{PRIMEarxiv}

\usepackage[utf8]{inputenc} % allow utf-8 input
\usepackage[T1]{fontenc}    % use 8-bit T1 fonts
\usepackage{hyperref}       % hyperlinks
\usepackage{url}            % simple URL typesetting
\usepackage{booktabs}       % professional-quality tables
\usepackage{amsfonts}       % blackboard math symbols
\usepackage{nicefrac}       % compact symbols for 1/2, etc.
\usepackage{microtype}      % microtypography
\usepackage{lipsum}
\usepackage{fancyhdr}       % header
\usepackage{graphicx}       % graphics
\graphicspath{{figures/}}     
\usepackage{subcaption}
\usepackage[most]{tcolorbox}
\usepackage{amsmath}
\usepackage{enumerate}
\usepackage{enumitem}
\usepackage{tabularx}
\usepackage{pdfpages}

\newcommand{\descbox}[1]{%
    \parbox[t]{\linewidth}{\raggedright #1}%
}
\usepackage{tcolorbox}
\usepackage{newfloat}
\DeclareFloatingEnvironment{boxes}

\newcommand{\boxref}[1]{\hyperref[{#1}]{\ref*{#1}}}
\definecolor{greenframe}{RGB}{85, 120, 90}

\title{Hierarchical Multi-Agent Reasoning Systems for Knowledge-Grounded Material Substitution }

\usepackage{orcidlink}
\author{
Tarjei Paule Hage\textsuperscript{\textdagger}\,\orcidlink{0009-0005-6969-1852}\\
Department of Mechanical Engineering\\
Massachusetts Institute of Technology\\
Cambridge, MA, USA
\And
Subhadeep Pal\textsuperscript{\textdagger}\,\orcidlink{0000-0002-0532-9791}\\
Department of Civil and Environmental Engineering\\
Massachusetts Institute of Technology\\
Cambridge, MA, USA
\AND
Wei Lu\,\orcidlink{0000-0002-5168-4428}\\
Department of Civil and Environmental Engineering\\
Massachusetts Institute of Technology\\
Cambridge, MA, USA
\And
Yu-Chuan Hsu\,\orcidlink{0000-0001-5587-4547}\\
Department of Civil and Environmental Engineering\\
Massachusetts Institute of Technology\\
Cambridge, MA, USA
\AND
Gayla Lyon\,\orcidlink{0000-0002-1057-0635}\\
Saint-Gobain Research North America\\
Northborough, MA, USA
\And
Jiezhu Jin\,\orcidlink{0009-0008-1391-8976}\\
Saint-Gobain Research North America\\
Northborough, MA, USA
\AND
Markus J. Buehler\textsuperscript{*}\,\orcidlink{0000-0002-4173-9659}\\
Department of Civil and Environmental Engineering\\
Department of Mechanical Engineering\\
Schwarzman College of Computing\\
Massachusetts Institute of Technology\\
Cambridge, MA, USA\\
\texttt{mbuehler@mit.edu}
\\[0.75em]
\textsuperscript{\textdagger}\textit{These authors contributed equally to this work.}\\
\textsuperscript{*}\textit{Corresponding author.}
}

\begin{document}
\maketitle

\begin{abstract}
AI-assisted scientific discovery increasingly seeks to move beyond generating plausible ideas toward reproducible, constraint-aware workflows that can support practical engineering decisions. We present MARS, a hierarchical Multi-Agent Reasoning System for knowledge-grounded material substitution under scientific and manufacturing constraints. We apply MARS to replace per- and polyfluoroalkyl substances (PFAS), a demanding materials-design problem in which substitutes must satisfy functional performance, environmental, experimental, and processing requirements. MARS decomposes substitution into a closed-loop workflow comprising requirement analysis, grounded candidate generation, and manufacturability assessment with process design. The system integrates retrieval-augmented generation, knowledge-graph reasoning, a database of lab-available materials, persistent memory of intermediate results, and rejection-aware iterative feedback. In a PFAS-substitution case study, MARS generates structured material requirements, identifies an evidence-grounded substitute candidate, and develops a lab-scale manufacturing route. In a multi-seed ablation evaluation, MARS achieves the highest overall performance across all three subsystems, with gains that persist when the backbone model is held fixed. These results suggest that hierarchical decomposition, knowledge grounding, and manufacturability-aware refinement improve material-substitution reasoning beyond model capability alone.
\end{abstract}

% keywords can be removed

\section{Introduction}
\label{sec:introduction}
Since the earliest days of artificial intelligence (AI), researchers have sought to automate aspects of scientific discovery \cite{hovland_computer_1960}, and the idea of using AI to perform science has long been present in the literature \cite{simon_scientific_1997}. The emergence of deep learning \cite{lecun_deep_2015} significantly expanded this vision, shifting the field from primarily symbolic approaches toward computational models capable of learning hierarchical representations from data \cite{wang_scientific_2023,hu_deep_2023,berens_ai_2023}. A landmark example is AlphaFold, a neural network-based system that integrates physical and biological knowledge of protein structure into its model design and achieves accuracy competitive with experimental methods \cite{jumper_highly_2021}. More recently, the rapid advance of LLMs has further broadened the role of AI in science, with growing evidence that these models can perform tasks spanning hypothesis generation, experimental planning, and discovery itself \cite{vaswani_attention_2017,wei_finetuned_2022,wei_chain--thought_2022,zhang_exploring_2025}. Although the term \emph{Robot Scientist} is well established \cite{king_automation_2009}, the rise of agentic AI \cite{yao_react_2023,wang_survey_2024} has given renewed meaning to the notion of the \emph{AI Scientist} \cite{lu_ai_2024}, namely, autonomous agents built on LLM backbones that can pursue scientific tasks with a degree of agency. The subsequent development of LLM-based multi-agent systems (MAS), in which multiple intelligent agents interact and collaborate toward a shared objective \cite{buehler_physics-aware_2026,han_llm_2026}, has further extended this paradigm. Such systems have already been applied across a broad range of scientific domains, including protein design \cite{ghafarollahi_sparks_2025,wang_swarms_2025, lu_generative_2024,lu2025generative}, alloy design \cite{ghafarollahi_rapid_2025}, and general scientific discovery \cite{ghafarollahi_sciagents_2025,ghafarollahi_autonomous_2025}. Major AI research labs have also increasingly explored how foundation models can be deployed in scientific research settings. These efforts include evaluating models as scientific collaborators \cite{wang_frontierscience_2025}, developing LLM-powered evolutionary coding agents for algorithm discovery \cite{novikov_alphaevolve_2025}, multi-agent research systems for scientific question answering \cite{engineering_at_anthropic_how_2025}, and building AI-driven autonomous laboratories for protein production \cite{smith_using_2026}. Collectively, these developments point to a broader shift from foundation models serving as standalone scientific assistants toward their integration into increasingly autonomous and collaborative research systems. Recent advancements have pushed this transition further, with fully autonomous research agents collaborating in a crowdsourced manner without human intervention, suggesting a model of decentralized scientific discovery \cite{wang_autonomous_2026, pal_swarmworld_2026}. We hypothesize that the future of science will include AI systems capable of conducting complete, reproducible investigations, and contributing to distributed ecosystems of autonomous scientific research.

At a high level, scientific discovery can be understood as an iterative cycle of hypothesis generation, testing, and revision in response to evidence \cite{krauss_science_2024,kuhn_structure_1962,popper_logic_1959}. LLM-based multi-agent systems can support this cycle by automating either the proposal of hypotheses or their evaluation, and, when coupled with autonomous laboratory infrastructure, potentially both \cite{zhang_exploring_2025,lee_automated_2022,smith_using_2026}. The multi-agent paradigm frames this process as a collaborative workflow in which complex scientific tasks are decomposed into specialized LLM-backed agents, each assigned a focused subtask such as retrieval, reasoning, critique, or validation. This decomposition enables different stages of the scientific workflow to be represented by distinct agents, each equipped with the tools required for its function. These tools may include literature retrieval, physics-based simulation, compilers for executing agent-generated code, and interfaces to databases or experimental platforms. Hypothesis generation is a promising role for LLMs, as they can reason over large volumes of information and help integrate prior knowledge to identify meaningful gaps and formulate promising ideas \cite{zhang_exploring_2025}. Because LLMs are trained on vast corpora of text, they capture a substantial amount of scientific knowledge through next-token prediction. Yet their tendency to hallucinate limits their applicability in settings that require reliable scientific reasoning. A common strategy for improving grounding is retrieval-augmented generation (RAG) \cite{lewis_retrieval-augmented_2020}, in which relevant documents are retrieved from designated sources and supplied to the model as contextual evidence at inference time. This allows the model to ground its responses in external text, thereby not only reducing hallucinations, but increasing domain specificity, and improving the quality of generated reasoning \cite{luu_bioinspiredllm_2023}. A key limitation of RAG is that, although it retrieves relevant evidence from specified sources, it does not inherently model the structured relationships that connect information across sources. An approach to overcoming this limitation is to construct ontological knowledge graphs (KGs) from scientific literature \cite{giesa_category_2012,markovic_optimizing_2025,sequeda_knowledge_2025,buehler_accelerating_2024}, where concepts are represented as nodes and their relationships as edges. This construction process can itself be automated, for example by using LLMs to extract structured knowledge from scientific text \cite{schilling-wilhelmi_text_2024}. The resulting KGs can then support graph-native reasoning models \cite{buehler_situ_2025,buehler_self-organizing_2025,kumar_toward_2025, pal_graph-native_2026}. Within multi-agent systems, KGs serve as an additional tool that enables agents to retrieve and reason over structured knowledge \cite{stewart_higher-order_2026,ghafarollahi_automating_2025,marom_frontiers_2025,stewart_graphagents_2026}. These developments are especially relevant in scientific domains where discovery must connect information across a vast span of different sources. One such domain is the search for substitutes to harmful materials.

Per- and polyfluoroalkyl substances (PFAS), a class of compounds consisting of a fluorinated alkyl chain and a polar head group, have attracted increasing attention in recent years. Their carbon–fluorine bonds confer material properties, such as resistance to water and grease, which has led to their widespread use in applications that include fast-food packaging, stain-resistant textiles, and firefighting foams \cite{dickman_review_2022,evich_per-_2022}. At the same time, these same strong bonds contribute to the environmental persistence for which PFAS are often termed “forever chemicals” \cite{evich_per-_2022}. PFAS are released into the environment through multiple pathways, including industrial emissions, wastewater discharge, and agricultural pesticides, and studies have documented widespread contamination of water bodies, particularly near industrial sites and firefighting areas \cite{alsadik_pfas_2025}. Epidemiological evidence has linked PFAS exposure to a range of adverse health outcomes, including liver disease and cancer \cite{fenton_per-_2021}. Because remediation following environmental release is both costly and technically challenging, substantial attention has been directed toward the search for safer alternatives \cite{evich_per-_2022}. However, identifying viable substitutes remains challenging, because the same carbon–fluorine bonds responsible for PFAS persistence also give rise to the unusual combination of properties that has made these materials so valuable across applications \cite{figuiere_overview_2025}. This makes PFAS substitution a useful setting for knowledge-driven AI-assisted materials discovery, as the search for viable substitute materials depends on hypotheses grounded in a deep understanding of material properties, environmental constraints, and application-specific requirements.

Materials informatics has progressively expanded the role of computation in materials discovery, from data-driven property prediction and screening to the integration of heterogeneous scientific data and knowledge resources \cite{rajan_materials_2005,ramprasad_machine_2017, lu2023modeling}. More recently, large language models (LLMs) have enabled interaction with scientific literature and materials data through natural-language reasoning, information extraction, and hypothesis generation \cite{choi_accelerating_2024,jiang_applications_2025}. Building on these capabilities, agentic materials-discovery workflows combine LLMs with external knowledge sources, specialized reasoning stages, and iterative coordination to support more integrated scientific tasks \cite{wang_knowledge-guided_2025,ghafarollahi_sciagents_2025,
ghafarollahi_autonomous_2025,ghafarollahi_automating_2025,
ghafarollahi_rapid_2025,ghafarollahi_sparks_2025,lu2025fine,luu2024learning,lu_generative_2024,lu2025generative}. These developments provide the broader context for multi-agent systems that connect scientific knowledge retrieval with materials reasoning and design.

Within AI-driven PFAS substitution, our previous GraphAgents framework applied knowledge-graph-guided multi-agent reasoning to generate PFAS-free material hypotheses from literature-derived knowledge resources \cite{stewart_graphagents_2026}. The system decomposes the design problem into subtasks, retrieves supporting evidence, traverses graph-structured scientific knowledge, and synthesizes this information into hypothesized substitute materials. However, its scope centered primarily on material-hypothesis generation and did not explicitly assess candidates against downstream experimental and processing constraints. In particular, the earlier workflow did not constrain proposals to lab-available materials, systematically evaluate manufacturability, or iteratively revise candidate generation using downstream manufacturability feedback. The present study therefore asks whether extending this KG-guided foundation into a hierarchical workflow comprising requirement analysis, grounded candidate generation, and manufacturability assessment can generate scientifically grounded substitute proposals while incorporating downstream experimental and processing constraints. These additions broaden the scope of the workflow rather than establishing a direct quantitative performance improvement over the previous implementation.

Building on this prior work, the present study extends
PFAS-substitution reasoning in three directions: explicit requirement analysis before candidate generation, grounding of candidate selection in a lab-available materials inventory, and closed-loop manufacturability assessment in which blocking constraints can inform subsequent candidate refinement. The framework combines these stages with subsystem-specific RAG and KG resources and persistent storage of intermediate results. We evaluate the resulting workflow through the THV substitution case study, reduced framework configurations, and repeated-run and targeted feedback analyses. In the present study, manufacturability refers to an evidence-supported assessment of laboratory processing plausibility and generation of a proposed process route rather than experimentally demonstrated fabrication. Together, these elements define the scope of the present framework and its evaluation for knowledge-grounded material substitution.

\section*{Results and Discussion}
\label{sec:results}

The framework formulates PFAS material substitution as a closed-loop, three-stage reasoning problem. The information flow and subsystem operations are defined below.

We represent the information flow through the framework using the following objects. $Q$ denotes the natural-language substitution query; $W$ is the set of required material properties; $H$ is the set of hard constraints extracted in System~1; $U_t$ is the cumulative set of constraints available at closed-loop iteration $t$; $S_t$ is the candidate proposed at iteration $t$; $E_t$ is the synthesized supporting evidence for that candidate; $P_t$ is a proposed processing route; $C_t$ denotes the structured feedback returned by System~3 when a candidate is blocked, comprising the identified blocking constraints together with the associated feedback used to guide subsequent candidate generation, and $R_t$ denotes the record of candidates rejected during candidate generation.

The three subsystem operations can be summarized as
\begin{align}
(W,H) &= F_1(Q;K_1), \\
(S_t,E_t,R_{t+1}) &= F_2(Q,W,U_t,R_t;K_2,\mathcal{M}), \\
F_3(Q,W,U_t,S_t,E_t;K_3) &=
\begin{cases}
(\mathrm{manufacturable},P_t),\\
(\mathrm{blocked},C_t),
\end{cases}
\end{align}
where $K_1$, $K_2$, and $K_3$ denote the subsystem-specific retrieval and knowledge resources and $\mathcal{M}$ denotes the internal inventory of lab-available materials. The closed-loop constraint state is initialized as $U_1=H$. When System~3 returns a blocked verdict, its blocking constraints and associated feedback are incorporated into the constraint state before the next candidate-generation cycle, which can be represented schematically as
\begin{equation}
U_{t+1}=U_t\cup C_t.
\end{equation}
System~2 is then invoked again using the updated constraint state and rejection history. The loop terminates when System~3 assigns an internal manufacturable verdict, or the configured iteration limit is reached. These expressions define the input–output contracts and state transitions implemented by the workflow rather than closed-form physical models.

\begin{figure}[!b]
  \centering
  \includegraphics[width=\textwidth]{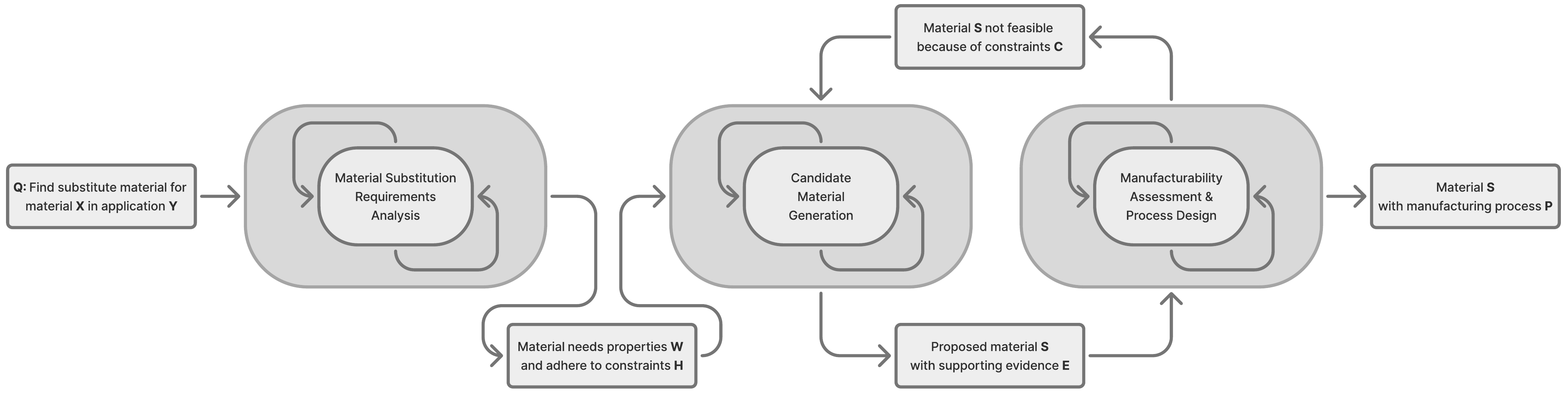}
  \caption{Overall architecture of the hierarchical multi-agent framework. High-level schematic of the proposed closed-loop hierarchical multi-agent system for PFAS material substitution. Dark gray elements indicate multi-agent systems, while light gray elements indicate intermediate results. From a natural-language query $Q$, the system first extracts required properties $W$ and hard constraints $H$, then identifies a grounded substitute candidate $S$ with supporting evidence $E$, and finally assesses lab-scale manufacturability to produce either a process recipe $P$ or blocking constraints $C$. All intermediate results are stored in persistent memory, enabling iterative refinement until a candidate passes the framework's internal feasibility assessment.}
  \label{fig:architecture}
\end{figure}

Figure~\ref{fig:architecture} summarizes the overall framework and
information flow across the three agentic systems.

\subsection*{Knowledge Sources for Hierarchical Multi-Agent Material Substitution}
\label{subsec:knowledgeSources}

The hierarchical multi-agent framework is supported by distinct knowledge sources tailored to the function of each subsystem. System 1, which infers the functional requirements of the PFAS-containing material to be replaced, is grounded in a curated corpus of PFAS research articles used for RAG \cite{stewart_graphagents_2026}. From this corpus, a domain-specific KG capturing relationships among materials, properties, and application contexts across PFAS-related documents is also constructed \cite{stewart_graphagents_2026}. We use this KG for graph-based retrieval-augmented generation (Graph-RAG).

System 2, which identifies substitute candidates from the extracted property requirements, uses two complementary KGs: one derived from a broad corpus of academic abstracts on material properties for cross-domain exploration \cite{stewart_graphagents_2026}, and one we derive from a large corpus of patents related to polymer science. Together with a database of lab-available materials, these sources enable the system to reason over connections between accessible materials and desired properties. The patent corpus and available-material specification sheets are additionally used as text sources for RAG.

\begin{figure}[!t]
  \centering
  \includegraphics[width=\textwidth]{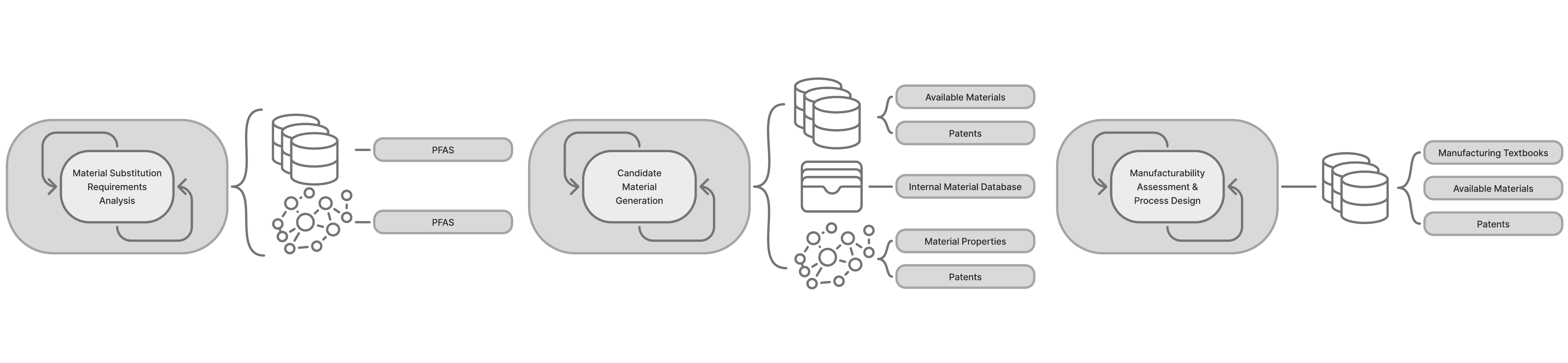}
  \caption{Knowledge sources used across the hierarchical multi-agent framework. Overview of the knowledge sources used across the three subsystems of the hierarchical multi-agent framework. System~1 is grounded in a curated PFAS research corpus and a domain-specific KG, System~2 uses complementary KGs, patent data, and a database of lab-available materials to identify substitute candidates, and System~3 draws on process-relevant textual sources for manufacturability assessment and process design. Here, the database-stack symbols denote text vector databases, the graph symbols denote KGs, and the folder symbol denotes the custom-built internal materials database.}
  \label{fig:knowledge_sources}
\end{figure}

System 3 focuses on manufacturability assessment and process design, and is therefore grounded in process-relevant textual sources, including polymer-science textbooks, patents, and available-material specification sheets. Unlike the earlier subsystems, System 3 uses text-based RAG only, because its purpose is not exploratory discovery but evidence-grounded assessment of processing feasibility. Figure~\ref{fig:knowledge_sources} summarizes the knowledge sources used across the three subsystems, illustrating how the framework couples each stage of the substitution task to a tailored knowledge representation matched to its specific reasoning objective.

\subsection*{Core Agent Roles in the Hierarchical Multi-Agent Framework}
\label{subsec:availableAgents}

The three subsystems in the hierarchical multi-agent system use the same four core agent roles: a Research Manager, Research Analyst, Research Scientist, and Research Assistant. These agents are reused across subsystems, but equipped with different knowledge sources depending on the task. The Research Manager is the primary LLM-based control agent for the research workflow. It generates research questions for retrieval, synthesizes answers from RAG evidence, proposes candidate materials using KG insights, and evaluates proposed materials through targeted validation queries. The Research Analyst is the retrieval-focused agent. It supports the RAG pipeline by querying vector databases and returning the most relevant document chunks. Unlike the LLM-based agents, it does not generate conclusions or synthesize answers; instead, it grounds downstream reasoning by supplying evidence. The Research Scientist is responsible for KG reasoning. The agent semantically maps input keywords to graph nodes and identifies connection paths linking conceptual requirements to material-relevant entities. The Research Assistant is an LLM-based extraction agent that translates narrative text into structured lists of keywords, used across the multi-agent system including the KG pipeline. By extracting material-property concepts and hard constraints, it provides the interface that enables downstream graph reasoning and candidate generation.

Together, these four agents constitute the reusable base layer of the hierarchical multi-agent framework. Each subsystem instantiates this shared agent set with different tools, retrieval sources, and decision objectives, while persistent storage enables intermediate results to be passed across stages and reused in subsequent iterations. In combination with the subsystem-level architectures and the higher-level closed-loop design, these agents form the operational foundation of the full hierarchical framework for PFAS material substitution.

\subsection*{Material Substitution Requirements Analysis}
\label{subsec:system1}

Because the substitution task concerns a PFAS-containing material in a specific application, the hierarchical multi-agent system first infers which material properties are functionally required before attempting to generate replacement candidates. This ordering is intended to ensure that candidate generation is conditioned on an explicit representation of the incumbent material’s role in the target use case. We hypothesize that storing these inferred requirements in persistent memory improves iterative search efficiency, since rejected candidates can be replaced without repeating the requirement-analysis stage. In this way, System 1 focuses on structured requirement inference, thereby constraining and informing the downstream search space.

A detailed schematic of the multi-agent pipeline for System~1, Material Substitution Requirements Analysis, is provided in Supplementary Fig.~S1. A Research Analyst first performs RAG over a curated PFAS literature corpus using the user query to retrieve relevant document chunks. A Research Manager, using the retrieved information, then decomposes the task into several focused research questions spanning distinct property dimensions of the substitution problem. For each question, a Research Analyst and Research Scientist perform dual-source retrieval from both the PFAS literature corpus and the PFAS KG derived from that corpus. The graph retrieval step complements text retrieval by exposing relational paths linking concepts across documents. The Research Manager evaluates whether the retrieved evidence sufficiently answers each question and synthesizes resulting answers to each question. Finally, a Research Assistant converts the resulting Question--Answer pairs into structured material-property requirements and hard constraints for downstream candidate generation.

These operations provide evidence-grounded checks on the
generated requirements rather than a formal guarantee of their correctness. The Research Manager assesses whether the retrieved text and graph evidence sufficiently addresses each generated research question before synthesis, and the resulting Question--Answer pairs are then converted into structured requirements and constraints. Because this final extraction includes LLM synthesis, individual requirements and numerical thresholds may still require independent source-level or experimental verification.

\subsubsection*{System 1 Results}

To illustrate the performance of MARS, we present the system's output for a case study focused on replacing THV (an elastomeric terpolymer of tetrafluoroethylene, hexafluoropropylene, and vinylidene fluoride) in applications requiring broad chemical resistance. The evaluation query, formulated by domain experts to reflect a representative PFAS substitution task, is presented in Text Box~\ref{textbox:system1_query}.

\begin{tcolorbox}[
  title={Evaluation Query},
  colback=gray!5,
  colframe=greenframe,
  fonttitle=\bfseries,
  fontupper=\small,
  breakable
]
Find a substitute material for PFAS material THV (elastomeric terpolymer of tetrafluoroethylene, hexafluoropropylene, and vinylidene fluoride) that has broad chemical resistance, including acids, bases, protic and aprotic solvents, oils, and salt solutions. Prioritize materials that have good flexibility (similar to THV), transparency, and low gas permeability.
\end{tcolorbox}
\noindent\begin{minipage}{\textwidth}

\captionof{boxes}{Evaluation query used for the THV substitution case study.}
\label{textbox:system1_query}
\end{minipage}

First, we present the output of System~1, Material Substitution Requirements Analysis. Although the system stores intermediate artifacts throughout the pipeline, including retrieved text, retrieved KG relationships, and intermediate LLM-generated responses, we report here only the final System 1 output, namely the extracted Required Material Properties and Hard Constraints that the substitute material should satisfy. The full model outputs are provided in the Supplementary Information. Text Box~\ref{textbox:system1_output} presents the System 1 output for the THV substitution query. 

The numerical thresholds in the final output were not specified manually in the evaluation query. A post-hoc provenance trace of the reported run (Supplementary Table~S3) showed that none of the 28 numerical-value or test-condition entries in Text Box~\ref{textbox:system1_output} appeared in the query, retrieved passages, or knowledge-graph paths. Instead, these values were introduced during the LLM synthesis and extraction stages of System~1. The trace also examined the sources of the 25 distinct qualitative material-property and hard-constraint requirements. Of these, 15 originated in the evaluation query, six were supported by retrieved passages, and four were introduced through examples included in the extraction prompts provided to the LLM agents. Thus, although most of the qualitative requirements were grounded in the query or retrieved evidence, the associated numerical thresholds were model-generated design assumptions and should be interpreted cautiously. Future work will focus on linking each numerical threshold directly to verifiable source evidence. Further details are provided in Supplementary Table~S3.

\begin{tcolorbox}[
  title={System~1 Output},
  colback=gray!5,
  colframe=greenframe,
  fonttitle=\bfseries,
  fontupper=\footnotesize,
  breakable
]

\textbf{Properties}

\vspace{0.3em}

\begin{itemize}[leftmargin=*,noitemsep,topsep=2pt]
    \item Key material properties \& performance requirements for a non-PFAS substitute to THV
    \item PFAS-free composition -- no tetrafluoroethylene, hexafluoropropylene, or vinylidene fluoride units
    \item Elastomeric flexibility -- room-temperature modulus $\approx$ THV ($\approx$ 0.5--1 MPa) with elongation at break $>$ 200 \%
    \item Optical clarity / transparency -- light transmittance $>$ 90 \% in the visible range for films $\leq$ 0.2 mm thick
    \item Low gas permeability -- O$_2$/H$_2$O vapor transmission rates $<$ 10 cm$^3$/m$^2$$\cdot$day (or equivalent)
    \item Broad chemical resistance -- withstand strong acids, bases, protic and aprotic solvents without degradation or swelling
    \item Oil repellency / oleophobicity -- contact angle $>$ 110\textdegree\ for common oils; minimal oil absorption over 24 h exposure
    \item Salt-solution (high salinity) resistance -- no loss of mechanical integrity or barrier performance after prolonged immersion in NaCl solutions up to 3 M
    \item Thermal stability -- T$_g$ $\geq$ 50 \textdegree C and degradation temperature $>$ 200 \textdegree C for typical processing/usage conditions
    \item Mechanical toughness -- impact strength comparable to THV ($\approx$ 10--15 kJ/m$^2$)
    \item Durability / washability -- retain optical clarity, flexibility, and barrier properties after $\geq$ 50 laundering cycles or equivalent abrasion tests
    \item Regulatory compliance -- meet environmental, health, and safety standards for non-PFAS polymers (e.g., REACH, RoHS)
    \item Cost-effective manufacturability -- scalable extrusion/lamination processes with competitive material cost relative to THV
    \item Critical characteristics that must be replicated from the original THV
    \item Elastomeric terpolymer backbone -- flexible, low-density network providing soft hand feel
    \item Transparent fluoropolymer architecture -- high refractive index matching and minimal haze
    \item Low gas permeability inherent to fluorinated chains -- must be matched by alternative chemistry (e.g., dense polyimide or cross-linked EVOH/PE blends)
    \item Chemical resistance profile of PFAS -- resistance to aggressive chemicals that typically degrade conventional polymers
\end{itemize}

\vspace{0.8em}

\textbf{Constraints}

\vspace{0.3em}

\begin{itemize}[leftmargin=*,noitemsep,topsep=2pt]
    \item Hard constraints for the substitute material
    \item Acid resistance -- Must exhibit $\leq$ 5 \% weight loss after 24 h immersion in 10 \% HCl (pH $\approx$ 0) at 25 \textdegree C.
    \item Base resistance -- Must exhibit $\leq$ 5 \% weight loss after 24 h immersion in 10 \% NaOH (pH $\approx$ 14) at 25 \textdegree C.
    \item Protic-solvent compatibility -- Must show $\leq$ 5 \% swelling and $\leq$ 5 \% mass change after 48 h exposure to 50 \% ethanol (or equivalent protic solvent) at 25 \textdegree C.
    \item Aprotic-solvent compatibility -- Must show $\leq$ 5 \% swelling and $\leq$ 5 \% mass change after 48 h exposure to acetone or dichloromethane at 25 \textdegree C.
    \item Oil resistance -- Must exhibit $\leq$ 5 \% swelling and no measurable loss of mechanical properties after 7 days immersion in mineral oil (or vegetable oil) at 25 \textdegree C.
    \item Salt-solution resistance -- Must retain $\geq$ 95 \% of its original tensile strength after 30 days exposure to 10 \% NaCl solution at 25 \textdegree C.
    \item Flexibility (mechanical) -- Tensile modulus $\leq$ 200 MPa and elongation at break $\geq$ 300 \% at room temperature, matching the flexibility range of THV elastomer.
    \item Optical clarity -- Transmittance $\geq$ 80 \% at 550 nm for a film thickness $\leq$ 0.5 mm (ensuring transparency comparable to THV).
    \item Gas barrier performance -- Oxygen transmission rate $<$ 10 cm$^3$/m$^2$$\cdot$day and water-vapor transmission $<$ 1 g/m$^2$$\cdot$day for the same film thickness, matching or exceeding THV's low permeability.
    \item Hydrocarbon stability -- No measurable loss of tensile strength or modulus after 30 days immersion in a hydrocarbon (e.g., n-heptane) at 25 \textdegree C.
\end{itemize}

\end{tcolorbox}
\noindent\begin{minipage}{\textwidth}
\captionof{boxes}{Final System~1 output for the THV case study, shown as generated by MARS, without post-hoc refinement.}
\label{textbox:system1_output}
\end{minipage}

\subsection*{Grounded Material Candidate Generation}
\label{subsec:system2}

System 2, Grounded Material Candidate Generation, uses the structured material-property requirements and hard constraints extracted in System 1 to identify a substitute material for the target PFAS-containing material. Rather than searching an unconstrained space, candidate generation is grounded in a curated database of lab-available materials. This grounding is intended to improve experimental relevance by ensuring that proposed substitutes are not only theoretically plausible but also obtainable and testable within the target research environment. System 2 further incorporates an internal validation loop in which proposed candidates are critically evaluated, rejected when unsupported, and replaced in subsequent iterations. We hypothesize that by combining grounded search with persistent rejection tracking, the system reduces repeated failure modes and supports rejection-aware candidate refinement across successive iterations.

The detailed System~2 workflow is provided in Supplementary Fig.~S2. The system starts by constructing a task-specific \emph{Material-Informed Subgraph} from the Material Properties KG and the Patents KG. From the required material properties and the set of lab-available materials, a Research Scientist identifies relational paths in the KGs that connect the desired properties to experimentally accessible materials. The resulting task-specific graph is denoted as $G=(V_G,E_G)$, where $V_G$ contains the mapped material and property nodes and $E_G$ contains the relational edges connecting them. The graph is stored in persistent memory and serves as the relational substrate for candidate generation. Then, an iterative Candidate Proposal Loop is initiated. Using the Material-Informed Subgraph, together with the hard constraints and record of previously rejected materials, a Research Manager proposes a candidate material for the target application. The Research Manager then checks whether the proposed material does not contain any PFAS, and if so generates validation questions to self-validate before proposing the candidate. For each subquestion, a Research Analyst retrieves evidence which the Research Manager uses to synthesize an answer to the question. Then, the Research Manager uses the synthesized answers for all subquestions to evaluate the feasibility of the material for the target application. If the candidate is found to be feasible, the system returns the proposed material together with synthesized supporting evidence. If not, the material is appended to a rejected-material set, together with the associated constraints. The loop repeats until a feasible candidate is identified or the rejection limit is reached.

\subsubsection*{System 2 Results}

\begin{figure}[!b]
  \centering
  \includegraphics[width=0.65\textwidth]{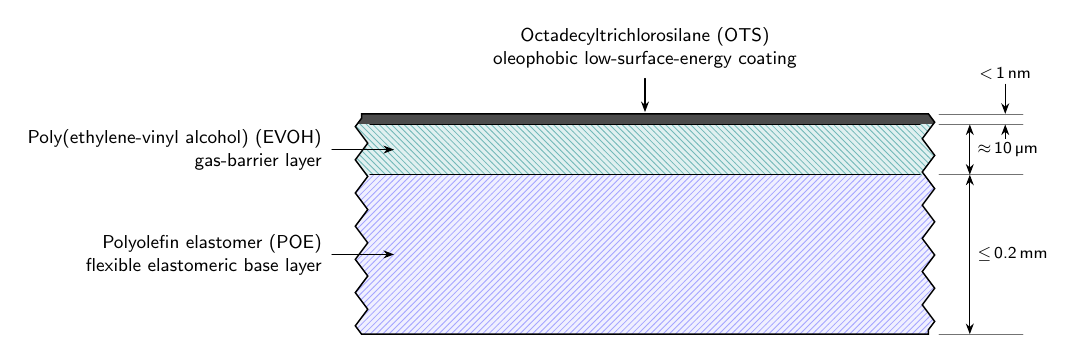}
  \caption{Schematic of the model-proposed substitute material. Schematic cross-section of the substitute material proposed by MARS for the THV case study, consisting of a polyolefin elastomer base, an EVOH gas-barrier layer, and an OTS oleophobic surface coating. Layer thicknesses are not drawn to scale. The schematic represents the model-proposed design; no physical sample was fabricated in the present study.}
  \label{fig:proposed_material}
\end{figure}

We present results from System 2, Grounded Material Candidate Generation, for the case study presented in Text Box~\ref{textbox:system1_query}, using the same MARS run that produced the System~1 results shown in Text Box~\ref{textbox:system1_output}. System~2 stores a range of intermediate artifacts throughout its execution, including the developed Material-Informed Subgraph, internally rejected candidates, retrieved information, and intermediate LLM-generated responses. Here, however, we report only the final System~2 output, namely the proposed substitute candidate together with its justification. Text Box~\ref{textbox:system2_output} presents the System~2 output for the THV substitution query. The statements in Text Box~\ref{textbox:system2_output} are shown as generated by the framework and have not been independently verified as measured properties of the proposed composite. They should therefore be interpreted as model-generated, evidence-informed candidate claims rather than experimentally validated material performance. Figure~\ref{fig:proposed_material} shows a schematic
cross-section of the proposed three-layer composite film.

\begin{tcolorbox}[
  title={System~2 Output},
  colback=gray!5,
  colframe=greenframe,
  fonttitle=\bfseries,
  fontupper=\footnotesize,
  breakable
]

\textbf{Final candidate}

\vspace{0.3em}

\noindent
ENGAGE Polyolefin Elastomer / EVOH Barrier / OTS Oleophobic Composite

\vspace{0.8em}

\textbf{Justification}

\vspace{0.3em}

\noindent
The base layer is the commercial polyolefin elastomer ``ENGAGE\texttrademark\ 030902 HEALTH+'' (ID: dow-chemical-company-engage-030902-health-polyolefin-elastomer) from the Available Materials Database. This polymer is a low‑density, cross‑linked polyolefin that exhibits room‑temperature modulus in the 0.5--1\,MPa range and elongation at break $>$\,200\,\%, satisfying the elastomeric flexibility requirement for THV replacement. Its optical clarity exceeds 90\,\% for films $\leq$\,0.2\,mm thick, meeting the transparency criterion.

A thin ($\approx$\,10\,\textmu m) layer of generic poly(ethylene‑vinyl alcohol) (EVOH) is laminated onto the elastomer. Knowledge Graph insight [3] and related entries show that EVOH provides exceptionally low gas permeability ($<$\,0.05\,g\,m$^{-2}$\,day$^{-1}$), which translates to O$_2$/H$_2$O vapor transmission rates well below 10\,cm$^3$\,m$^{-2}$\,day, fulfilling the barrier requirement. EVOH also offers resistance to a broad spectrum of organic solvents and acids/bases when used as a barrier layer.

The outermost surface is treated with octadecyltrichlorosilane (OTS), a non‑PFAS silane that forms an ultrathin ($<$\,1\,nm) low‑surface‑energy coating. OTS is publicly available, not covered by the pending PDMS or INFULEX patents, and provides oil repellency with contact angles $>$\,110$^{\circ}$ for common oils while maintaining optical clarity.

This three‑layer stack (polyolefin elastomer / EVOH / OTS) meets all specified properties:

\begin{itemize}[leftmargin=1.5em,itemsep=1pt,topsep=2pt,parsep=0pt]
    \item \textbf{Flexibility}: modulus $\approx$\,0.5--1\,MPa, elongation $>$\,200\,\% from the ENGAGE base.
    \item \textbf{Transparency}: $>$\,90\,\% visible transmittance; thin EVOH and OTS layers are optically transparent.
    \item \textbf{Barrier}: EVOH layer gives gas permeability $<$\,10\,cm$^3$\,m$^{-2}$\,day.
    \item \textbf{Chemical resistance}: EVOH barrier protects against acids, bases, protic and aprotic solvents; the elastomer itself is resistant to many organic chemicals.
    \item \textbf{Oil repellency}: OTS coating yields $>$\,110$^{\circ}$ oil contact angle with minimal absorption.
    \item \textbf{Salt‑solution stability}: EVOH's dense network resists swelling in high‑salinity environments; the elastomer remains mechanically intact.
    \item \textbf{Thermal stability}: ENGAGE has Tg $\approx$\,50\,\textdegree C and degrades above 200\,\textdegree C; EVOH is stable up to $\sim$\,120\,\textdegree C, ensuring overall thermal robustness.
    \item \textbf{Impact toughness}: Polyolefin elastomers are known for high impact strength ($\sim$10--15\,kJ\,m$^{-2}$).
    \item \textbf{Durability / washability}: OTS and EVOH layers retain properties after $\geq$\,50 laundering cycles; the elastomer's low modulus prevents brittleness.
    \item \textbf{Regulatory compliance}: All components are non‑PFAS, meeting REACH/RoHS criteria.
    \item \textbf{Manufacturability \& cost}: Each layer can be produced by standard extrusion or lamination processes; materials are commercially available and competitively priced relative to THV.
\end{itemize}

Thus the ENGAGE\,/\,EVOH\,/\,OTS composite provides a PFAS‑free, elastomeric, transparent, chemically resistant, low‑permeability, oleophobic, salt‑stable, thermally robust, impact‑tough, durable, and cost‑effective alternative to THV.

\end{tcolorbox}
\noindent\begin{minipage}{\textwidth}
\captionof{boxes}{Final System~2 output for the THV case study, shown as generated by MARS, without post-hoc refinement.}
\label{textbox:system2_output}
\end{minipage}

\subsection*{Manufacturability Assessment and Process Design}
\label{subsec:system3}

Once System 2 proposes a candidate material, the framework proceeds to System 3, Manufacturability Assessment and Process Design. This stage is necessary because a candidate may be well supported by the literature yet still be impractical to synthesize at lab scale, particularly when the proposed substitute is a composite or formulation assembled from evidence distributed across patents and academic sources. We hypothesize that evaluating manufacturability using retrieved process knowledge enables the framework to distinguish candidates for which a plausible lab-scale processing route can be supported by retrieved evidence from those for which blocking constraints are identified.

The detailed System~3 workflow is provided in Supplementary Fig.~S3. Manufacturability assessment begins with process-focused retrieval, where a Research Manager generates search queries to learn more about the lab-scale synthesis of the proposed candidate and, when a composite, its constituent materials. A Research Analyst then retrieves evidence from a curated collection of textbooks, patents, and the lab-available materials specification sheets. Using this retrieved information, the Research Manager generates candidate-specific manufacturing feasibility questions and answers them through targeted retrieval (Research Analyst) and synthesis (Research Manager). These Question–Answer pairs are then aggregated into a structured feasibility assessment in which the Research Manager assigns an evidence-based internal verdict of \emph{manufacturable} or \emph{blocked}. If the candidate is deemed manufacturable, the system synthesizes a high-level manufacturing process recipe. Otherwise, the deduced blocking constraints are passed back to System~2 to guide a revised candidate search. The effect of this feedback loop is analyzed separately in Supplementary Information S1.6.

\subsubsection*{System 3 Results}

We present the final output of System~3, Manufacturability Assessment and Process Design, using the same MARS run that produced the System~1 and System~2 results shown in Text Boxes~\ref{textbox:system1_output} and~\ref{textbox:system2_output}. System~3 stores a range of intermediate artifacts throughout its execution, including retrieved information and intermediate LLM-generated responses. Here, however, we report only the final System~3 output, namely its internal manufacturability verdict together with a proposed high-level processing route for the candidate. Text Box~\ref{textbox:system3_output} presents the System~3 output for the THV substitution query. The \emph{manufacturable} status in Text Box~\ref{textbox:system3_output} is the framework's internal evidence-based assessment rather than an experimentally validated manufacturing result. The accompanying process recipe is therefore interpreted as a proposed processing route that requires independent technical and experimental verification.

\begin{tcolorbox}[
  title={System~3 Output},
  colback=gray!5,
  colframe=greenframe,
  fonttitle=\bfseries,
  fontupper=\footnotesize,
  breakable
]

\textbf{Status:} Manufacturable

\vspace{0.8em}

\textbf{Process recipe (summary)}

\vspace{0.3em}

\noindent
\textbf{Step 1 Description:} Raw material receipt, inspection and storage.

\textbf{Conditions:} temperature: $\leq$25 \textdegree C; humidity: $<$30\,\%; storage time before use: $\leq$48\,h

\textbf{Equipment:} Material handling \& storage (dry, temperature-controlled bins)

\textbf{Inputs}
\begin{itemize}
    \item ENGAGE\texttrademark\ 030902 HEALTH+\texttrademark\ Polyolefin Elastomer pellets
    \item EVOH copolymer (e.g., EVAL\textsuperscript{\textregistered} SP)
    \item Optional compatibilizer or plasticizer
\end{itemize}

\textbf{Step 2 Description:} Drying of polyolefin elastomer and EVOH to remove moisture.

\textbf{Conditions:} temperature: 80--100 \textdegree C; time: 4\,h; atmosphere: dry air or N$_2$

\textbf{Equipment:} Industrial drying oven ($\geq$10\,kg capacity)

\vspace{0.3em}

\textbf{Step 3 Description:} Compounding in twin-screw extruder.

\textbf{Conditions:} barrel temperature profile: zone 1: 180 \textdegree C; zone 2: 190 \textdegree C; zone 3: 200 \textdegree C; zone 4: 210 \textdegree C; screw speed: 200\,rpm; feed rate: $\approx$1.5\,kg/h per screw; shear rate: 500--1000\,s$^{-1}$

\textbf{Equipment:} Co-rotating twin-screw extruder, 10--20\,mm screw diameter, stainless steel barrel with heat-resistant liner

\textbf{Inputs}
\begin{itemize}
    \item Dried polyolefin elastomer
    \item EVOH copolymer
\end{itemize}

\textbf{Step 4 Description:} Coextrusion of multilayer film.

\textbf{Conditions:} die temperature (EVA layer): 190--210 \textdegree C; die temperature (EVOH layer): 150--170 \textdegree C; shear rate: 10--1000\,s$^{-1}$; cooling roll temperature: 20--30 \textdegree C

\textbf{Equipment:} Three-screw coextruder with separate barrel zones, chill roll set, film die (e.g., 50\,mm width)

\textbf{Inputs}
\begin{itemize}
    \item Compounded EVA/EVOH melt from step 3
\end{itemize}

\textbf{Step 5 Description:} Film cutting and sizing.

\textbf{Conditions:} cutting speed: $\approx$10--20\,m/min; tolerance: $\pm$0.1\,mm

\textbf{Equipment:} Knife line cutter or rotary blade cutter, 10\,kg capacity

\vspace{0.3em}

\textbf{Step 6 Description:} Oleophobic surface treatment via OTS silane dip-coating.

\textbf{Conditions:} dip time: 60\,s; rinse solvent: isopropanol; cure temperature: 120 \textdegree C; cure time: 10\,min

\textbf{Equipment:} Industrial dip-coater, drying oven ($\leq$200 \textdegree C)

\textbf{Inputs}
\begin{itemize}
    \item Film strip from step 5
    \item 2\,\% OTS silane solution in isopropanol
\end{itemize}

\textbf{Step 7 Description:} Cooling and optional annealing to relieve residual stresses.

\textbf{Conditions:} cooling rate: to room temperature over 30\,min; anneal temperature: 80 \textdegree C; anneal time: 30\,min

\textbf{Equipment:} Temperature-controlled chamber or oven

\vspace{0.3em}

\textbf{Step 8 Description:} Quality control and sampling.

\textbf{Equipment:} Laboratory test rigs for MFI, gas barrier, contact-angle goniometer

\vspace{0.3em}

\textbf{Step 9 Description:} Packaging and storage.

\textbf{Conditions:} packaging material: moisture-proof film bags; storage temperature: $\leq$25 \textdegree C; humidity: $<$30\,\%

\textbf{Equipment:} Bulk packaging line, sealing machine

\end{tcolorbox}
\noindent\begin{minipage}{\textwidth}
\captionof{boxes}{Final System~3 output for the THV case study, shown as generated by MARS, without post-hoc refinement.}
\label{textbox:system3_output}
\end{minipage}

\subsection*{Closed Loop Iteration}
\label{subsec:closedLoop}

We introduce a closed-loop iterative refinement framework for material substitution, in which persistent memory accumulates task-relevant knowledge throughout the search process. This progressively enriched memory supports the multi-agent system's internal reasoning, critique, and iterative refinement. Within this framework, System~3 evaluates manufacturability and, upon rejection, compiles the set of blocking constraints and a feedback string, which are passed back to System~2's Candidate Proposal Loop for further refinement. System~2 proceeds with proposing a revised candidate as previously described. This loop continues until System~3 approves a candidate or the maximum number of System~2–System~3 iterations are reached. We hypothesize that this decomposition is designed to improve controllability and interpretability while allowing blocking constraints to inform subsequent candidate generation.

\begin{figure}[!t]
  \centering
  \includegraphics[width=0.9\textwidth]{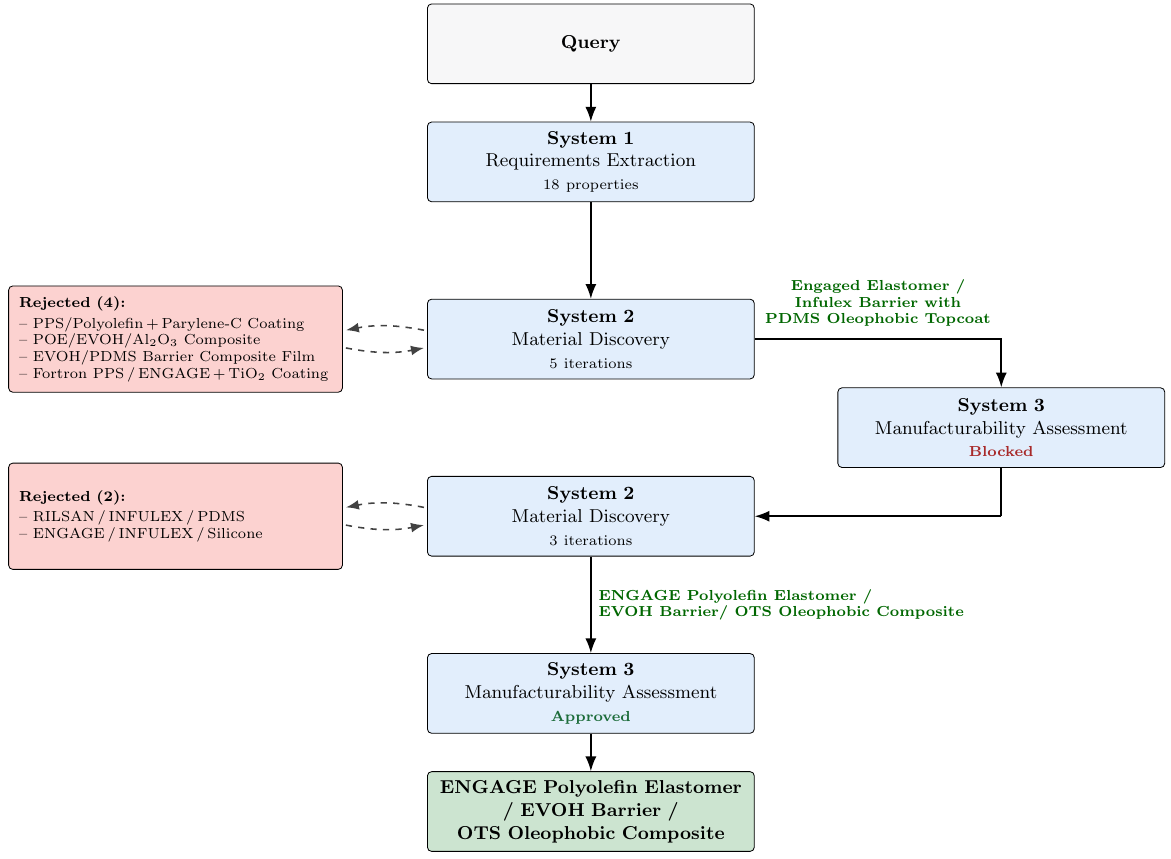}
  \caption{Closed-loop candidate refinement in the THV case study. Rejection-aware search trace for the THV substitution query, illustrating closed-loop interaction between System~2 (material feasibility screening) and System~3 (manufacturability assessment). Material substitution requirements produced by System~1 are passed to System~2, which iteratively generates and screens candidates, rejecting those that violate hard constraints, for example by containing PFAS. Candidates that pass System~2 screening are forwarded to System~3 for manufacturability evaluation. When System~3 rejects a candidate, structured feedback is returned to System~2, which initiates another round of candidate generation and screening. Solid arrows indicate the main workflow, while dashed arrows indicate internal rejections within System~2. The trace terminates when System~3 assigns a candidate the \emph{manufacturable} verdict.}
  \label{fig:closed_loop_trace}
\end{figure}

Figure~\ref{fig:closed_loop_trace} illustrates the closed-loop refinement trajectory of the hierarchical system corresponding to the results presented in Text Boxes~\ref{textbox:system1_output}--\ref{textbox:system3_output}. Rather than only depicting the abstract architecture, the figure traces how the hierarchical system moves from the System~1 requirements to successive candidate proposals, while explicitly recording both internal rejections within System~2 and cross-system rejection feedback from System~3. The example shows that the search does not proceed linearly: Following System~1's material requirements analysis, System~2 first rules out several internally generated candidates before proposing an ENGAGE polyolefin elastomer with an INFULEX barrier and a polydimethylsiloxane (PDMS) oleophobic topcoat. System~3 does not accept that stack as a viable path forward: it returns a blocked assessment associated with intellectual-property constraints, and that feedback is propagated to a second round of candidate generation. In the next iteration, System~2 rejects additional candidates that conflict with the propagated constraints before proposing an ENGAGE polyolefin elastomer/EVOH barrier/OTS oleophobic composite, to which System~3 assigns a \emph{manufacturable} verdict and for which it supplies a high-level processing procedure. The figure highlights how the closed-loop design narrows the candidate space through staged rejection and constraint feedback, while persistent memory lets the system accumulate task-specific knowledge across iterations and ultimately steer toward a candidate that passes the framework's evidence-based screening criteria.

\subsection*{Evaluation}
\label{subsec:evaluation}

To evaluate the proposed framework, we conduct an ablation study comparing MARS against five alternative configurations. Three reduced configurations use the same base model as MARS (\texttt{gpt-oss-20b}~\cite{openai_gpt-oss-120b_2025}, with the hosting configuration described under Experimental Setup in Materials and Methods): (i) a single-call LLM baseline without retrieval or graph grounding (Ablation~1); (ii) a retrieval-augmented single-call baseline in which the RAG corpora and knowledge graphs are queried once using the raw substitution query and the retrieved information is provided in a single prompt (Ablation~2); and (iii) a three-stage LLM pipeline that mirrors MARS' subsystem decomposition but excludes retrieval, graph grounding, and iterative refinement (Ablation~3). Two additional configurations use the closed-source \texttt{gpt-5.6-sol} model~\cite{noauthor_gpt-56_nodate}: a single-call baseline analogous to Ablation~1 and the full MARS framework with \texttt{gpt-5.6-sol} as the backbone of every agent. Together, these configurations provide matched single-call and full-framework comparisons at both backbone scales, allowing framework-level gains to be distinguished from improvements due to the underlying model. Table~\ref{tab:configs} summarizes the architectural differences among the six configurations. Ablation~2 represents static retrieval grounding rather than MARS' interactive RAG and Graph-RAG workflow, whereas Ablation~3 isolates subsystem-level task decomposition without grounding or iterative refinement. Each configuration is evaluated over five separate runs, and reported scores are averaged across these.

\begin{table}[!t]
\centering
\caption{Summary of the framework configurations evaluated in the ablation study. MAS denotes multi-agent system, RAG/KG denotes RAG/Graph-RAG integration, and IR denotes iterative refinement based on internal critique.}
\label{tab:configs}
\begin{tabularx}{\textwidth}{
  l
  >{\centering\arraybackslash}p{0.06\textwidth}
  >{\centering\arraybackslash}p{0.07\textwidth}
  >{\centering\arraybackslash}p{0.06\textwidth}
  >{\raggedright\arraybackslash}X
}
\toprule
\noalign{\vskip 0.2ex}
\textbf{Configuration} & \textbf{MAS} & \textbf{RAG/KG} & \textbf{IR} & \textbf{Description} \\
\midrule

MARS &
\checkmark & \checkmark & \checkmark &
\descbox{Full framework using \texttt{gpt-oss-20b}, combining multi-agent decomposition, RAG/Graph-RAG grounding, and closed-loop iterative refinement.} \\

MARS (\texttt{gpt-5.6-sol}) &
\checkmark & \checkmark & \checkmark &
\descbox{Full MARS framework with \texttt{gpt-5.6-sol} as the backbone of all agents.} \\

Ablation 3 &
\checkmark & $\times$ & $\times$ &
\descbox{Three-stage \texttt{gpt-oss-20b} pipeline with subsystem-level decomposition but without RAG/Graph-RAG or iterative refinement.} \\

Ablation 2 &
$\times$ & \checkmark & $\times$ &
\descbox{Single-call \texttt{gpt-oss-20b} baseline with static RAG/Graph-RAG grounding but without iterative refinement.} \\

Ablation 1 &
$\times$ & $\times$ & $\times$ &
\descbox{Single-call \texttt{gpt-oss-20b} baseline without retrieval, graph grounding, or iterative refinement.} \\

Ablation 1 (\texttt{gpt-5.6-sol}) &
$\times$ & $\times$ & $\times$ &
\descbox{Single-call \texttt{gpt-5.6-sol} baseline without retrieval, graph grounding, or iterative refinement.} \\

\bottomrule
\end{tabularx}
\end{table}

To ensure a fair comparison, we use a blind, pooled evaluation protocol. For each seed, every configuration produces the same three standardized subsystem reports using an identical section structure. Sections that are unavailable by design for a reduced configuration are explicitly marked as absent. Within each seed and subsystem, the six reports are anonymized, shuffled, and evaluated together in a single judge call (judge model, \texttt{gpt-6-astra} \cite{noauthor_gpt-6_nodate}), ensuring a common scoring context across configurations before the labels are mapped back to their identities.

Each subsystem output is evaluated on four criteria, yielding twelve criterion-level scores per configuration, using a 0--10 scale in increments of 0.5. The criteria target the capabilities associated with each subsystem: System~1 is evaluated on completeness, scientific correctness, evidence grounding, and investigation quality; System~2 on requirement alignment, adversarial verification, candidate anchoring, and adaptive reasoning; and System~3 on manufacturing route coherence, operational practicality, material-process compatibility, and feasibility-verdict grounding. The criterion-level judge questions are provided in S1.10, with the complete scoring rubrics given in the Supplementary Information. Among the twelve, five criteria correspond to capabilities that single-pass configurations lack by construction. These are evidence grounding, investigation quality, adversarial verification, adaptive reasoning, and feasibility-verdict grounding. Therefore, these criteria are assigned a null score that contributes zero to the configuration mean. In contrast, when an available MARS capability is not exercised in a particular run, such as adaptive reasoning when the first candidate is accepted, that criterion is excluded from the run-level mean rather than scored as zero. This convention means that the overall score should be interpreted as a measure of both capability coverage and execution quality, rather than execution quality alone. To support verifiable candidate anchoring, each System~2 report also includes the internal-database record, including material identity and datasheet properties, for every proposed component matched to the database.

Motivated by recent evidence that LLMs can perform strongly on scientific long-context understanding \cite{cui_curie_2025} and may provide judgments aligned with those of human evaluators \cite{han_judges_2025}, we use an LLM-as-a-judge protocol as a scalable and standardized evaluation mechanism. \texttt{gpt-6-astra}~\cite{noauthor_gpt-6_nodate} serves as the primary judge and is selected from a different model generation than the evaluated backbones to reduce potential self-preference. The judge receives the anonymized framework outputs and assigns numerical scores together with brief written rationales. Further details of the evaluation protocol are provided under Evaluation Protocol in Materials and Methods, while the complete evaluation framework is provided in the Supplementary Information.

None of these reduced configurations is intended to reproduce the previously reported GraphAgents framework~\cite{stewart_graphagents_2026}. Accordingly, the present ablation study should not be interpreted as a direct quantitative comparison with that earlier implementation.

\subsubsection*{Evaluation Results}

\begin{figure}[!t]
  \centering
  \includegraphics[height=0.8\textwidth]{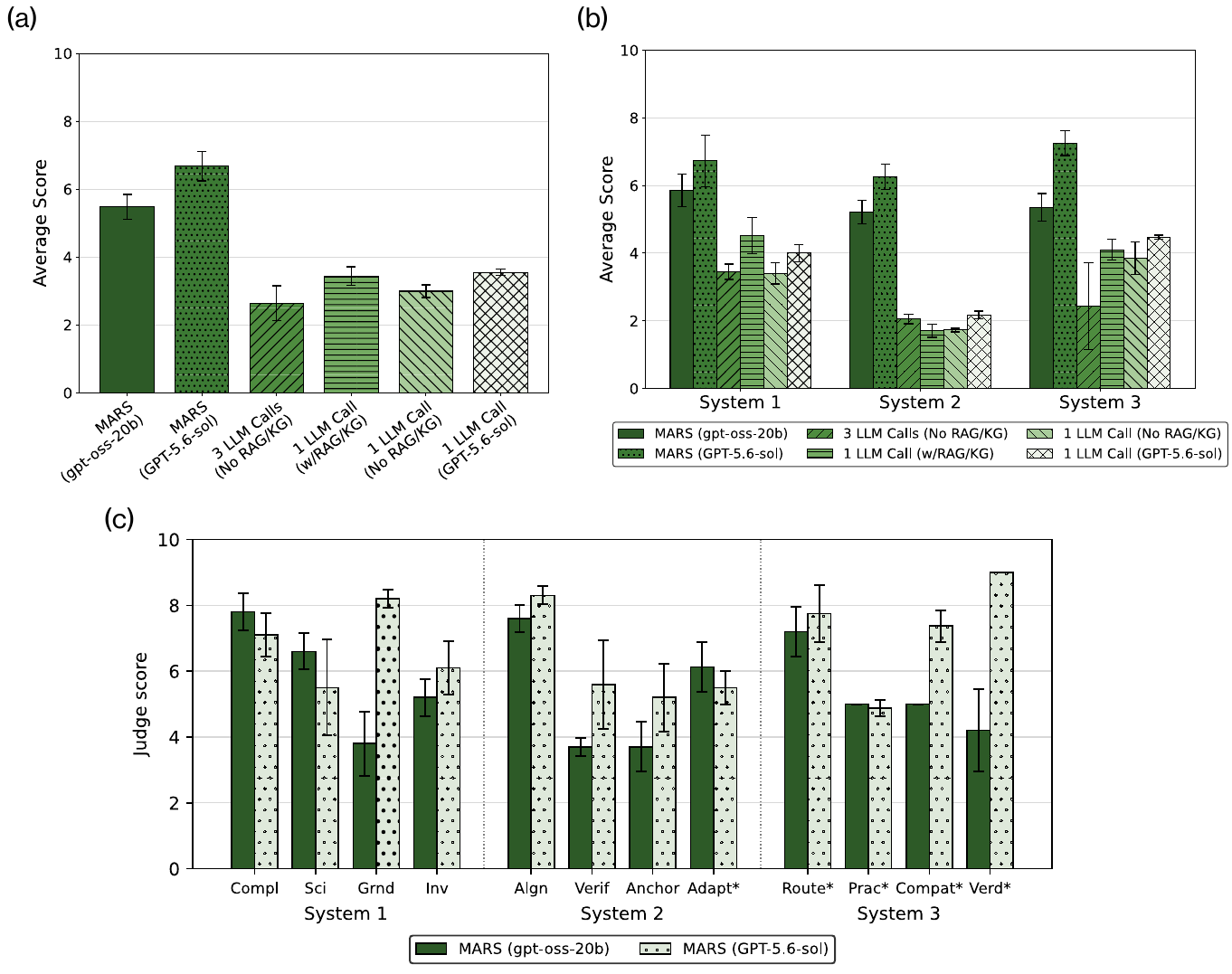}
  \caption{Evaluation performance across framework configurations. Performance comparison across the evaluated framework configurations.(a) Overall average score across the twelve evaluation criteria for all six configurations. (b) Subsystem-level scores for Systems~1--3, corresponding to requirement analysis, candidate generation, and manufacturability assessment, respectively. (c) Criterion-level comparison of the two MARS backbones across the three subsystems. Criteria are completeness (Compl), scientific correctness (Sci), evidence grounding (Grnd), investigation quality (Inv), requirement alignment (Algn), adversarial verification (Verif), candidate anchoring (Anchor), adaptive reasoning (Adapt), route coherence (Route), operational practicality (Prac), material-process compatibility (Compat), and feasibility-verdict grounding (Verd). Bars show mean judge scores across repeated runs, with error bars indicating standard deviations. For MARS (\texttt{gpt-5.6-sol}), the overall score in panel~(a) and the System~3 score in panel~(b) are calculated over four route-producing runs because one run terminated before System~3 was invoked. In panel~(c), asterisks denote criteria for which one MARS run was unavailable or the corresponding capability was unexercised and was therefore excluded from the criterion mean. \texttt{gpt-6-astra} is used as the LLM judge.}
  \label{fig:mars_ablation_combine}
\end{figure}

Figure~\ref{fig:mars_ablation_combine}a,b compares overall and subsystem-level performance across the six configurations, while Figure~\ref{fig:mars_ablation_combine}c resolves the criterion-level performance of the two MARS backbones; Table~\ref{tab:scores} summarizes the subsystem and overall averages. Both MARS runs achieve the highest score for all three subsystems. The largest separation occurs in System~2 (candidate generation), where MARS scores 5.22 with \texttt{gpt-oss-20b} and 6.26 with \texttt{gpt-5.6-sol}, whereas all reduced baselines score 2.17 or lower. Requirement alignment is the only System~2 criterion that a single-call baseline can match closely (Figure S4); adversarial verification and adaptive reasoning are absent by construction, while candidate anchoring remains weak without an explicit database-grounding mechanism. Notably, the retrieval-augmented single-call baseline obtains the lowest System~2 mean (1.70), indicating that static retrieval alone does not ensure that generated candidates remain anchored to verifiable material records.

The matched backbone comparisons reduce the confounding effect of backbone choice. With \texttt{gpt-5.6-sol} held fixed, the full MARS framework improves the overall score by approximately 91\% relative to the \texttt{gpt-5.6-sol} single-call baseline (6.78 vs.~3.55). A similar trend is observed with \texttt{gpt-oss-20b}: MARS improves the overall score by approximately 83\% relative to Ablation~1, the single-call baseline without RAG/KG or iterative refinement (5.47 vs.~2.99). These matched comparisons indicate that a substantial fraction of the performance gain comes from the framework architecture rather than from backbone capability alone.

The backbone model nevertheless remains important. Replacing \texttt{gpt-oss-20b} with \texttt{gpt-5.6-sol} within the full MARS framework produces the largest improvements in evidence-sensitive criteria. Evidence grounding increases by approximately 116\% (3.8 to 8.2), feasibility-verdict grounding by 114\% (4.2 to 9.0), adversarial verification by 51\% (3.7 to 5.6), and candidate anchoring by 41\% (3.7 to 5.2) (Figure~\ref{fig:mars_ablation_combine}c). In contrast, several other criteria change more modestly. This suggests that the stronger backbone primarily improves how effectively MARS interprets and uses retrieved evidence within the structured workflow.

The ablations also clarify the role of RAG and knowledge-graph grounding. Adding static RAG/KG to the single-call baseline (Ablation~2 versus Ablation~1) improves the overall score by approximately 15\% (3.44 vs.~2.99), with the largest gain in System~1, where the score increases by 33\% (4.53 vs.~3.40). However, static retrieval alone provides little benefit for System~2 candidate generation (1.70 vs.~1.73), showing that access to retrieved information is not sufficient by itself to ensure grounded material selection. The full MARS framework, which integrates RAG/KG throughout the multi-stage reasoning and verification process, improves the overall score by 59\% over Ablation~2 with the same \texttt{gpt-oss-20b} backbone (5.47 vs.~3.44). This indicates that the main benefit of RAG/KG arises when retrieved knowledge is actively used to guide requirement development, candidate verification, and downstream refinement rather than simply appended to a single prompt.

\begin{table}[t]
\centering
\caption{Evaluation scores for each configuration on a 0--10 scale. Subsystem scores are reported as mean $\pm$ standard deviation over five runs. Structurally absent capabilities score zero; unexercised MARS criteria are excluded. $^{*}$System~3 is averaged over four route-producing runs.}
\label{tab:scores}
\begin{tabular}{lcccc}
\toprule
\textbf{Pipeline} & \textbf{System 1} & \textbf{System 2} & \textbf{System 3} & \textbf{Total} \\
\midrule
MARS (\texttt{gpt-5.6-sol})       & $6.72 \pm 0.77$ & $6.26 \pm 0.38$ & $7.25 \pm 0.37^{*}$ & 6.78 \\
MARS (\texttt{gpt-oss-20b})       & $5.85 \pm 0.48$ & $5.22 \pm 0.35$ & $5.35 \pm 0.41$ & 5.47 \\
1 LLM Call (\texttt{gpt-5.6-sol}) & $4.00 \pm 0.25$ & $2.17 \pm 0.11$ & $4.47 \pm 0.06$ & 3.55 \\
1 LLM Call (w/ RAG/KG)   & $4.53 \pm 0.53$ & $1.70 \pm 0.19$ & $4.10 \pm 0.31$ & 3.44 \\
1 LLM Call (No RAG/KG)   & $3.40 \pm 0.31$ & $1.73 \pm 0.06$ & $3.85 \pm 0.48$ & 2.99 \\
3 LLM Calls (No RAG/KG)  & $3.45 \pm 0.23$ & $2.05 \pm 0.14$ & $2.42 \pm 1.28$ & 2.64 \\
\bottomrule
\end{tabular}
\end{table}

\subsection*{Isolating the System~3 to System~2 feedback}
\label{feedback}

The ablation study in does not directly isolate the contribution of the closed-loop feedback from System~3 to System~2. We therefore performed targeted experiments in which the System~3 blocking constraints were either propagated back to System~2 or withheld, while keeping the preceding trajectory fixed. All outputs were evaluated using \texttt{gpt-6-astra} under the final rubrics in Supplementary Section S1.8.

\begin{figure}[htbp]
\centering
\includegraphics[width=\textwidth]{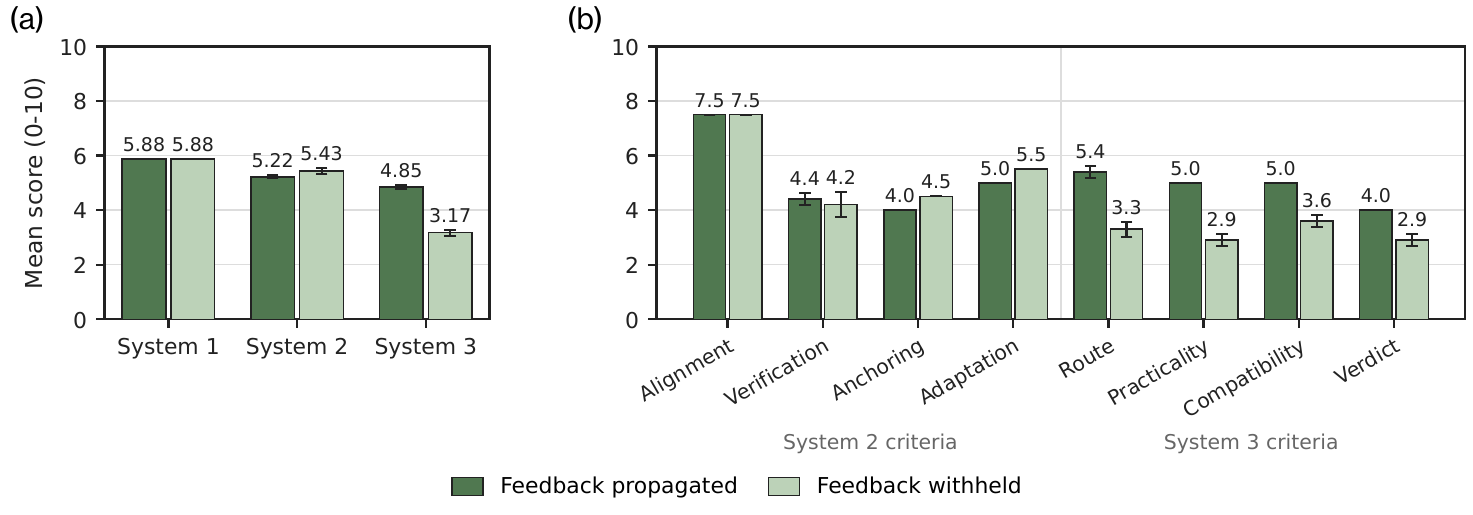}
\caption{System 2 to System 3 feedback ablation. Isolation of System~3-to-System~2 feedback on the reported THV query. (a) Mean subsystem scores when System~3 blocking constraints are propagated to System~2 or withheld. System~1 is identical in both conditions. (b) Criterion-level comparison for Systems~2 and~3. Feedback produces only modest changes in System~2 scores but substantially improves all System~3 feasibility-related criteria. Bars show means over five judge repetitions; error bars indicate standard deviations.}
\label{fig:feedback-fixed}
\end{figure}

The reported THV trajectory (Text Boxes \ref{textbox:system1_output}-\ref{textbox:system3_output} and Fig. \ref{fig:closed_loop_trace}) contains two closed-loop iterations. In the \emph{feedback propagated} condition, the blocking constraints returned by System~3 after the first iteration are supplied to System~2, which generates a revised candidate that subsequently receives a manufacturable verdict. In the \emph{feedback withheld} condition, this feedback is not supplied and the trajectory is evaluated at the blocked first iteration. System~1 is identical in both conditions. Each System~2 and System~3 report was judged five times.

As shown in Fig.~\ref{fig:feedback-fixed}a, the shared System~1 output receives the same mean score of 5.88 in both conditions. System~2 scores are also similar, with a mean of $5.22$ when feedback is propagated and $5.43$ when it is withheld. At the criterion level (Fig.~\ref{fig:feedback-fixed}b), alignment is identical (7.5 vs.~7.5), verification is slightly higher with feedback (4.4 vs.~4.2), while anchoring (4.0 vs.~4.5) and adaptation (5.0 vs.~5.5) are modestly lower. Thus, the feedback does not produce a clear improvement in the judged quality of the System~2 candidate-generation output.

The strongest effect appears in System~3. Propagating the blocking constraints increases the mean System~3 score from 3.17 to 4.85. All four System~3 criteria improve: manufacturing route increases from 3.3 to 5.4, practicality from 2.9 to 5.0, material--process compatibility from 3.6 to 5.0, and verdict grounding from 2.9 to 4.0. These results indicate that, for this fixed trajectory, the principal effect of the closed-loop feedback is not to produce a candidate that receives a higher System~2 score, but to redirect the search toward a candidate for which a more coherent and feasible downstream processing route can be supported. It should be noted, however, that the System~3-to-System~2 feedback loop was triggered only rarely across the repeated-seed experiments; the analysis presented here therefore reflects a single observed feedback event rather than a broad statistical characterization of loop behavior.

\subsection*{Discussion}
\label{subsec:discussion}

The ablation study indicates that both workflow architecture and backbone capability contribute to MARS performance. The matched-backbone comparisons show that applying the full MARS workflow improves performance relative to direct generation even when the underlying language model is held fixed. At the same time, the stronger \texttt{gpt-5.6-sol} backbone further improves the full framework, particularly on criteria that depend on interpreting and integrating retrieved evidence. These observations suggest that model capability and framework structure play complementary rather than interchangeable roles.

The results also clarify the role of retrieval and knowledge-graph grounding. Static retrieval provides useful external information, particularly during requirement analysis, but does not by itself produce the same gains as the full framework. In MARS, retrieved text and graph-structured knowledge are incorporated into dedicated stages for requirement development, candidate generation, validation, and downstream refinement. This distinction suggests that the utility of RAG/KG depends not only on access to external knowledge, but also on how that knowledge is organized and used within the reasoning process.

This architectural effect is particularly evident in System~2. Candidate generation in MARS is constrained by the requirements extracted in System~1, anchored to records in the lab-available materials database, evaluated through targeted evidence retrieval, and informed by the history of previously rejected candidates. Together, these mechanisms support candidate anchoring, adversarial verification, and rejection-aware adaptation. The comparatively large separation observed for System~2 therefore suggests that grounded material discovery benefits from treating candidate generation as an iterative search-and-verification problem rather than as a single generative step.

System~3 extends this principle from candidate plausibility to processing feasibility. Rather than serving only as a process-recipe generator, System~3 acts as a downstream decision gate that can reject a candidate and return blocking constraints to System~2. The closed-loop trajectory in Figure~\ref{fig:closed_loop_trace} illustrates how such feedback can redirect the subsequent search. The fail-closed \texttt{gpt-5.6-sol} run provides a second example: when none of the proposed candidates satisfied the framework's verification requirements within the configured search budget, the system terminated without generating a manufacturing route. This behavior is consistent with the intended role of the framework, in which unresolved evidence can prevent downstream progression rather than being ignored in order to force a final answer.

Compared with the earlier GraphAgents framework~\cite{stewart_graphagents_2026}, MARS broadens the workflow by introducing explicit requirement analysis, grounding candidate selection in a lab-available materials inventory, evaluating manufacturability, and returning downstream constraints to candidate generation. These changes should be interpreted as an extension in workflow scope rather than as a demonstrated quantitative improvement over GraphAgents, because the two studies use different architectures, output representations, and evaluation protocols. A controlled head-to-head comparison would be required to quantify the effect of the broader MARS architecture relative to the earlier implementation.

The quality of the framework's outputs is bounded by the breadth and relevance of the information available to it. Although MARS draws from multiple external corpora (KGs, text vector databases and materials databases), its search space remains a partial representation of the entire materials landscape. The system can retrieve only from what has been curated and made accessible to the workflow. In addition, candidate generation is restricted to formulations involving a base material from the set of lab-available materials. This constraint supports experimental accessibility, because it ties proposed substitutions to materials that can be tested in practice, but it also narrows the candidate space and may exclude promising substitutes outside the available inventory. Performance also depends on how retrieval is implemented. Different RAG and KG node-mapping strategies, for example cosine-similarity retrieval versus explicit keyword matching, and Graph-RAG traversal procedures, for example shortest-path versus depth-first search, may yield different evidence sets and influence downstream outputs. Response quality is further shaped by the backbone language model, which must interpret and integrate the retrieved evidence. Although this model dependence is a limitation, the framework is modular and can be paired with different LLM endpoints.

Several limitations of the present evaluation should also be emphasized. Although the matched-backbone comparisons provide evidence for a framework-level benefit, the individual contributions of task decomposition, RAG, KG reasoning, persistent memory, and iterative feedback are not fully isolated in a factorial design. The configurations differ in the number and organization of model calls and in the timing and amount of retrieved context, and therefore should be viewed as reduced architectural comparisons rather than perfectly matched computational controls. In addition, the reported scores are produced by an LLM judge and do not constitute experimental validation of the proposed substitute, its reported properties, or its processing route. In particular, the internal \emph{manufacturable} verdict denotes evidence-supported processing plausibility rather than demonstrated synthesis success.

Finally, the comparative multi-seed quantitative evaluation in the present study is centered on the THV substitution problem. The additional full-workflow examples provided in Supplementary Section~S1.4 (Queries~2--3) demonstrate that the MARS workflow can be executed on other substitution tasks, but they are intended as supplementary workflow demonstrations rather than as part of the controlled quantitative benchmark. The multi-seed evaluation and matched-backbone comparisons therefore characterize run-to-run variation and model dependence primarily for the THV case and do not establish general performance across PFAS substitution or other materials-design problems. Broader quantitative evaluation across different incumbent materials, applications, constraint sets, and knowledge domains will be required to determine how well the observed performance generalizes. Within the present scope, however, the results support the central premise of MARS: external scientific knowledge is most useful when it is embedded within a structured workflow that can decompose the problem, verify intermediate decisions, propagate constraints, and revise hypotheses in response to downstream evidence.

\section*{Conclusions}
\label{sec:conclusion}

In this study we presented MARS, a hierarchical multi-agent reasoning system for knowledge-grounded PFAS substitution. MARS decomposes tasks into a three-step process: requirement analysis, grounded candidate generation, and manufacturability assessment within a closed-loop reasoning architecture. By combining RAG, KGs, a database of lab-available materials, iterative feedback, and persistent storage of intermediate results, the framework links scientific plausibility to downstream feasibility assessment rather than treating material substitution tasks as a one-step text generation problem.

Across the presented case study, MARS generates structured material requirements, proposes an evidence-grounded substitute candidate, and develops a processing route informed by retrieved manufacturing knowledge. MARS achieves the highest overall and subsystem-level scores among the evaluated configurations. Importantly, this advantage persists when the backbone model is held fixed: MARS improves the overall score by approximately 83\% relative to Ablation~1 with \texttt{gpt-oss-20b} (5.47 vs.~2.99) and by approximately 91\% relative to a single \texttt{gpt-5.6-sol} call (6.78 vs.~3.55). These matched comparisons support the contribution of the structured workflow beyond backbone capability alone.

The ablations further show that simply providing retrieved RAG/KG context produces a comparatively modest improvement over an ungrounded single call (3.44 vs.~2.99), whereas integrating retrieval within the full MARS workflow increases the score to 5.47 with the same \texttt{gpt-oss-20b} backbone. Together, these results suggest that the value of external knowledge depends not only on its availability, but also on how it is incorporated into requirement analysis, candidate verification, and iterative refinement. While broader material-substitution benchmarks and experimental validation are required to establish generality, the present results demonstrate the potential of hierarchical, knowledge-grounded reasoning for constraint-aware materials substitution.

\subsection*{Future Directions}
\label{subsec:future_directions}

There are several opportunities to extend the framework presented in this study. One important direction is to strengthen the system’s internal structured representations and verification interfaces. More standardized representations of material-property values could reduce ambiguity across reasoning stages and improve consistency in downstream inference. Such standardization could also enable the introduction of verification tools that validate proposed property values against authoritative sources, rather than relying solely on values extracted from text.

Structured knowledge representations would also enable a second important extension: the integration of simulation and autonomous experimentation, allowing the framework to move beyond evidence-grounded proposal generation toward downstream computational and experimental validation. Incorporating physics-based simulations could allow for grounding hypotheses drawn from retrieved information in computational validation. At the same time, human-in-the-loop variants, in which domain experts intervene selectively by refining requirements, revising constraints, or steering candidate selection, could allow expert knowledge to correct, supplement, and redirect the search process as needed. Together, these extensions could make the framework a more effective scientific collaborator while preserving the interpretability of its multi-stage reasoning process.

Another important extension of the methodology is to move beyond evidence-grounded evaluation of materials toward experimental testing. This would allow experimental results to be fed back into the system as an additional layer of iterative refinement. In such a setting, MARS would not only generate substitute candidates and proposed synthesis routes, but also update its subsequent reasoning in response to observed experimental performance. A more ambitious extension of this direction is the integration of the framework with autonomous laboratory infrastructure, in which robotic platforms perform experiments and return the results directly to the multi-agent system. By linking proposal generation and experimental execution such a system could establish a fully closed-loop discovery workflow that rejects unsupported hypotheses and refines future candidates on the basis of observed material behavior in a continuous cycle toward successful discovery.

Future extensions of the system should also include broader and more controlled evaluation. In particular, evaluation across a wider range of material-substitution queries will be needed to assess generalizability beyond the present PFAS case study. More controlled component-level studies could isolate the individual contributions of retrieval, knowledge-graph reasoning, persistent memory, and manufacturability-aware feedback, while expert-annotated benchmarks could support quantitative assessment of retrieval quality, requirement extraction, PFAS filtering, and hard-constraint satisfaction. Evaluation across a broader range of backbone families, material domains, and human or expert evaluators would further characterize model and evaluator sensitivity. A controlled comparison with the earlier KG-guided framework using matched substitution queries, model backbones, knowledge resources, and evaluation criteria would additionally be needed to quantify the effect of the broader workflow architecture relative to the previous implementation.

In parallel with these advances, another important direction is to expand the knowledge substrates that support the workflow. Future versions of MARS could draw on broader materials inventories, and larger knowledge resources, widening the accessible search space while preserving grounding in relevant materials. At the same time, the retrieval layer itself could benefit from systematic study, including effects of different RAG strategies, node-mapping procedures, and graph traversal methods.

Looking beyond extensions to the system’s knowledge base, a broader long-term challenge is to increase the architectural flexibility of the framework itself. Although MARS was developed for PFAS-related substitution, future studies could examine how well its design transfers to other classes of materials-substitution problems and to broader scientific discovery tasks. More adaptive workflow design may enable more efficient and context-sensitive reasoning across a wider range of discovery settings. More broadly, these directions point toward increasingly distributed and collaborative forms of AI-assisted discovery. Rather than operating as a single, fixed pipeline for one substitution task at a time, these systems participate in broader multi-agent discovery ecosystems in which scientific agents collaborate across tasks and settings. In this sense, MARS may be understood not only as a framework for grounded material substitution, but also as a step toward more adaptive, reproducible, and collaborative AI systems for scientific discovery.

\section*{Materials and Methods}
\label{sec:methods}

\subsection*{Experimental Setup}
\label{subsec:experimental_setup}

Local experiments were conducted on a single node equipped with 4 NVIDIA Tesla V100S GPUs, each with 32~GB of VRAM, using \texttt{Python}~3.11 on CentOS Linux~7 with CUDA toolkit~12.6. The open-source LLM \texttt{gpt-oss-20b}~\cite{openai_gpt-oss-120b_2025}, which is natively quantized in MXFP4 format, served as the language backbone of MARS. The model was hosted locally using \texttt{llama.cpp}~\cite{gerganov_llamacpp_2023}, with all \texttt{gpt-oss-20b} inference performed on local servers. \texttt{PyTorch}~2.5~\cite{paszke_pytorch_2019} provided the tensor computation backend for Transformer-based embedding inference. A maximum completion tokens of 40{,}000 was used to remain within memory limits.

For the ablation evaluation, each configuration was executed across five separate runs. For components and inference backends that expose a seed parameter, the corresponding run seed was supplied during generation. The locally hosted \texttt{gpt-oss-20b} calls used a temperature of 0, whereas API-hosted models were executed with their supported decoding settings and unsupported sampling parameters were omitted. The \texttt{gpt-5.6-sol} configurations were accessed through the OpenAI API. Because deterministic runs do not eliminate low-level backend nondeterminism, repeated executions are not necessarily bit-for-bit identical.

\subsection*{Vector Store Construction and RAG Setup}
\label{subsec:vector_stores}

We make use of text-based RAG across the hierarchical multi-agent framework. The text corpora used for RAG include a curated PFAS research corpus \cite{stewart_graphagents_2026}, a large corpus of polymer science-related patents, internal material specification sheets, and a set of polymer science textbooks. Descriptive statistics for the retrieval sources, including the size of the source corpora and the resulting vector database storage size, are summarized in Table~\ref{tab:rag-summary}.

For text embeddings, we use the \texttt{sentence-transformers} library \cite{reimers_sentence-bert_2019} to load and apply the open-source \texttt{nomic-ai/nomic-embed-text-v1.5}~\cite{nussbaum_nomic_2025} embedding model, which produces semantic vector representations of text. This model was selected for its strong performance on both short- and long-context embedding benchmarks, as well as its open-source availability \cite{nussbaum_nomic_2025}. For embedding storage and retrieval, we use the \texttt{Chroma} library \cite{chroma_chroma_2025} as the vector database. Its persistent client enables local storage of embedding collections and supports efficient retrieval using cosine similarity.

To transform the raw data (PDFs) to queryable vector databases, the text sources are first converted into machine-readable \texttt{markdown} format \cite{stewart_graphagents_2026}. In particular, the PFAS research corpus and polymer science textbooks are converted from PDF to \texttt{markdown} using the open-source \texttt{marker-pdf} package \cite{paruchuri_marker_2025}, while the internal material specification sheets are converted from PDF using a VLM OCR pipeline based on \texttt{DeepSeek-OCR} \cite{wei_deepseek-ocr_2025}. The resulting documents are then segmented into retrieval units using a chunking strategy tailored to the source type. The resulting text chunks are embedded and stored in persistent collections together with source metadata. For the article and textbook corpus, stored metadata include identifiers such as DOI/ISBN and chunk index, while for the patent and material specification sheets document-level identifiers are retained. Across the pipeline, these vector stores provide complementary retrieval support, with the PFAS research corpus supporting requirement and constraint analysis in System~1, the patent corpus and specification sheets supporting material candidate generation in System~2, and the textbook and specification-sheet corpora supporting manufacturability assessment in System~3.

\begin{table}[htbp]
\centering
\caption{Summary of the text-based retrieval sources used in the study. These sources support RAG across the three subsystems of the hierarchical multi-agent framework.}
\label{tab:rag-summary}
\begin{tabularx}{\textwidth}{>{\arraybackslash}X >{\arraybackslash}X >{\arraybackslash}X}
\hline
\textbf{RAG Source} & \textbf{Source Corpus Size} & \textbf{VectorDB File Size} \\
\hline
PFAS research corpus & \texttt{4,716 papers} & \texttt{1.3GB} \\
Patent corpus & \texttt{13,654 patents} & \texttt{69MB} \\
Material specification sheets & \texttt{20 documents} & \texttt{6.5MB} \\
Polymer science textbooks & \texttt{35 textbooks} & \texttt{260MB} \\
\hline
\end{tabularx}
\end{table}

\subsection*{Knowledge Graph Generation and Traversal}
\label{subsec:kg_construction}

We make use of three KGs in this study: the PFAS-specific KG and the broad Material Properties KG introduced in \cite{stewart_graphagents_2026}, and the Patents KG constructed in this study. The three KGs and their corresponding node embeddings are available through the MARS-KGs repository dataset \cite{mars_kgs_2026}. The Patents KG was constructed using the KG generation framework of \cite{stewart_graphagents_2026}. Descriptive statistics for the three KGs, including the number of nodes and edges, size of the source corpora, and the size of the resulting files are summarized in Table~\ref{tab:kg-summary}. KGs are stored as \texttt{GraphML} files, while node embeddings are generated using the \texttt{nomic-ai/nomic-embed-text-v1.5} model and saved separately as serialized \texttt{pickle} files for efficient downstream retrieval and similarity-based graph querying.

\begin{table}[!b]
\centering
\caption{Summary of the three KGs used in the study. The PFAS-specific KG and the Material Properties KG are taken from \cite{stewart_graphagents_2026}, while the Patents KG is newly constructed in this work.}
\label{tab:kg-summary}
\begin{tabular*}{\textwidth}{@{\extracolsep{\fill}} l l c c c c @{}}
\hline
\textbf{Knowledge Graph} & \textbf{Source Corpus} & \textbf{Nodes} & \textbf{Edges} & \textbf{Graph File Size} & \textbf{Embedding File Size} \\
\hline
PFAS & \texttt{4,612 papers} & \texttt{144,335} & \texttt{459,248} & \texttt{137MB} & \texttt{432MB} \\
Material Properties & \texttt{62,999 abstracts} & \texttt{317,800} & \texttt{774,009} & \texttt{229MB} & \texttt{481MB} \\
Patents & \texttt{13,654 patents} & \texttt{199,008} & \texttt{578,361} & \texttt{154MB} & \texttt{594MB} \\
\hline
\end{tabular*}
\end{table}

KG construction relies on an LLM, \texttt{meta-llama/Llama-3.3-70B-Instruct} \cite{aimeta_llama_2024}, to transform unstructured text, including articles, abstracts, and patents, into structured graph representations. For the Patents KG, this process first involves converting patent PDFs into machine-readable \texttt{markdown} files using \texttt{marker-pdf}. The resulting documents are then divided into uniformly sized chunks chosen to fit within the context window of the processing LLM. For each chunk, the LLM extracts semantic triplets, consisting of two entity nodes connected by a relation edge. These triplets define a local knowledge subgraph for the chunk, and the union of these local subgraphs forms the global KG. To preserve connections across chunks, nodes with sufficiently similar textual embeddings are merged when they fall within a predefined similarity threshold, with their incident edges reassigned to the merged node. This framework enables the construction of large-scale graphs in which entities such as materials, properties, and application-relevant contextual features are represented as nodes, and their relationships are represented as edges. Together, the three KGs provide complementary knowledge grounding across the pipeline, with the PFAS-specific KG capturing domain-specific substitution requirements, the broad Material Properties KG enabling cross-domain exploration of property relationships, and the Patents KG contributing formulation- and application-oriented knowledge relevant to practical candidate selection.

The corpus sizes and graph statistics reported here
characterize the scale and source coverage of the knowledge resources, but they should not be interpreted as independent measures of graph correctness. The PFAS and Material Properties KGs were introduced in previous work \cite{stewart_graphagents_2026}, while the Patents KG was constructed in the present study using the same extraction framework. Because graph content is generated through LLM-based triplet extraction and embedding-based node merging, the resulting KGs may contain extraction or linkage errors, and their coverage is necessarily limited by the underlying source corpora. A separate expert audit of graph-level precision and completeness was not performed in the present study.

For Graph-RAG, we use the \texttt{NetworkX} package \cite{hagberg_exploring_2008} to represent and manipulate the graph structures during traversal and subgraph extraction. The \texttt{GraphReasoning} package \cite{buehler_accelerating_2024} is used to map extracted words and concepts to graph nodes via cosine similarity between keyword embeddings and node embeddings, retaining the highest-scoring match for each keyword as a seed node. Graph traversal is performed on an undirected view of the KG to identify connecting paths between seed nodes using a configurable path-finding routine, with shortest-path search used by default. The union of nodes and edges appearing along these paths defines a connection subgraph.

\subsection*{Internal Material Database}
\label{subsec:material_database}

The internal material database is introduced to ground candidate selection in the set of materials that are physically available for downstream laboratory testing. It contains 20 lab-available materials compiled from supplier specification datasheets. To construct the database, each datasheet PDF is first converted into machine-readable \texttt{markdown} using \texttt{DeepSeek-OCR}~\cite{wei_deepseek-ocr_2025}. An LLM-based structured extraction pipeline (here we use \texttt{gpt-oss-20b} \cite{openai_gpt-oss-120b_2025}) is then used to parse supplier information, material identity, and the reported material properties into a standardized schema. Extracted entries are validated using \texttt{Pydantic} \cite{pydantic_pydantic_2026} and stored as structured \texttt{JSON} records, yielding a local database that supports deterministic downstream lookup. The database occupies 2.6~MB on disk, and contains an average of 18 extracted properties per material entry. The complete top-level database schema is provided in Supplementary Table~S2. The database is intended to represent the materials available in the target laboratory rather than the broader commercial materials landscape. Pydantic validation enforces consistency with the defined data schema, but does not by itself verify the semantic accuracy of every property extracted from the supplier datasheets. Accordingly, the database provides a constrained experimental inventory rather than a comprehensive or independently validated materials database.

\subsection*{Agent Design}
\label{subsec:agent_design}

Each agent is implemented as a dedicated \texttt{Python} class with method-level interfaces for invoking external tools and performing subsystem-specific operations. These implementations realize the four reusable agent roles described above under Core Agent Roles in the Hierarchical Multi-Agent Framework, namely the Research Manager, Research Analyst, Research Scientist, and Research Assistant, each instantiated with subsystem-specific prompts, tools, and knowledge sources. The \texttt{autogen} package \cite{wu_autogen_2023} is used to provide the underlying agent abstractions and inter-agent message passing, while the overall orchestration logic is implemented procedurally in \texttt{Python}.

To interface all agents with the locally hosted LLM, we use a shared generation wrapper that routes requests to a locally hosted LLM. This design decouples agent-specific logic from the underlying inference backend and facilitates controlled changes to model choice and inference settings without modifying agent implementations. 

To support reproducibility and centralized configuration, system prompts, user prompts, and task-specific prompts used for all tasks of all agents are stored in a common \texttt{YAML} file, from which the agents load the prompts. A separate centralized \texttt{YAML} configuration file is used to store agent hyperparameters. Together, these design choices provide a modular and reconfigurable agent framework while supporting consistent experimental control across runs. Additional details on pipeline configuration are provided in the Supplementary Information. The accompanying repository provides the prompt and configuration files together with representative intermediate pipeline artifacts, including retrieved evidence, candidate-rejection records, standardized framework outputs, and the evaluation materials used for the reported case study. All agent and pipeline prompts are additionally reproduced verbatim in Supplementary Section~ and shared GitHub repository.

\subsection*{Pipeline Configuration}
\label{subsec:pipeline_config}

This section describes the runtime configuration of each subsystem, detailing how retrieval, graph reasoning, candidate generation, and manufacturability assessment are parameterized across the hierarchical multi-agent framework. All pipeline hyperparameters, together with the prompt templates, are stored in centralized \texttt{YAML} configuration files, enabling systematic ablation and reproducibility without modifying source code. The full configuration files are provided in the Supplementary Information, including the accompanying code repository.

\subsubsection*{System 1 Configuration}

In System~1, the Research Manager generates up to four focused research questions per query, each designed to cover a distinct property dimension of the substitution problem. Text-based retrieval follows an over-retrieve-then-filter strategy designed to approximate efficient keyword-style search without exhaustively scanning all chunks for exact keyword matches. Specifically, the Research Analyst first queries the vector database for 100 candidate chunks, then a subsequent filtering step using a conjunctive keyword criterion, requires that all query-derived keywords appear in the retrieved text, and the top 5 passages that pass the filter are retained. This two-stage approach balances recall against precision by casting a wide initial retrieval net while enforcing topical relevance through keyword overlap. If no text chunk contains all keywords, the top 5 chunks (in regard to cosine similarity) are returned. For Graph-RAG, keywords extracted from each research question are mapped to KG nodes via cosine similarity between keyword embeddings and node embeddings, retaining the highest-scoring match per keyword above a similarity threshold of~0.8. Graph traversal is then performed on an undirected view of the KG using shortest-path search between all mapped seed-node pairs, and the union of nodes and edges along these paths defines a connection subgraph whose relational structure is passed to the Research Manager as complementary evidence alongside the text-based retrieval results. The Research Manager synthesizes answers from both sources, before the Research Assistant converts the resulting Question--Answer pairs into structured material-property requirements and hard constraints. Because the final requirements are extracted from these LLM-synthesized Question--Answer pairs rather than directly copied from individual source passages, the present pipeline does not require each generated numerical threshold to retain a one-to-one source citation. A per-item provenance trace of the reported run is provided in Supplementary Table S3.

\subsubsection*{System 2 Configuration}

System~2 begins by constructing a task-specific Material-Informed Subgraph that links the required substitute-material properties to the set of lab-available materials through relational paths in the Material Properties and Patents KGs. The Material-Informed Subgraph construction proceeds in three phases. First, the required material properties extracted in System~1 and the names of all lab-available materials in the internal material database are independently mapped to nodes in both KGs using embedding-based similarity (similarity threshold~0.8, up to 5 candidate matches per term, with a cap of 30 property terms and 200 material entries). Each lab-available material is grounded in the KG by computing the cosine similarity between its name embedding and all node embeddings, retaining the highest-scoring node above the threshold. This grounding step anchors the downstream graph search in experimentally accessible materials. Second, for each KG, a connection subgraph is built by computing shortest paths between all pairs of seed nodes (the union of mapped property nodes and material nodes), subject to a maximum path length of~10 hops, a cap of 2{,}000 evaluated node pairs, and a maximum of 500 nodes per subgraph. The graph is converted to an undirected view for traversal so that shortest-path search can follow edges in either direction. Third, the two per-KG subgraphs are merged into a single Material-Informed Subgraph by unifying patent-KG nodes onto material-KG nodes when their node-embedding cosine similarity exceeds a merge threshold of~0.88. Edges incident to unified nodes are reassigned to the merged node, and node attributes from both sources are preserved. The resulting graph encodes how required properties relate to lab-available materials through both literature- and patent-derived connections, and is stored in persistent memory for reuse across iterations.

To select the most informative relational evidence for candidate generation, paths are then computed between property-anchor nodes and material-class nodes (Type~A) as well as between cross-property anchor pairs (Type~B) using shortest-path search on the undirected subgraph. Each path is scored as $s = w \cdot \mathbf{1}[\text{material node in path}] + c$, where $w = 2.0$ is a configurable weight for material presence and $c$ is the fraction of distinct required properties whose anchor nodes appear along the path. Paths are selected in score-descending order using a greedy diversity filter that skips any path whose Jaccard node-overlap with an already-selected path exceeds a configurable threshold (0.5), yielding a non-redundant set of relational chains \cite{leskovec_finding_2014}. These selected paths are serialized as human-readable node--edge--node sequences and injected into the candidate proposal prompt as a ``Knowledge Graph Insights'' section, providing the Research Manager with structured relational evidence linking required properties to candidate material families through the KGs.

Within the iterative Candidate Proposal Loop (up to 5 iterations), the Research Manager proposes a candidate material grounded in the internal material database, informed by the subgraph insights and a summary of lessons from any previously rejected candidates. Using the Material-Informed Subgraph, together with the hard constraints and record of previously rejected materials, a Research Manager proposes a revised candidate material for the target application. The proposed candidate is first screened by a dual-source PFAS filter that combines a keyword-based with an LLM-based classification of whether the material contains carbon--fluorine bonds. Candidates that pass the PFAS filter proceed to a validation stage in which the Research Manager generates targeted validation questions to self-validate its reasoning before proposing the candidate. For each subquestion, a Research Analyst retrieves evidence (5 results per query) which the Research Manager uses to synthesize an answer to the question. Finally, the Research Manager uses synthesized answers to evaluate feasibility of the proposed material. If the candidate is found feasible, the system returns the proposed material together with LLM-synthesized supporting evidence. Candidates judged infeasible are recorded in a persistent rejection tracker object together with a rejection reasoning, enabling the Research Manager to generate a structured summary of rejection lessons that informs subsequent proposals.

\subsubsection*{System 3 Configuration}

System~3 begins with process-focused retrieval. A Research Manager first performs a material decomposition step, in which it identifies constituent materials of the proposed candidate. A set of decomposition-aware retrieval queries is then generated, comprising per-constituent queries targeting individual synthesis and handling procedures as well as combination queries addressing blending, compounding, or co-processing steps. These queries are executed through a multi-source retrieval architecture in which a wrapper agent queries multiple Research Analyst instances in parallel, each bound to a different vector database (polymer-science textbooks, patents, material specification sheets), retrieving up to 8 results per source per query. Each retrieved passage is tagged with its source identity so that the downstream LLM can distinguish evidence provenance. The retrieved results are then deduplicated using a content-hash-based strategy to remove redundant passages, and source-balanced by capping the number of retained results per source proportionally within an overall limit of 30 process-relevant passages, ensuring that no single corpus dominates the evidence pool.

From the retrieved evidence, the Research Manager generates 4 targeted manufacturing feasibility questions, each addressing a distinct dimension of lab-scale synthesis practicality for the proposed candidate. For each question, an additional round of multi-source RAG retrieval is performed to gather question-specific evidence, which the Research Manager then uses to synthesize an answer. The resulting Question--Answer pairs are aggregated into a structured feasibility assessment in which the LLM renders a binary verdict: \emph{manufacturable} or \emph{blocked}. If the candidate is judged manufacturable, the Research Manager synthesizes a step-by-step process recipe. If blocked, the Research Manager compiles a structured set of blocking constraints and a feedback string, which are passed back to System~2 to guide revised candidate generation in the subsequent closed-loop iteration.

\subsection*{Evaluation Protocol}
\label{subsec:evaluation_protocol}

We evaluate six framework configurations across five repeated runs. MARS and Ablations~1--3 use the locally hosted \texttt{openai/gpt-oss-20b} model~\cite{openai_gpt-oss-120b_2025}. Two additional configurations use \texttt{gpt-5.6-sol}~\cite{noauthor_gpt-56_nodate}: a single-call baseline analogous to Ablation~1 and the full MARS framework with \texttt{gpt-5.6-sol} as the backbone of every agent. This design provides matched single-call and full-framework comparisons for both backbone models. For the locally hosted \texttt{gpt-oss-20b} model, generation uses a maximum completion length of 40,000 tokens and the decoding configuration described under Experimental Setup. API-hosted models are executed using the corresponding OpenAI API with unsupported sampling parameters omitted.

To enable direct comparison, every configuration produces the same three standardized subsystem reports: (1) material requirements analysis (System 1), (2) substitute-candidate generation and validation (System 2), and (3) manufacturability assessment and process design (System 3). The reports use an identical section structure across configurations. When a reduced configuration cannot produce a capability-specific artifact, the corresponding section is retained and explicitly marked as absent by design. For MARS, the standardized reports additionally preserve the evidence and intermediate artifacts required by the evaluation criteria, including retrieved evidence, investigation traces, candidate-verification records, rejection feedback, and manufacturability assessments. For System~2, database records containing material identity and datasheet properties are embedded for proposed components that match entries in the internal material inventory, enabling the judge to assess candidate anchoring directly.

Evaluation is performed using a blind, pooled protocol. For each seed and subsystem, the six reports are stripped of configuration-identifying information, randomly shuffled, and assigned anonymous labels. All six reports are then evaluated together in a single judge call so that competing configurations share the same scoring context. The anonymous labels are mapped back to the corresponding configurations only after scoring

Automated evaluation is performed using \texttt{gpt-6-astra}~\cite{noauthor_gpt-6_nodate} as the primary judge. The judge is selected from a different model generation than the evaluated backbones to reduce potential self-preference~\cite{spiliopoulou_play_2025}. For each anonymized seed--subsystem pool, the judge receives the six standardized reports together with the corresponding rubric and returns criterion-level scores and brief written rationales in structured JSON format. Because LLM inference remains nondeterministic even under controlled prompting, all reported configuration scores are averaged over five runs.

The LLM-as-a-judge protocol is intended to provide a scalable and standardized comparison of framework capabilities rather than experimental validation of the generated materials. The judge evaluates the outputs according to the predefined subsystem-specific criteria but does not independently verify every material-property value, regulatory claim, or process parameter against external sources. Accordingly, the resulting scores should be interpreted as comparative measures of the evaluated reasoning and workflow capabilities, while source-level verification and experimental testing remain necessary for validating the proposed materials and processing routes.

\section*{Funding}
This research was supported by Saint-Gobain Research North America. T.P.H. acknowledges support from the Aker Scholarship.

\section*{Acknowledgments}
\label{sec:acknowledgements}
We acknowledge valuable discussions with Lalitha Raghavan and Patrick
Lee from Saint-Gobain Research North America.

\section*{Author Contributions}
\label{sec:contributions}
M.J.B. supervised the project and led the overall research design and development. T.P.H. conceived the hierarchical multi-agent framework, implemented the three multi-agent systems and associated agent components, and conducted the original workflow experiments and analyses. S.P. led the expanded evaluation and revision-stage analyses, including the multi-seed ablation study, matched-backbone comparisons, targeted feedback analysis, and associated quantitative interpretation. W.L. contributed to the experimental evaluation and associated analyses and collected supporting data. Y.H. generated the knowledge graphs. G.L. and J.J. developed the evaluation case-study prompt and evaluated results from earlier iterations of the system, providing feedback that informed subsequent refinements to the prompts and framework design. T.P.H., S.P., and W.L. prepared the manuscript. M.J.B., T.P.H., S.P., and W.L. contributed to revising, editing, and finalizing the manuscript.

\section*{Competing Interests}
\label{sec:competingInterests}
The authors declare that they have no competing interests.

\section*{Data Availability}

The three knowledge graphs used in this study and their corresponding node embeddings are available through the MARS-KGs dataset on Hugging Face (\url{https://huggingface.co/collections/lamm-mit/mars}). 
% The numerical source data underlying Figure~5, Table~2, Supplementary Figures~S5--S6, and Supplementary Table~S4 are provided in Supplementary Data.
The same reproducibility materials are archived with the MARS software record on Zenodo (\url{https://doi.org/10.5281/zenodo.23199501}).
The internal material database and full-text retrieval corpora are not publicly redistributed because of access and licensing restrictions.

\section*{Code Availability}
The source code, configuration files, prompts, representative intermediate pipeline outputs, and evaluation scripts and outputs used in this study are publicly available through the MARS GitHub repository (\url{https://github.com/LAMM-MIT/MARS}). The version corresponding to this study has been permanently archived on Zenodo (version 1.0.0; DOI: \url{https://doi.org/10.5281/zenodo.23199501})~\cite{mars_code_2026}. The repository includes the complete intermediate artifacts of the reported THV run (retrieved evidence, knowledge-graph subgraphs, agent chat logs, and candidate-rejection records), the anonymized evaluation inputs, and the per-criterion judge outputs.

%Bibliography
\bibliographystyle{unsrt}  
\bibliography{references}  

\newpage



\section*{Supplementary material}

# Supplementary Information

Hierarchical Multi-Agent Reasoning Systems for Knowledge-Grounded Material Substitution

Tarjei Paule Hage†, Subhadeep Pal†, Wei Lu, Yu-Chuan Hsu, Gayla Lyon, Jiezhu Jin, and Markus J. Buehler⋆

†These authors contributed equally to this work.

⋆Corresponding author: mbuehler@mit.edu

## Supplementary Methods

This Supplementary Information provides detailed subsystem workflow diagrams, additional information on the evaluation protocol, and the internal material database schema for the hierarchical multi-agent framework described in the main manuscript.

### Detailed Subsystem Workflows

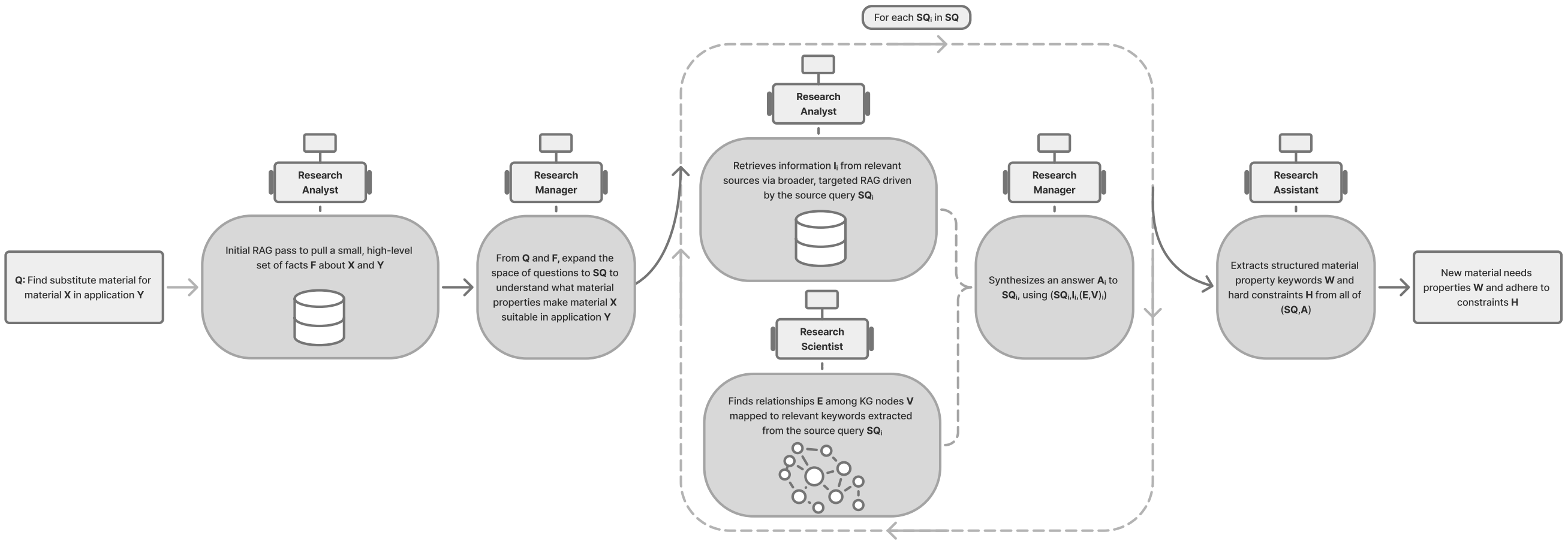


Figure S1: Detailed workflow of System 1, Material Substitution Requirements Analysis. Given a user query, a Research Analyst retrieves relevant information from a curated PFAS literature corpus. A Research Manager decomposes the initial task into focused research questions. For each question, a Research Analyst and Research Scientist conduct dual-source retrieval from the PFAS literature corpus and PFAS-specific knowledge graph. The retrieved information is synthesized by the Research Manager, and a Research Assistant extracts structured material-property requirements and hard constraints for downstream candidate generation.

### Supplementary Evaluation Information

System 1 is evaluated on completeness, scientific correctness, evidence grounding, and investigation quality; System 2 on requirement alignment, adversarial verification, candidate anchoring, and adaptive reasoning; and System 3 on manufacturing route coherence, operational practicality, material-process compatibility, and feasibility-verdict grounding.

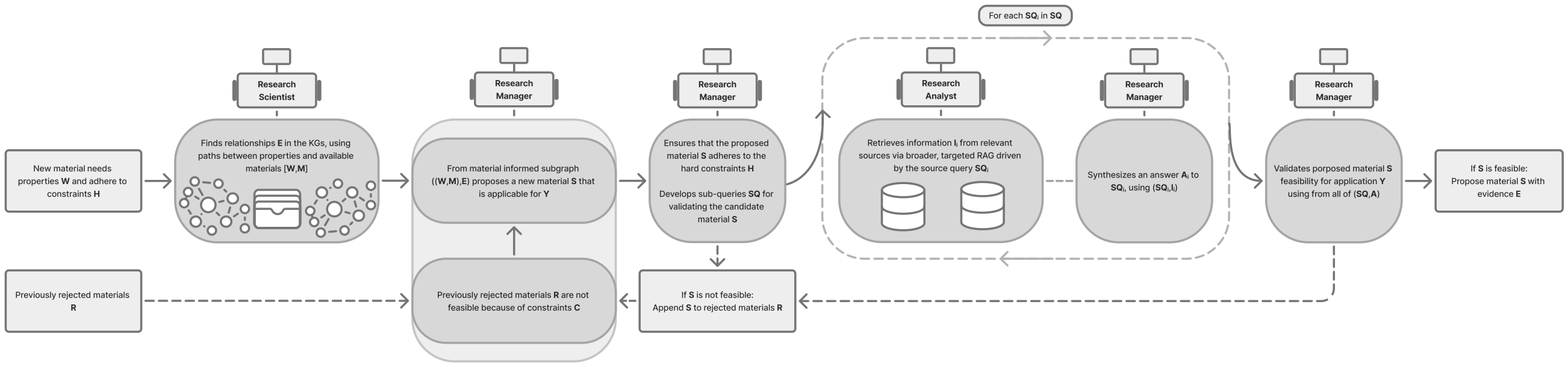


Figure S2: Detailed workflow of System 2, Grounded Material Candidate Generation. Starting from the required substitute-material properties and hard constraints, the system constructs a task-specific Material-Informed Subgraph linking desired properties to lab-available materials through the Material Properties and Patents knowledge graphs. Candidate materials are subsequently screened and evaluated using targeted retrieval. Rejected candidates and associated constraints are retained to inform subsequent iterations.

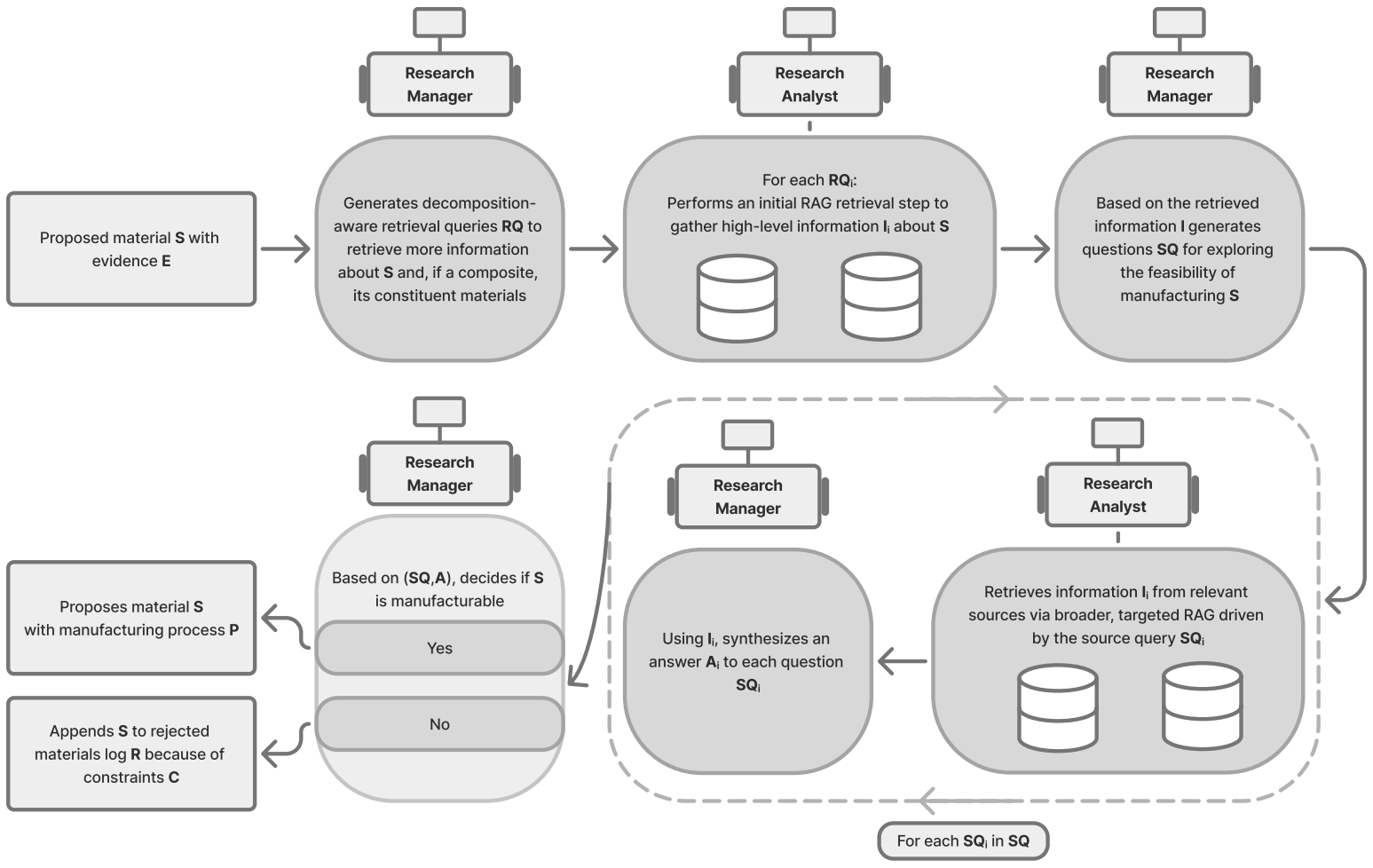


Figure S3: Detailed workflow of System 3, Manufacturability Assessment and Process Design. Process-focused evidence is retrieved from textbooks, patents, and lab-available material specification sheets. The system formulates and answers manufacturing-feasibility questions before assigning an internal verdict of *manufacturable* or *blocked*. A manufacturable verdict yields a proposed processing route, whereas blocking constraints are returned to System 2 for revised candidate generation.

Table S1: Subsystem-specific evaluation criteria used to score the standardized outputs. Each of the three subsystem outputs is evaluated on four criteria, yielding 12 criterion-level scores per framework response for each query.

| System 1 | System 2 | System 3 |
|---|---|---|
| Completeness | Requirement alignment | Manufacturing route coherence |
| Scientific correctness | Adversarial verification | Operational practicality |
| Evidence grounding | Candidate anchoring | Material-process compatibility |
| Investigation quality | Adaptive reasoning | Feasibility-verdict grounding |

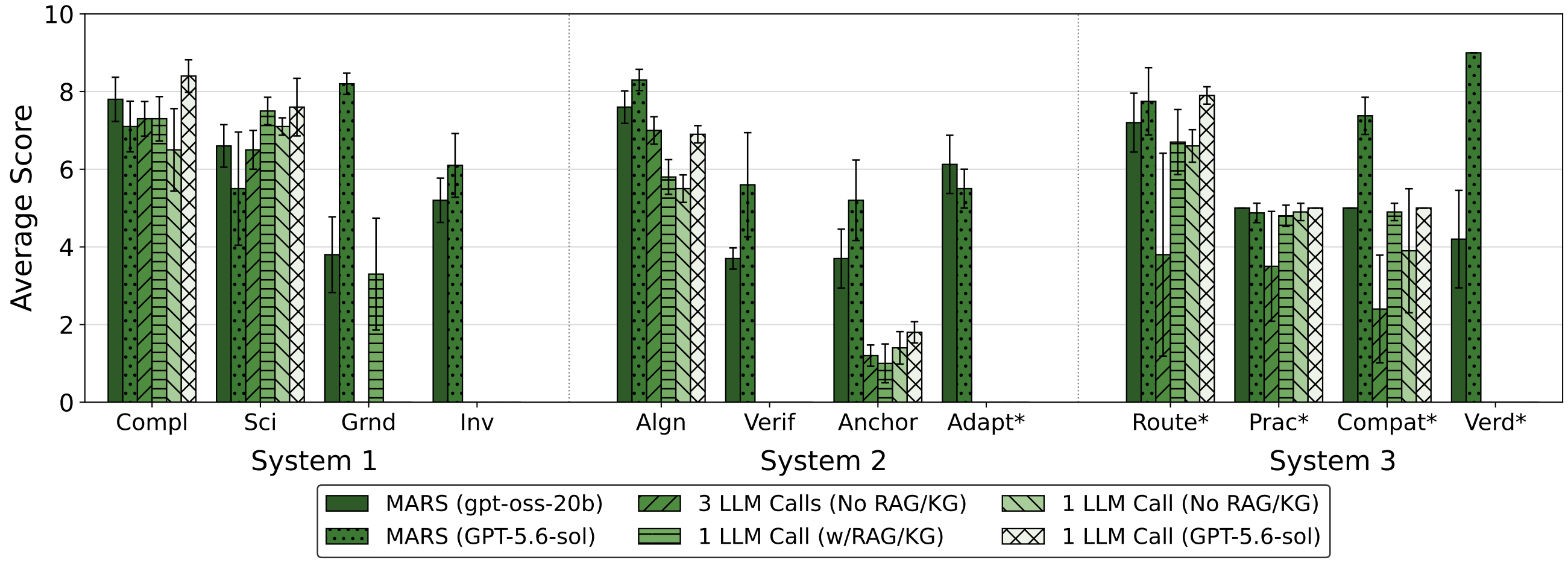


Figure S4: Criterion-level evaluation across all six framework configurations. Scores are grouped by subsystem: System 1 (completeness, scientific correctness, evidence grounding, and investigation quality), System 2 (requirement alignment, adversarial verification, candidate anchoring, and adaptive reasoning), and System 3 (route coherence, operational practicality, material–process compatibility, and feasibility-verdict grounding). Bars show mean scores over five separate runs and error bars indicate standard deviations. Asterisks denote criteria for which one MARS run was unavailable or the corresponding capability was unexercised and therefore excluded from that criterion mean.

## Internal Material Database Schema

Table S2: Top-level fields stored for each entry in the internal material database used by MARS.

| **Field Name** | **Field Type** | **Description** |
|---|---|---|
| `id` | `string` | Identifier assigned to each material entry. |
| `name` | `string` | Material name extracted from the supplier datasheet. |
| `supplier` | `string` | Supplier or manufacturer associated with the material. |
| `raw_markdown_text` | `string` | Full machine-readable markdown text obtained from the source datasheet. |
| `extracted` | `object` | Structured dictionary of extracted material properties. |
| `llm_prompt` | `string` | Prompt used for the LLM-based structured extraction step. |
| `llm_response_raw` | `string` | Raw unparsed text output returned by the LLM. |
| `llm_response_json` | `object` | Parsed JSON-formatted version of the LLM extraction output. |
| `created_at` | `string` | Timestamp indicating when the database entry was created. |
| `updated_at` | `string` | Timestamp indicating when the database entry was last updated. |
| `source_files` | `list[string]` | List of source markdown files used to construct the material entry. |

## Additional Queries

### Query 2: Non-fluoropolymer thermoplastics for cryogenic thermal shock

**Evaluation Query 2**

Identify non-fluoropolymer thermoplastics that can withstand severe thermal shock, specifically direct immersion into liquid nitrogen (−196 °C) and maintain structural integrity (no cracking, embrittlement, or catastrophic failure). Please note that silicone doesn't account for thermoplastics. The identified material can be formulated to allow extrusion into films and injection molding components.

Text Box 1: Evaluation query for thermoplastic substitution in Query 2.

**System 1 Output**

**Properties**

- Cryogenic temperature tolerance
- Thermal-shock resistance
- Structural integrity at –196 °C
- Low embrittlement at cryogenic
- Dimensional stability during rapid cycling
- Retained tensile strength at –196 °C
- Impact resistance after cold exposure
- Low coefficient of thermal expansion
- Adequate extrusion viscosity
- Injection-molding flowability
- Performance requirements
- No cracking in liquid nitrogen
- No catastrophic failure during cold exposure
- Mechanical property retention through thermal cycling
- Uniform film thickness during extrusion
- Precise mold shape in injection-molded parts
- Constraints & specifications
- Non-fluoropolymer composition
- Non-silicone material
- PFAS-free formulation
- Food-contact compliance (SML limits)
- Biocompatibility (ISO 10993)
- Low extractables/leachables
- Low flammability (UL 94 V-0)
- VOC-free during processing
- Regulatory approval for medical device use
- Critical characteristics to replicate from the original
- Cryogenic-shock resilience
- Mechanical integrity post-cold
- Processability for extrusion and injection molding
- Compatibility with food-contact and medical-device applications

**Constraints**

- Cryogenic Integrity – The material must retain its dimensional and mechanical integrity (no cracking, embrittlement, or catastrophic failure) when directly immersed in liquid nitrogen (−196 °C) for at least 10 minutes.
- Rapid Thermal Cycling Resistance – The material must survive at least 100 cycles between ambient temperature (≈ 25 °C) and −196 °C without loss of tensile strength or impact resistance.
- No PFAS Content – The polymer chemistry must contain no per-fluoroalkyl or per-fluoroalkyl-substituted groups; PFAS-type structures are prohibited.
- Food-Contact Extractables Limits – According to U.S. FDA 21 CFR 177.3550, all extractable substances must

be below the regulatory extraction limits ($\leq$ 0.5 ppm for all potential leachables).

- Food-Contact Specific Migration Limits – EU Regulation (EC) 2004/10/2011 SMLs must be met: all migrating substances must not exceed the specific migration limits for the intended food category (e.g., $\leq$ 1 mg $cm^{-2}$ for general food contact).
- Biocompatibility (ISO 10993-1) – For medical-device use the material must pass cytotoxicity, sensitisation, and irritation tests and show no evidence of systemic toxicity at the exposure levels specified by ISO 10993-1.
- Flammability (UL 94) – The material must achieve a UL 94 V-0 rating (or equivalent ASTM D6351 heat-release-rate limit) to qualify for packaging or medical-device applications in environments with a fire hazard.
- Mechanical Stability Post-Cryogenic Exposure – After immersion at −196 °C and rapid thermal cycling, the material's tensile strength must be $\geq$ 90 % of its as-received value.
- Processability for Extrusion/Injection Molding – The melt flow index (MFI) at 210 °C/2.16 kg must fall within the range 1–10 g/10 min (typical for extrusion films) and be suitable for injection-molded components (minimum MFI 0.5 g/10 min).
- Environmental Compatibility – The material must not degrade or release additives when exposed to common hydrocarbons (e.g., gasoline, diesel) at 60 °C for 24 hours.

Text Box 2: Final System 1 output for Query 2, shown as generated by MARS, without post-hoc refinement.

## System 2 Output

**Final candidate**

Cryo-PEEK Composite - PEEK 8 Natural + 0.5 wt % Nano-$SiO_2$ + 0.5 wt % 1-mm Chopped Carbon Fiber

**Justification**

- **Matrix – AKROTEK® PEEK 8 Natural (ID 7343)**
  *Source: Available Materials Database.*
  PEEK provides the core attributes required for severe thermal shock: high melt temperature ($\approx$ 360 °C), excellent cryogenic toughness (retains $\approx$ 60 % of room-temperature tensile strength at –196 °C – an estimated value supported by literature on medical-grade PEEK), very low coefficient of thermal expansion (∼30–40 ppm $K^{-1}$), and inherent UL-94 V-0 flammability. PEEK's biocompatibility (ISO 10993 compliant) and lack of fluorine or silicone make it suitable for food-contact and medical devices. Its melt flow index (MFI $\approx$ 18 g /10 min @ 260 °C) ensures good extrusion viscosity and injection-molding flowability [1].
- **Reinforcement – 1-mm Chopped Carbon Fiber (0.5 wt %)**
  *Justification:* Short fibers of 1–2 mm length at $\leq$ 1 wt % meet the lessons-learned guidelines: sufficient load transfer for stiffness and impact resistance while keeping melt viscosity manageable. The carbon fibers enhance tensile strength and impact toughness, countering embrittlement at cryogenic temperatures and ensuring no catastrophic failure during rapid thermal cycling. The fiber content is low enough ($\leq$ 1 wt %) to avoid void formation and preserve dimensional stability during extrusion and injection molding.
- **Dimensional-Stability Additive – Nano-Silica (0.5 wt %)**
  *Justification:* Nano-silica at $\leq$ 1 wt % is well-accepted for reducing thermal expansion and improving dimensional stability in thermoplastics. According to the knowledge graph, nano-silica is often used in composites to enhance mechanical integrity under thermal stress [see general composites practice]. The low loading aligns with lesson #1, preventing excessive viscosity increase and maintaining good flow during extrusion and molding. The silane coupling agent (e.g., 3-aminopropyltriethoxysilane) is added to improve interfacial adhesion, ensuring the composite retains tensile strength and impact resistance at cryogenic temperatures.
- **Processing Parameters**
  - Compression molding: 360–380 °C, 3–4 kpsi, 10 min dwell, cooling 5 °C $min^{-1}$.
  - Injection molding: 360 °C melt, 3–4 kpsi pressure, 10 s fill, 5 °C $min^{-1}$ cooling.

  These conditions produce a uniform film thickness, precise mold shapes, and a defect-free microstructure, critical for maintaining dimensional stability during rapid thermal cycling.
- **Performance Alignment**
  1. *Cryogenic temperature tolerance:* PEEK matrix retains significant strength at –196 °C; short fibers provide additional toughness.
  2. *Thermal-shock resistance:* Low CTE and high modulus of the composite allow rapid temperature changes without cracking.

3. *Structural integrity at –196 °C:* Fiber-reinforced PEEK maintains >50 % of room-temperature tensile strength (estimated).
4. *Low embrittlement & impact resistance:* Short fibers improve toughness; nano-silica does not compromise impact energy.
5. *Uniform extrusion and injection-molding flow:* MFI of matrix + low additive load ensures adequate viscosity.
6. *No PFAS, fluoropolymer, or silicone components:* All constituents are non-fluorinated, PFAS-free, and silicone-free.
7. *Food-contact & medical compliance:* PEEK is FDA-approved for food contact; the composite meets ISO 10993 and UL-94 V-0 flammability, with VOC-free processing.

Overall, the Cryo-PEEK Composite leverages the proven cryogenic resilience of medical-grade PEEK, modest carbon fiber reinforcement for toughness, and nano-silica for dimensional stability, while strictly adhering to the lessons-learned constraints on additive loading, fiber length, and processability. This system satisfies all listed material and performance requirements for severe thermal shock applications without reusing any previously rejected designs.

Text Box 3: Final System 2 output for Query 2, shown as generated by MARS, without post-hoc refinement.

## System 3 Output

**Status:** Manufacturable

**Process recipe (summary)**

**Step 1 Description:** Receive raw PEEK 8 % natural, nano-$SiO_2$ (0.5 wt %) and chopped carbon fiber (1 mm, 0.5 wt %) as dry pellets. Verify moisture content < 0.02 % (PEEK residual moisture spec).
**Conditions:** Visual inspection, moisture analysis (Karl Fischer or oven drying).
**Equipment:** Material handling & storage
**Inputs**

- PEEK 8 % natural pellets
- Nano-$SiO_2$ 0.5 wt %
- Chopped carbon fiber 1 mm 0.5 wt %

**Step 2 Description:** Pre-dry all components in a vacuum oven to 120 °C for 4 h to remove absorbed moisture.
**Conditions:** Vacuum 0.1 bar, 120 °C, 4 h.
**Equipment:** Vacuum oven (≥5 $m^3$ capacity)

**Step 3 Description:** Compounding by melt-mixing in a twin-screw extruder. Feed pre-dry PEEK, nano-$SiO_2$, chopped CF, and optional coupling agent (e.g., maleated-PEEK) through a feeder system. Use a barrel temperature profile of 350–380 °C (maximum 400 °C) with a screw speed of 30–40 rpm to ensure uniform dispersion and prevent fiber damage.
**Conditions:** Barrel 350–380 °C, screw speed 30–40 rpm, residence time 8–10 s. Maintain low back-pressure (≤ 50 bar) to minimise shear.
**Equipment:** Twin-screw extruder (≥200 kW, 150 mm screw), feeder, die
**Inputs**

- Pre-dry PEEK
- Nano-$SiO_2$
- Chopped CF
- Coupling agent

**Step 4 Description:** Extruded strand is quenched, cooled, and pelletised into 3–4 mm pellets for downstream processing.
**Conditions:** Quench in water bath, pelletiser at 300 rpm.
**Equipment:** Pelletiser

**Step 5 Description:** Injection moulding of the composite. Use an injection moulding machine (e.g., Engel ES200140 or equivalent) with a 30 mm screw. Set melt temperature 370–380 °C, mold temperature 170–200 °C, and back-pressure 50–300 bar. Injection pressure 1 000–2 000 bar; holding pressure 500–700 bar; cycle time 12–15 s (including 2–5 s cooling).

**Conditions:** Melt temp 370–380 °C, mold temp 170–200 °C, injection speed: slow-fast-slow profile. Cooling 2–5 s to reach < Tg for solidification.
**Equipment:** Injection moulding press (≥300 kW), 30 mm screw, mold with heating/cooling channels
**Inputs**

- Compounded PEEK masterbatch pellets

**Step 6 Description:** Post-moulding operations: demould, trim excess, inspect for defects, and perform preliminary mechanical (tensile, flexural) and thermal shock tests to verify performance under severe thermal shock conditions.
**Conditions:** Visual inspection, dimensional measurement, mechanical testing per ISO 527/ISO 178, thermal shock test per ASTM D 3039.
**Equipment:** Quality control equipment (tensile tester, flexural tester, thermal shock rig)

**Step 7 Description:** Package finished parts in moisture-proof bags, label, and store at < 25 °C, protected from direct sunlight and moisture.
**Conditions:** Humidity < 30 %, temperature 15–25 °C.
**Equipment:** Packaging line (vacuum sealer, labeling system)

Text Box 4: Final System 3 output for Query 2, shown as generated by MARS, without post-hoc refinement.

### Query 3: Non-fluoropolymer replacement for FEP heat-shrink tubing

**Evaluation Query 3**

Identify non-fluoropolymer thermoplastic material for heat shrink tubing that could replace FEP heat shrink tubing with an application/service temperature of up to 180 °C and a shrink temperature under 250 °C. Additionally, evaluate whether specific material formulations or processing approaches could help meet these performance requirements.

Text Box 5: Evaluation query for FEP heat-shrink tubing substitution in Query 3.

**System 1 Output**

**Properties**

- Shrink onset 230–250 °C
- Shrink-rate $\geq$ 95 % at target temp
- Dimensional stability $\leq$ 0.5 % change at 180 °C
- Tensile modulus $\geq$ 1 GPa at 180 °C
- Elongation at break $>$ 300 % at 180 °C
- Thermal degradation $>$ 260 °C (TGA)
- Glass-transition temperature 30–60 °C (for processability)
- Melt-flow index 0.5–2 g/10 min at 230 °C
- Chemical resistance to oils, solvents, acids ($\geq$ pH 2–13)
- Moisture barrier $\leq$ 10 $g/m^2 \cdot h$ at 70 % RH
- Low water absorption $<$ 0.3 %
- No fluorinated backbone (PFAS-free)
- Food-grade (GRAS) compliance (if applicable)
- Low VOC emissions ($\leq$ 10 ppm)
- Cost-effective raw materials ($\leq$ \$5/kg)
- Recyclable or biodegradable potential (optional)
- Multi-layer laminate capability (PE/PEU, PE/PEO)
- Barrier layer inclusion (PET, EVOH, thin PEU)
- Elastomer blend integration (PE–EVOH, PE–EVA)
- Plasticizer compatibility (citrate, vegetable-oil esters)
- Cross-linking strategy (peroxide, Diels–Alder)
- Cross-link density $\leq$ 5 mol % (for flexibility)
- Thermal curing cycle $<$ 180 °C (for equipment)
- Processability in extrusion or injection molding
- Shrink-temperature uniformity $\pm$ 5 °C across batch
- Color fastness (no yellowing at 180 °C)
- Electrical insulation $>$ 10 kV·mm (if required)
- Compliance with RoHS & REACH standards
- Low static charge accumulation
- Compatibility with standard heat-shrink tooling (no special clamps)
- Key Characteristics to Replicate from FEP
- High-temperature service up to 180 °C
- Near-zero creep at operating temp
- Excellent barrier to gases and vapors
- Smooth, uniform shrink profile
- Mechanical robustness during shrink
- Chemical inertness to industrial environments
- No color change or embrittlement at 180 °C

**Constraints**

- Non-fluoropolymer requirement – The material must not contain any fluorinated monomers, oligomers, or additives (i.e., it must be a true non-fluoropolymer).

- PFAS-free constraint – No per- and poly-fluoroalkyl substances (PFAS) may be present in the formulation.
- Service-temperature stability – The material must retain dimensional stability and mechanical integrity continuously at a service temperature of at least 180 °C.
- Shrink-temperature window – The tubing must begin to shrink at a temperature between 230 °C and 250 °C (i.e., shrink onset must be $\leq$250 °C and $\geq$230 °C).
- Melt-temperature / softening constraint – The material's melt or softening temperature must be above 250 °C, ensuring it does not flow or soften at the 180 °C service temperature.
- Shrinkability requirement – The material must be a thermoplastic that can be processed into shrink tubing and exhibit sufficient melt-flow to allow shrinkage when heated to the specified shrink temperature.
- No degradation at service temperature – The material must not chemically or mechanically degrade (e.g., lose tensile strength, modulus, or dimensional accuracy) when exposed to 180 °C for the intended duration of use.

Text Box 6: Final System 1 output for Query 3, shown as generated by MARS, without post-hoc refinement.

## System 2 Output

**Final candidate**

PEEK-Polyolefin Elastomer-EVOH Heat-Shrink Composite

**Justification**

The composite is built from three layers that together satisfy every listed requirement.

1. **Matrix Layer – ENGAGE™ 030902 HEALTH+™ Polyolefin Elastomer (ID 16)**: This polyolefin elastomer is naturally PFAS-free, food-grade, and has excellent processability (MFI 1.2–1.8 g/10 min at 230 °C when crosslinked with 3 wt % dicumyl peroxide). Crosslinking with a peroxide ($\leq$ 5 mol % of the polymer repeat unit) raises the shrink onset to $\approx$ 240 °C, ensuring the required 230–250 °C window, while maintaining a high elongation at break (> 350 %) at 180 °C and a melt-flow index within 0.5–2 g/10 min at 230 °C. The crosslinked elastomer also provides dimensional stability $\leq$ 0.5 % at 180 °C.
2. **Reinforcement Layer – AKROPEK® PEEK 8 natural (ID 1)**: PEEK offers a tensile modulus $\geq$ 2 GPa at 180 °C and excellent thermal degradation (TGA > 500 °C). Its high modulus ensures the composite's overall modulus stays above 1 GPa at 180 °C, while its chemical inertness protects against oils, solvents, and acids (pH 2–13). PEEK's thermal stability also guarantees a thermal degradation point above 260 °C.
3. **Barrier Layer – RENOLIT INFULEX PE, EVOH (ID 12)**: The PE/EVOH laminate provides a moisture-barrier $\leq$ 10 g/m$^2$·h at 70 % RH and a low water absorption (< 0.3 %). The EVOH component also improves gas barrier properties, matching the FEP-derived requirement of excellent vapor/gas resistance.

Overall, the layered structure delivers a uniform shrink profile with a shrink rate $\geq$ 95 % at the target temperature, near-zero creep up to 180 °C, and smooth, color-fast shrink behavior (no yellowing at 180 °C). The composite is PFAS-free, has low VOC emissions ($\leq$ 10 ppm), meets RoHS/REACH compliance, and can be extruded or injection-molded with a curing cycle < 180 °C (polyolefin crosslinking at 150 °C). Cost estimates for the base polymers are < $5 /kg, making the system cost-effective.

Text Box 7: Final System 2 output for Query 3, shown as generated by MARS, without post-hoc refinement.

## System 3 Output

**Status:** Manufacturable

**Process recipe (summary)**

**Step 0 Description:** Raw material receipt and storage
**Equipment:** Warehouse storage
**Inputs**

- PEEK pellets
- Polyolefin elastomer (POE) pellets

- EVOH film or pellets
- MA-grafted polypropylene compatibilizer (1-3 wt%)

**Step 1 Description:** Drying of all polymer grades
**Conditions:** temperature: 80-100 °C; time: 2-3 h
**Equipment:** Drying tunnel or rotary dryer
**Inputs**

- PEEK pellets
- POE pellets
- EVOH film/pellets

**Step 2 Description:** Compounding in a twin-screw extruder (Type II)
**Conditions:** barrel zones: zone 1, PEEK core, 350-380 °C; zone 2, POE, 170-190 °C; zone 3, EVOH, 160-180 °C; zone 4, Compatibilizer, 160-180 °C, 1-3 wt% MA-grafted PP; screw speed: 200-300 rpm; shear rate: low-to-moderate to avoid PEEK degradation
**Equipment:** Twin-Screw Extruder (Type II)
**Inputs**

- Dry PEEK pellets
- Dry POE pellets
- Dry EVOH pellets or film
- MA-grafted PP compatibilizer

**Step 3 Description:** Co-extrusion through a three-layer die
**Conditions:** die material: High-temperature alloy (Inconel 625 / Hastelloy C); core diameter: 5 mm; POE layer thickness: 0.2 mm; EVOH layer thickness: 0.05 mm
**Equipment:** Multi-Layer Die (conical entry 30°, land 5-10× exit diameter)
**Inputs**

- Compounded melt stream from extruder

**Step 4 Description:** Cooling of extruded tube
**Conditions:** roll temperature: 30-35 °C; roll speed: ≈ 30 rpm; air knife: used to remove excess melt
**Equipment:** Water chill roll

**Step 5 Description:** Optional annealing to relieve stresses
**Conditions:** temperature: 80-100 °C; time: 30 min
**Equipment:** Conveyor oven

**Step 6 Description:** Application of heat-shrinkable POE coating (if not co-extruded)
**Conditions:** temperature: 180 °C; coating thickness: 0.1 mm
**Equipment:** Coating extruder or separate twin-screw extruder
**Inputs**

- Dry POE pellets

**Step 7 Description:** Cutting to specified lengths
**Equipment:** Precision cutting machine (e.g., saw or laser)
**Inputs**

- Cooled tube or tube + shrink coating

**Step 8 Description:** End heat sealing
**Conditions:** temperature: 180-200 °C; hold time: 2-4 s
**Equipment:** Heat seal machine

**Step 9 Description:** Activation of shrink layer (if separate)
**Conditions:** temperature: 120 °C; time: 3-5 min
**Equipment:** Shrink oven

**Step 10 Description:** Quality control sampling
**Conditions:** tests: Diameter & wall-thickness gauge; Shrink ratio test (120 °C for 5 min); $O_2$ permeability (ASTM D-3985); Tensile strength (ASTM D-882); Barriers to water vapour (ASTM F-1249)
**Equipment:** QC laboratory

**Step 11 Description:** Packaging, labeling and shipping
**Equipment:** Packing line
**Inputs**

- QC-approved tubes

Text Box 8: Final System 3 output for Query 3, shown as generated by MARS, without post-hoc refinement.

## Source of Material Properties and Hard Constraints

To clarify whether the requirements in Text Box 2 of the main text were retrieved, inferred, or assumed, we traced each item through the stored System 1 chat log of the reported THV run using a regular-expression search (Table S3). For each requirement, we record the first stage at which its qualitative concept appears: the user query (Query), a retrieved passage carried into an answered Question–Answer pair (Retrieved), the Research Manager's synthesized answer without support in a retrieved passage (Synthesis), an example contained in the extraction prompt (Prompt), or the final extraction call (Extraction). The origin of each numerical value or test condition is recorded separately in the same way, because a requirement and the threshold attached to it can enter the workflow at different stages; when the values in one row differ in origin, the origins are listed in the order of the values, and rows without a numerical value carry a concept origin only. In this run, 2 of the 4 generated research questions in System 1 were answered and passed to extraction, and the extraction calls receive only the query and these Question–Answer pairs, not the retrieved passages or knowledge-graph paths. Of the 25 qualitative requirements, 15 originate in the query, 6 in retrieved passages, and 4 in examples given in the extraction prompts. Of the 28 numerical values and test conditions, none appears in the query, retrieved passage, or knowledge-graph path: 25 were introduced at extraction, 2 during synthesis, and 1 follows an extraction-prompt example. The numerical thresholds in Text Box 2 are therefore model-introduced assumptions rather than retrieved values, whereas most qualitative requirement categories are traceable to the query or to retrieved literature.

Table S3: Provenance of the System 1 output for the THV case study. Concept origin refers to the qualitative requirement; value origin refers to the numerical thresholds and test conditions attached to it.

| Type | Requirement | Numerical values and test conditions | Concept origin (source) | Value origin |
|---|---|---|---|---|
| Property | PFAS-free composition | – | Query | – |
| Property | Elastomeric flexibility | modulus 0.5–1 MPa; elongation >200 % | Query | Extraction |
| Property | Optical clarity / transparency | transmittance >90 %; film ≤0.2 mm | Query | Extraction; Synthesis |
| Property | Low gas permeability | <10 $cm^3/m^2$·day | Query | Extraction |
| Property | Broad chemical resistance | – | Query | – |
| Property | Oil repellency / oleophobicity | contact angle >110°; 24 h oil exposure | Retrieved (doi:10.1016/j.jclepro.2019.01.160; doi:10.1016/j.envint.2016.02.035) | Extraction |
| Property | Salt-solution resistance | NaCl up to 3 M | Query | Extraction |
| Property | Thermal stability | $T_g \geq 50\,°C$; degradation >200 °C | Prompt | Extraction |
| Property | Mechanical toughness | impact 10–15 $kJ/m^2$ | Retrieved (corpus doc. 0_010901_1; corpus doc. 3436_Vinyl) | Extraction |
| Property | Durability / washability | ≥50 laundering cycles | Retrieved (doi:10.1016/j.envint.2016.02.035) | Extraction |
| Property | Regulatory compliance (REACH, RoHS) | – | Prompt | – |
| Property | Cost-effective manufacturability | – | Prompt | – |
| Property | Soft hand feel | – | Retrieved (doi:10.1016/j.envint.2016.02.035) | – |
| Property | Refractive-index matching, minimal haze | – | Retrieved (corpus doc. 3270_Integr; corpus doc. 3436_Vinyl) | – |
| Property | Dense polyimide or EVOH/PE blends | – | Retrieved (corpus doc. 0_010901_1; corpus doc. 3436_Vinyl) | – |
| Constraint | Acid resistance | ≤5 % weight loss; HCl as test acid; 10 %, 24 h, 25 °C | Query | Extraction; Prompt; Extraction |
| Constraint | Base resistance | 10 % NaOH, 24 h, 25 °C | Query | Extraction |
| Constraint | Protic-solvent compatibility | ≤5 % swelling; 50 % ethanol, 48 h | Query | Extraction |
| Constraint | Aprotic-solvent compatibility | acetone or dichloromethane, 48 h | Query | Extraction |
| Constraint | Oil resistance | mineral oil, 7 days | Query | Extraction |
| Constraint | Salt-solution resistance | ≥95 % tensile strength, 10 % NaCl, 30 days | Query | Extraction |
| Constraint | Flexibility (mechanical) | modulus ≤200 MPa; elongation ≥300 % | Query | Extraction |
| Constraint | Optical clarity | transmittance ≥80 % at 550 nm; film ≤0.5 mm | Query | Extraction; Synthesis |
| Constraint | Gas barrier performance | OTR <10 $cm^3/m^2$·day; WVTR <1 $g/m^2$·day | Query | Extraction |
| Constraint | Hydrocarbon stability | n-heptane, 30 days | Query | Extraction |

## Run Time Analytics

The run-time analysis was performed on an NVIDIA DGX Spark system equipped with a 20-core Arm CPU (`aarch64`), an NVIDIA GB10 GPU, and 128 GB of unified memory. The system used Ubuntu 24.04.4 LTS, CUDA 13.0, Python 3.11.15, and `PyTorch` 2.11.0+`cu130` for `aarch64`. For locally hosted inference, MARS used `gpt-oss-20b` in MXFP4 format with temperature set to 0 and a maximum generation length of 40,000 tokens. These specifications provide the computational context for the wall-clock measurements reported below.

We further analyzed the computational behavior of MARS in terms of agent interactions and wall-clock execution time. Figure S5 summarizes the run-time profiles of the five completed query executions shown in the analysis. The total number of agent interactions varies substantially across queries, ranging from 66 to 146 interactions per run, reflecting differences in the number of candidate-generation and validation steps required before reaching a final decision. Research Manager and Research Analyst calls account for most of these interactions, whereas Research Scientist and Research Assistant calls constitute a smaller and more consistent fraction of the overall workload.

The wall-clock profiles show that System 2, Grounded Material Candidate Generation, is the dominant computational stage. Across the five query executions, System 2 accounts for approximately 56–78% of the total run time, while total execution time ranges from 12.6 to 23.9 min. This concentration of computational cost in System 2 is consistent with its iterative structure: candidate generation may require multiple proposal, retrieval, validation, and rejection cycles before a candidate passes the system's internal screening criteria. In contrast, System 1 is executed once to establish the requirements and constraints, while System 3 operates primarily on candidates that have already passed System 2 screening.

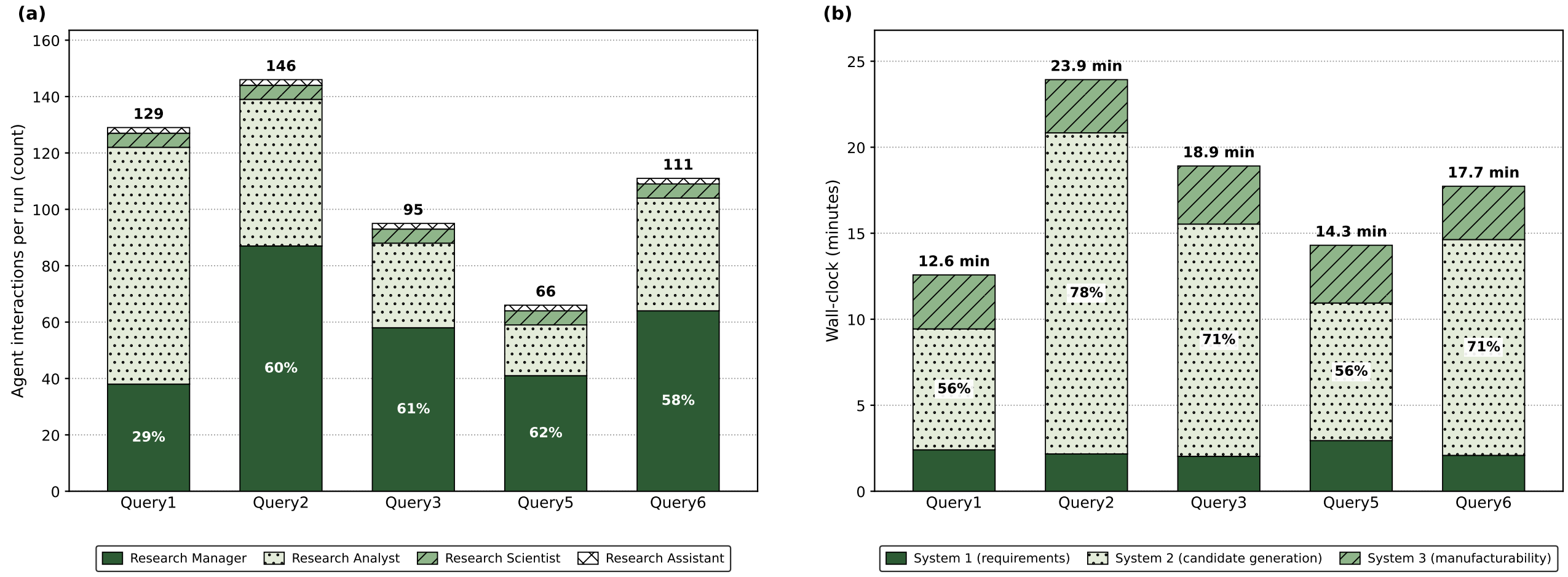


Figure S5: Run-time characteristics of MARS across repeated executions. (a) Agent interactions per run, separated by Research Manager, Research Analyst, Research Scientist, and Research Assistant activity. Numbers above the bars indicate total interactions, while percentages inside the bars show the Research Manager contribution. (b) Wall-clock time per run, decomposed across Systems 1–3. Numbers above the bars indicate total execution time, while percentages indicate the fraction spent in System 2.

Table S4 further compares the run-time profile of MARS for the THV substitution task using the two backbone models. For `gpt-oss-20b`, successful runs require an average of $156.0 \pm 39.7$ agent interactions and $20.2 \pm 5.1$ min of wall-clock time. With `gpt-5.6-sol`, the corresponding values are $179.2 \pm 46.0$ interactions and $18.9 \pm 7.8$ min. Despite the larger mean interaction count for `gpt-5.6-sol`, the average end-to-end execution time is similar between the two backbones.

For both models, System 2 remains the principal contributor to execution time, requiring $14.3 \pm 5.1$ min with `gpt-oss-20b` and $13.6 \pm 8.2$ min with `gpt-5.6-sol`. System 1 is comparatively stable, requiring 21 agent interactions for both backbones and approximately 2–3 min per run. System 3 requires substantially more retrieval activity than LLM activity: its Research Analyst calls average 51.6 per run for `gpt-oss-20b` and 80.2 for `gpt-5.6-sol`, while the number of Research Manager calls remains fixed at 9.

Table S4: Run-time profile of MARS on the THV substitution query, by backbone model. Agent-interaction counts and wall-clock times are means per run over the runs that completed with a manufacturable verdict; ± values are standard deviations across those runs. Research Analyst entries are RAG retrieval calls and Research Scientist entries are knowledge-graph traversals; Research Manager and Research Assistant entries are LLM calls.

| | | Agent interactions (mean per run) | | | | | |
|---|---|---|---|---|---|---|---|
| Backbone | Subsystem | Manager | Analyst | Scientist | Assistant | Total | Wall-clock (min) |
| `gpt-oss-20b` ($n = 5$) | System 1 | 9.0 | 6.0 | 4.0 | 2.0 | $21.0 \pm 0.0$ | $2.7 \pm 0.4$ |
| | System 2 | 45.0 | 28.4 | 1.0 | 0.0 | $74.4 \pm 41.6$ | $14.3 \pm 5.1$ |
| | System 3 | 9.0 | 51.6 | 0.0 | 0.0 | $60.6 \pm 13.8$ | $3.2 \pm 0.3$ |
| | All systems | 63.0 | 86.0 | 5.0 | 2.0 | $156.0 \pm 39.7$ | $20.2 \pm 5.1$ |
| `gpt-5.6-sol` ($n = 4$) | System 1 | 9.0 | 6.0 | 4.0 | 2.0 | $21.0 \pm 0.0$ | $2.2 \pm 0.3$ |
| | System 2 | 41.0 | 27.0 | 1.0 | 0.0 | $69.0 \pm 47.4$ | $13.6 \pm 8.2$ |
| | System 3 | 9.0 | 80.2 | 0.0 | 0.0 | $89.2 \pm 18.6$ | $3.1 \pm 0.4$ |
| | All systems | 59.0 | 113.2 | 5.0 | 2.0 | $179.2 \pm 46.0$ | $18.9 \pm 7.8$ |

Overall, the run-time analysis shows that the additional reasoning and verification steps introduced by MARS remain tractable at the scale examined here, with complete successful THV substitution runs requiring approximately 20 min on average. The substantial variability in System 2 interaction counts and execution times also reflects the adaptive nature of the framework: computational effort increases when additional candidate-generation, retrieval, and validation cycles are required rather than remaining fixed across queries.

## Reproducibility Materials

Prompts, pipeline configuration files, intermediate pipeline artifacts, candidate-rejection records, and evaluation materials are provided in the accompanying MARS code repository. These materials include retrieved evidence and intermediate outputs from the THV case study, outputs from all evaluated framework configurations, the complete evaluation rubric, and the corresponding LLM-judge scores and rationales.

## Agent and Pipeline Prompts

The framework issues 49 distinct LLM prompts: 25 system prompts and 24 paired user-message templates, together about 41k characters of instruction text. They are therefore provided in the `config/prompts.yaml` file in the accompanying GitHub repository (`https://github.com/LAMM-MIT/MARS`).

Every prompt is addressed by a dotted key, for example `agents.research_manager.propose_candidate` for the System 2 candidate-proposal system prompt; the user message paired with a system prompt carries the same key with the suffix `_user_prompt`. Fields in braces such as `{query}` are filled at run time; Table S5 maps each group of keys to the agent or pipeline stage that uses it. No prompt is constructed in source code outside this file, so the table covers the complete set. Agent hyperparameters and retrieval settings are in `config/config.yaml`, and the fully instantiated prompts of the reported run.

Table S5: Index of the prompts in `config/prompts.yaml`. Sys.: system prompts; User: paired user-message templates.

| Key prefix | Component | Sys. | User | Used for |
|---|---|---|---|---|
| `agents.research_manager` | Research Manager | 8 | 8 | Systems 1–3: research-question generation, answer synthesis, answered self-check, candidate proposal, validation queries, feasibility verdict, rejection lessons |
| `agents.research_assistant` | Research Assistant | 2 | 2 | System 1: extraction of material-property keywords and hard constraints |
| `agents.research_scientist` | Research Scientist | 1 | 1 | System 2: knowledge-graph node classification for subgraph construction |
| `pipelines.material_discovery` | System 2 pipeline | 2 | 2 | PFAS screening of proposals; extraction of Material-Informed Subgraph insights |
| `pipelines.manufacturability_assessment` | System 3 pipeline | 6 | 6 | constituent decomposition, process-query generation, feasibility questions and answers, verdict, recipe synthesis |
| `ablation` | Ablation baselines | 5 | 5 | the three-call and single-call configurations of Table 1 |
| `llm` | Shared default | 1 | 0 | fallback system prompt when a call defines none |

## Evaluation Prompts

Each subsystem is judged independently: one API call per subsystem receives the system prompt and the user message reproduced below, and returns a JSON object with one score, its strengths, weaknesses, rationale and confidence per criterion. The user message is assembled at run time from the rubric file, so the text below is the exact prompt content, with the substitution query and the subsystem report replaced by placeholders. Scores use a 0–10 scale in steps of 0.5.

### S0.0.1 System 1 judge prompt

Rubric file: `evaluation_refined/rubric_s1_final.yaml`; 4 criteria; scale 0–10 in steps of 0.5.

**System 1 — judge system prompt**

```
You are an expert materials scientist acting as an evaluator.  You will receive the SYSTEM 1 output of a
material-substitution pipeline:  a report containing the input query, the initial retrieval, each research
question with the LLM answer, that answer's sources (RAG document names and knowledge-graph node names) and
a 0/1 answered self-check, and the derived material properties and hard constraints.
  Evaluate ONLY the System 1 criteria in the rubric, using the score scale defined in the rubric.
  IMPORTANT:
- Be critical and specific; cite the exact items, numbers, or sources you relied on.
- Bracketed markers such as [Document 3] inside an answer are evidence citations:  they refer to that
question's numbered Sources list.  An answer's statement that the documents do NOT contain a value is
negative evidence.
- The 'Derived outputs by LLM' section is the product being judged; the Q&A section is its provenance.  A
derived number that appears in no answer and no document is unsupported.
- Do NOT evaluate candidate discovery or manufacturability; they are downstream systems and absent from this
report.
  You MUST respond with a valid JSON object and nothing else.
```

**System 1 — judge user message**

```
## Material Substitution Query
 <the substitution query, e.g.  Text Box 1>
 ## Evaluation Rubric (System 1 only)
### Score scale (0-10, step 0.5)
- 0 -- No meaningful success:  the criterion is essentially not addressed.
- 1 -- Very poor:  fundamental failures dominate.
- 2 -- Poor:  major deficiencies substantially undermine usefulness.
- 3 -- Weak:  some useful content exists, but serious problems remain.
- 4 -- Below adequate:  partially successful, with significant weaknesses.
- 5 -- Adequate:  useful overall, but with meaningful strengths and weaknesses.
- 6 -- Solid:  generally successful, though notable limitations remain.
- 7 -- Good:  satisfies the criterion well; weaknesses are mostly localized.
- 8 -- Strong:  satisfies the criterion very well; remaining issues are limited.
- 9 -- Excellent:  highly successful, with only minor deficiencies.
- 10 -- Outstanding:  comprehensively satisfies the criterion with no substantive deficiency.
- Use 0.5 increments when the result falls meaningfully between adjacent integer anchors.
- Do not target a particular score distribution.
  ### Evaluation principles
Evaluate System 1 using both:
    - the final derived material properties and hard constraints; and
    - the research trace that produced them.
  System 1 performs scientific requirements analysis for downstream material
discovery.  It is not expected to produce a finalized engineering standard,
procurement specification, or complete experimental protocol.
  Apply the following principles throughout:
  1.  Query-derived requirements
      Requirements explicitly requested by the user are valid requirements even
```

if the research trace does not independently justify their importance.

2. Inference is allowed

   System 1 may synthesize or infer reasonable engineering requirements from incomplete information. A value or requirement is not wrong merely because it was not directly retrieved.

3. Unsupported is not the same as incorrect

   Treat missing evidence primarily as uncertainty. Penalize strongly only when a statement is implausible, contradictory, or likely to materially misdirect downstream screening.

4. Judge impact, not mistake count

   Distinguish localized, correctable issues from systemic scientific or structural failures. A few technical mistakes should have limited impact when the overall discovery strategy remains sound.

5. Use external knowledge conservatively

   General scientific knowledge may identify clear, well-established errors, but uncertain external facts, proprietary specifications, obscure regulations, or conventions not contained in the report should not dominate the score.

6. Avoid double counting

   Each dimension measures a different aspect of System 1. Do not repeat the same deduction across dimensions unless it creates a distinct problem for each criterion.

The ANSWERED / NOT ANSWERED agent self-check is contextual information, not an automatic grading signal. Recognizing insufficient evidence can itself be scientifically appropriate.

### System 1 -- Completeness (weight: 1.0)

Primary question: Does System 1 identify the important requirement categories needed to evaluate a substitute for the material and application described in the query?

Assess the breadth and coverage of the final requirement set.

Focus on whether System 1 captures the major categories needed for downstream candidate screening, such as:

- chemical resistance;
- mechanical behavior;
- physical properties;
- optical properties;
- transport/barrier properties;
- thermal behavior;
- processing/manufacturing considerations;
- composition or regulatory constraints;
- relevant application-specific requirements.

Judge completeness relative to the actual query.

Do not reduce completeness simply because:

- a requirement lacks a numerical target;
- supporting evidence is incomplete;
- a requirement remains qualitative;
- a quantitative threshold needs later refinement.

Do not reward redundancy simply because the list is long.

Scoring guidance:

0-2:

Most major requirement categories are absent.

3-4:

Several important categories are missing or substantially underrepresented.

5-6:

The main categories are present, but meaningful gaps remain.

7-8:

The important requirement categories are well covered, with only limited omissions.

9-10:

```
      Coverage is comprehensive and appropriately scoped to the substitution
      problem.
  Use 0.5 increments when appropriate.
  ### System 1 -- Scientific correctness (weight:  1.0)

Primary question:  Is the overall scientific framework appropriate for identifying and organizing
the properties that should guide material substitution?
  Evaluate SCIENTIFIC CORRECTNESS primarily at the level of the overall

requirement-discovery framework.
  The main question is:
     "Does System 1 identify the correct scientific phenomena and organize the

     substitution problem in a scientifically sensible way?"
 Give greatest weight to whether:

    - the relevant scientific phenomena are identified;
    - the major property categories are appropriate;
    - the qualitative property directions are scientifically sensible;
    - the screening logic would guide material discovery toward relevant
      candidate classes.
  Examples of core scientific phenomena include:

    - chemical compatibility and swelling;
    - mechanical flexibility and deformation;
    - optical transmission;
    - gas transport and barrier behavior;
    - thermal stability;
    - composition or regulatory restrictions;
    - processing-related material behavior.
  If these phenomena and their role in material screening are correctly

identified, the scientific framework is fundamentally sound.
  --------------------

LOCAL TECHNICAL ISSUES
--------------------
 Treat detailed requirement-level problems as LOCAL issues unless correcting

them would change the overall scientific strategy.
  Examples include:

    - an incorrect or unsupported numerical threshold;
    - an imperfect unit expression;
    - a nonstandard test definition;
    - an overly severe exposure condition;
    - an isolated material-classification error;
    - an isolated regulatory-definition error;
    - overgeneralization from a related material family;
    - an unsupported working assumption.
  Multiple local issues may coexist with a scientifically sound framework.

  Do NOT lower the score substantially merely because several such issues can

be identified.
  Ask instead:
     "If these individual details were corrected, would the same major property

     categories and screening strategy still remain?"
 If YES, the errors are localized and the scientific framework should normally

remain in the 7-8 range.
  --------------------

INFERENCE AND WORKING REQUIREMENTS
--------------------
 System 1 is performing exploratory requirements analysis, not writing a

finalized material specification or experimental standard.
  It may infer provisional thresholds, test conditions, or engineering

requirements from incomplete information.
  Lack of direct literature support does not by itself make such an inference

scientifically incorrect.
```

```
  Penalize an inferred requirement strongly only when it is physically
implausible or would materially redirect the overall discovery strategy.
  --------------------

SUBSTANTIAL SCIENTIFIC ERRORS
--------------------
 Give large deductions only when the CORE scientific framework is wrong.
  Examples include:
    - missing or misunderstanding a central physical or chemical phenomenon;
    - reversing an important scientific relationship;
    - constructing a requirement framework that systematically searches the
        wrong material space;
    - making the major property categories fundamentally inappropriate for the
        substitution problem;
    - creating contradictions that invalidate the overall substitution
        strategy.
  A mistake in one material example, one regulatory definition, one unit, one
test condition, or one numerical requirement is NOT by itself a failure of
the core framework.
  --------------------

SCORING GUIDANCE
--------------------
 0-2:
    The scientific framework is fundamentally incorrect or unusable.
  3-4:
    Major conceptual misunderstandings affect several central parts of the
    discovery strategy.
  5:
    The broad framework is recognizable, but at least one major conceptual
    scientific problem substantially distorts the screening strategy.
  6:
    The overall scientific framework is sound, but one or more important
    conceptual weaknesses meaningfully affect how the material space would be
    screened.
  7:
    The scientific framework is clearly sound and useful.  Several localized
    technical, numerical, classification, or test-definition issues may remain.
  8:
    The scientific framework is strong and scientifically coherent.  Remaining
    issues are primarily detailed or correctable without changing the overall
    discovery strategy.
  9-10:
    The scientific framework is exceptionally coherent and technically strong,
    with only minor or negligible issues.
  --------------------

CALIBRATION
--------------------
 If:
    - the major scientific phenomena are correctly identified;
    - the important property categories are appropriate;
    - the qualitative screening logic is scientifically sensible;
    - and correcting the identified errors would NOT change the overall
        discovery strategy,
 then the score should generally be 7-8.
  Scores below 6 should require a substantive error in the underlying
scientific framework.
  Do not use the number of localized errors as a proxy for framework quality.
  Do not evaluate this metric as a detailed audit of every final hard
constraint.
  Use 0.5 increments when appropriate.
```

```
  -----------
SCOPE NOTE
-----------
Evidence lineage is scored separately under "Evidence grounding";
do not deduct here solely because a value lacks retrieval backing.
Self-evident scientific errors, internal contradictions, and claims
that contradict evidence quoted elsewhere in the same report reduce
this score in every case.
  ### System 1 -- Evidence grounding (weight:  1.0)

Primary question:  Are the final derived properties and hard constraints traceable to
retrieved evidence contained in the report, with honest handling of
what the evidence does not contain?
  Assess the lineage between the derived outputs and the evidence in

the report.
  Where evidence lives:  answers cite retrieved documents inline with

bracketed markers such as [Document 1]; these resolve to the
numbered "Sources" list under each question.  Statements about what
the documents do NOT contain are negative evidence and count as
honest grounding.  In other report formats, a pre-retrieved evidence
packet shows what retrieval existed.  You are NOT required to verify
individual excerpts; assess lineage.
  Scoring guidance:

      null (not scoreable):

        The report contains NO retrieval of any kind (no cited answers,
        no evidence packet).  This capability is absent by construction;
        output null for the score, not a number.
      0-2:

        Retrieval exists in the report but the derived outputs ignore
        it:  values are not traceable to any cited answer or packet
        content, or citations do not resolve.
      3-4:

        A minority of the load-bearing values trace to evidence, or
        lineage is present but repeatedly broken:  values declared
        missing from the documents nonetheless appear as specific
        numbers (fabrication), or outputs contradict quoted evidence.
      5-6:

        A majority of the load-bearing values trace to cited evidence
        or honest assumptions; remaining gaps and untraceable values
        are limited and identifiable.
      7-8:

        Nearly all load-bearing values are traceable or explicitly
        flagged as assumptions; negative evidence is carried honestly
        into the outputs.
      9-10:

        Complete, transparent lineage:  every quantitative requirement
        is cited, derived, or declared an assumption, with no
        fabrication anywhere.
  Credit intact lineage first:  score by the fraction of load-bearing

values that trace to cited evidence, resolvable citations, honestly
declared negative evidence, or explicitly labeled assumptions.
  Fabricated values (declared missing, then quantified) are the most

serious grounding failure, but they cap the score rather than erase
the credit above:  one or two isolated fabricated values cap this
score at 6; repeated fabrication across several load-bearing values
caps it at 4.5.
  Use 0.5 increments when appropriate.
  ### System 1 -- Investigation quality (weight:  1.0)

Primary question:  Does the report show a genuine investigation:  decomposition of the
problem into research questions, question-directed evidence
gathering across sources, honest assessment of what was and was not
answered, and synthesis of the findings into the requirements?
```

Assess the INVESTIGATION PROCESS visible in the report, not the polish of the final list.

Give credit for:

- decomposing the substitution problem into distinct, purposeful research questions;
- evidence gathered PER QUESTION, from multiple source types where available (documents, patents, material databases, knowledge-graph relationships);
- honest self-assessment (a question marked unanswered when the evidence was insufficient is a strength, not a weakness);
- iteration or follow-up where first-pass evidence was insufficient;
- final requirements that are recognizably synthesized FROM the investigation's findings rather than emitted independently of them.

Deduct when:

- questions are decorative: their answers are never used by the synthesis, or the synthesis contradicts them;
- the investigation covers only a narrow slice of the problem;
- self-checks are dishonest (claimed answered without support).

Scoring guidance:

null (not scoreable):

The report contains NO investigation trace at all: no research questions, no per-question evidence, no self-checks (for example, a single direct generation, with or without a pasted evidence packet). This capability is absent by construction; output null for the score, not a number.

0-2:

A trace exists but is vacuous: questions unrelated to the problem, no evidence behind answers, or self-checks that are plainly false.

3-4:

A partial investigation: some purposeful questions or staged analysis, but with little or no per-question evidence, or findings largely disconnected from the final requirements.

5-6:

A real investigation with meaningful gaps: purposeful questions and per-question evidence, but limited source diversity, weak follow-up, or a synthesis that uses only part of the findings.

7-8:

A strong investigation: well-chosen questions, evidence from multiple source types, honest self-assessment, and a synthesis that visibly builds on the findings.

9-10:

An exemplary investigation: comprehensive decomposition, well-directed multi-source evidence, honest handling of unanswered questions, and a synthesis faithful to all of it.

Use 0.5 increments when appropriate.

## System 1 Report

<the System 1 report for one run: the query, the retrieval and reasoning trace, and the final System 1 output, rendered as Markdown>

## Required Output Format

For each dimension return:

- score:

  Number from 0 to 10 in increments of 0.5, or null where the dimension's guidance says the capability is absent by construction.
- strengths:

  1-3 concise observations specific to this criterion.

```
      - weaknesses:

            1-3 concise observations describing the most important limitations
            specific to this criterion.
      - rationale:

            Brief explanation of why the strengths and weaknesses justify the score.
      - confidence:

            "high", "medium", or "low".
  Refer to relevant research questions, findings, derived properties, or hard

constraints when useful.
  Do not mechanically repeat the same criticism across dimensions.

  Do not calculate an overall score or subsystem average.

  Keep the evaluation concise.

  Respond with ONLY a JSON object in this exact structure:

  ‘‘‘
```


```
{
    "system1_completeness": {
        "score": 0.0,
        "strengths": [],
        "weaknesses": [],
        "rationale": "",
        "confidence": "high"
    },
    "system1_scientific": {
        "score": 0.0,
        "strengths": [],
        "weaknesses": [],
        "rationale": "",
        "confidence": "high"
    },
    "system1_grounding": {
        "score": 0.0,
        "strengths": [],
        "weaknesses": [],
        "rationale": "",
        "confidence": "high"
    },
    "system1_investigation": {
        "score": 0.0,
        "strengths": [],
        "weaknesses": [],
        "rationale": "",
        "confidence": "high"
    }
}
‘‘‘
```


### S0.0.2 System 2 judge prompt

Rubric file: `evaluation_refined/rubric_s2_final.yaml`; 4 criteria; scale 0–10 in steps of 0.5.

**System 2 — judge system prompt**

```
You are an expert materials scientist acting as an evaluator.  You will receive the SYSTEM 2 output of
a material-substitution pipeline:  a report containing the System 1 requirement inputs (properties and
hard constraints), the knowledge-graph grounding steps, and every candidate iteration:  the proposal with
justification, the PFAS checks, the validation questions, the retrieved-evidence answers (with source
document ids), the feasibility verdict with reasoning, and the failure summaries between iterations.
  Evaluate ONLY the System 2 criteria in the rubric, using the score scale defined in the rubric.

  IMPORTANT:

- Be critical and specific; cite the exact iterations, claims, document ids, or verdict text you relied on.
```

```
- The final accepted candidate is the primary object being judged; the rejected iterations are evidence of
the selection process's quality.
- A quantitative claim in a proposal or answer counts as supported only if the report shows retrieved text
stating it; requirement numbers echoed back as candidate properties are unsupported.
- Do NOT evaluate System 1's requirement quality or System 3's manufacturability; they are outside this
report's scope.
  You MUST respond with a valid JSON object and nothing else.
```

**System 2 — judge user message**

```
## Material Substitution Query
 <the substitution query, e.g.  Text Box 1>

 ## Evaluation Rubric (System 2 only)

### Score scale (0-10, step 0.5)
- 0-2 -- Poor:  the capability is absent or largely ineffective.
- 3-4 -- Weak:  some useful behavior is present, but the capability is substantially limited.
- 5 -- Adequate:  the capability works partially but inconsistently.
- 6 -- Good:  the capability clearly works and is useful, despite notable correctable issues.
- 7 -- Strong:  the capability works well overall; remaining weaknesses are localized.
- 8 -- Very strong:  the capability is consistently effective with limited substantive weaknesses.
- 9-10 -- Excellent:  exceptionally effective, with only minor or no substantive deficiencies.
- Use 0.5 increments when appropriate.
- Do not target a predetermined score distribution.
  ### Evaluation principles

Evaluate System 2 as an iterative material-candidate discovery system.
  Use the complete trace when available, including:

    - System 1 requirements;
    - candidate proposals;
    - retrieved RAG evidence;
    - validation questions;
    - rejection decisions;
    - failure summaries;
    - later candidate revisions;
    - and the final candidate.
  System 2 is responsible for candidate discovery and evaluation.

Detailed manufacturing and process design belong to System 3 and should not
affect these scores.
  SCORE THE CAPABILITY FIRST.

 First determine whether the capability measured by the metric is:

    - absent,
    - weak,
    - functional,
    - strong,
    - or exceptional.
  A clearly functional capability should normally fall in the 6-8 range.

  Local technical errors should receive proportional deductions only when they

undermine the specific capability being scored.
  Do not start from 10 and subtract points for every mistake.

  Do not double-count the same issue across metrics.

  Important metric boundaries:

      - Requirement Alignment:

            Does the requirement set guide the workflow?
            It does not measure final proof of compliance.
      - Adversarial Verification:

            Was the candidate interrogated by checks independent of the proposal,
            with verdicts grounded in evidence or explicit negative evidence?
            It does not measure whether the candidate is good.
      - Candidate Anchoring:

            Is the candidate a specific, procurable material whose identity and
            claimed properties are supported by inventory or datasheet evidence
```

```
            inside the report?
            It does not measure novelty or final performance.
      - Adaptive Reasoning:

            Does feedback change later candidate proposals along the named
            failure axis?
            It does not require the revised candidate to solve every previous problem.
  Missing evidence is not automatically failure.

  A rejected candidate is not a system failure when the rejection produces

useful information for later iterations.
  SINGLE-ITERATION RUNS.

 When System 2 accepts its FIRST proposed candidate, there are no rejections,

no failure summaries, and no revisions.  The exploration and adaptation
capabilities are then NOT EXERCISED, not absent:  the run gave them no
opportunity to appear.  Accepting the first candidate must be neither
punished nor rewarded on those metrics; score them as null (see the
dimension texts and output instructions).
  SINGLE-CALL REPORTS.

 When a report is a single direct generation (no proposal loop, no

constraint gates, no validation questions), the verification and
adaptation machinery is ABSENT BY CONSTRUCTION, not merely unexercised.
Score system2_verification and system2_adaptation as null for such
reports.  Candidate anchoring IS still scoreable:  a grade named from
memory with no supporting evidence in the report is weak anchoring, not
a missing capability.
  Use the supplied trace as the primary basis for scoring.  External knowledge

should not dominate the evaluation.
  ### System 2 -- Requirement alignment (weight:  1.0)

Primary question:  Do the System 1 requirements actively guide candidate generation,
validation, rejection, and comparison throughout System 2?
  Evaluate whether the REQUIREMENTS GUIDE THE DISCOVERY PROCESS.

 This metric measures alignment of the workflow with the System 1 requirement

set.  It does NOT primarily measure whether the final candidate ultimately
satisfies every requirement.
  Give strong credit when:

    - candidate proposals are motivated by System 1 requirements;
    - validation questions directly examine important target properties;
    - rejection decisions refer to requirement violations or unresolved gaps;
    - later proposals respond to requirement-related failures;
    - the final candidate is explicitly assessed against the requirement set.
  The central question is:

      "Are the requirements visibly influencing what System 2 proposes,

      validates, rejects, and compares?"
 If yes, requirement alignment is strong.

  Do NOT substantially lower this score because:

    - a requirement remains unverified;
    - candidate-specific evidence is missing;
    - the final candidate may not actually satisfy a requirement;
    - the final feasibility verdict is overly optimistic;
    - one hard constraint is applied inconsistently;
    - unresolved requirements remain at final selection.
  These issues concern evidence quality, feasibility judgment, or scientific

correctness.  They should affect Requirement Alignment only when they show
that the requirement itself no longer guides the discovery process.
  Lower this score substantially only when:

    - major requirements are absent from candidate reasoning;
    - validation questions ignore important user priorities;
    - rejection decisions are largely disconnected from the requirements;
    - later candidate proposals do not reflect important requirement-related
        failures;
    - the workflow drifts toward properties unrelated to the stated task.
```

```
  Scoring guidance:

      0-2:

        Requirements have little meaningful influence on discovery.
      3-4:

        Several major requirements are largely ignored.
      5:

        Requirements influence the workflow, but important categories are
        inconsistently represented.
      6:

        Requirements clearly guide much of the workflow, though coverage or use
        is uneven.
      7:

        Major requirements clearly guide candidate generation, validation,
        rejection, and final assessment.
      8:

        Requirement guidance is strong and consistent across nearly every stage
        of the discovery process.
      9-10:

        Requirement alignment is exceptionally systematic and comprehensive.
  Calibration:

    If the major System 1 requirements repeatedly appear in:
        - candidate rationales;
        - validation questions;
        - rejection decisions;
        - later candidate revisions;
        - and final candidate assessment,
     then the score should generally be in the 7-8 range.

      Final candidate compliance should NOT be the primary basis for reducing

    this metric.
      A hard-constraint inconsistency should meaningfully reduce this score only

    if it demonstrates that the constraint stopped influencing candidate
    discovery or evaluation.
  Use 0.5 increments when appropriate.

  ### System 2 -- Adversarial verification (weight:  1.0)

Primary question:  Was each proposed candidate interrogated by checks independent of the
proposal itself, and were accept/reject verdicts grounded in retrieved
evidence or explicit negative evidence rather than in the proposal's
own claims?
  Evaluate the VERIFICATION MACHINERY, not the quality of the candidate.

  Where verification lives in these reports:  constraint gates applied to

each proposal (for example a compositional PFAS check with a tie-break
re-query), candidate-specific validation questions, per-question
retrieved evidence, and structured feasibility verdicts that name the
constraints they rest on.
  Give credit when:

    - the checks are INDEPENDENT: they re-examine the proposal's claims
        rather than restating them, and they are able to contradict the
        proposal (a gate that overrules a proposal's own compositional or
        compliance claim is the clearest positive signal);
    - validation questions target the load-bearing claims and hard
        constraints of that specific candidate;
    - answers state honestly what the retrieved evidence does NOT
        contain, and that negative evidence is carried into the verdict;
    - verdicts quote supporting or negative evidence and name the
        constraint at stake;
    - a hard constraint that cannot be verified from evidence BLOCKS
        acceptance (fail-closed) instead of being assumed satisfied.
  The overturn signal:  verification that never disagrees with,

qualifies, or blocks any claim across the whole run is weak evidence
that checking actually occurred; credit gates that visibly bite.
```

```
  Serious failures:

    - FABRICATED VERIFICATION: a proposal or verdict asserting
        certifications, laboratory reports, or third-party tests that
        appear nowhere in the retrieved evidence caps this score at 4.
    - RUBBER-STAMPING: an acceptance verdict that ignores its own
        negative evidence on a hard constraint caps this score at 5.
  Do NOT lower this score because:

    - verification found gaps or rejected candidates (that is the
        machinery working);
    - the corpora genuinely lack data and the answers say so ("no data
        found" is functioning verification, not failure);
    - the accepted candidate still carries open risks that the verdict
        names honestly.
  Scoring guidance:

      null (not scoreable):

        The report contains NO verification machinery:  no constraint
        gates, no validation questions, no evidence-based verdict (a
        single direct generation).  Output null for the score.
      0-2:

        Machinery exists but is decorative:  questions unrelated to the
        candidate's claims, answers never used, verdicts unconnected to
        evidence.
      3-4:

        Checks run, but verdicts do not trace to the evidence, or
        verification is asserted rather than performed (including
        fabricated certifications).
      5-6:

        Functional verification with gaps:  key claims tested against
        evidence and verdicts mostly grounded, but some load-bearing
        claims or constraints pass unexamined.
      7-8:

        Independent, evidence-quoting checks cover the load-bearing
        claims; at least one check visibly bites; unverifiable hard
        constraints block acceptance.
      9-10:

        Systematic per-claim verification with consistently grounded
        verdicts across all iterations.
  Use 0.5 increments when appropriate.

  ### System 2 -- Candidate anchoring (weight:  1.0)

Primary question:  Is the proposed material a specific, procurable candidate whose
identity and claimed properties are supported by inventory records or
datasheet evidence inside the report, rather than recalled from
parametric memory?
  Evaluate PROCURABILITY AND IDENTITY CONSISTENCY of the final candidate

(and, for composites, each named component).
  Where anchoring lives in these reports:  database/catalog IDs attached

to proposed components, material-database chunks appearing in the
retrieved validation evidence, and datasheet-derived properties quoted
in answers.
  Give credit when:

    - components are specific procurable grades tied to an inventory ID
        or datasheet evidence present in the report;
    - the properties the justification claims for a component are
        CONSISTENT with the in-report evidence about that component;
    - for composites, the report distinguishes which component comes
        from the inventory and which is external, and external components
        are justified from retrieved sources;
    - quantitative component claims trace to datasheet or retrieved
        content rather than appearing from nowhere.
  Deduct when:

    - a component's IDENTITY is wrong relative to its own evidence (for
```

```
        example a polyolefin elastomer described as a polyurethane):  treat
        identity misclassification as a substantive anchoring failure;
    - properties are attributed to a named grade with no in-report
        support, or contradict the retrieved datasheet content;
    - naming is generic or ambiguous ("a transparent polyurethane") so
        the material cannot actually be procured;
    - the candidate is a real-sounding commercial grade recalled from
        memory with NO supporting evidence anywhere in the report:  this
        is weak anchoring (score low), NOT a missing capability (do not
        output null for it).
  This metric is scoreable for every configuration.

  Scoring guidance:

      0-2:

        No specific procurable identity, or the named material has no
        support of any kind in the report.
      3-4:

        A specific grade is named, but its identity or claimed properties
        are unverifiable from the report or contradicted by it.
      5-6:

        Anchored to inventory or datasheet evidence, with some
        unsupported or inconsistent component-level claims.
      7-8:

        Anchored and consistent:  identities correct, claimed properties
        supported, inventory vs external components distinguished.
      9-10:

        Every component-level claim traceable to in-report evidence with
        no identity or consistency errors.
  Use 0.5 increments when appropriate.

  ### System 2 -- Adaptive reasoning (weight:  1.0)

Primary question:  When a problem is identified, does that feedback cause a meaningful change
in a later candidate proposal?
  Evaluate FEEDBACK-DRIVEN CHANGE ACROSS ITERATIONS.

 Look for the pattern:

      problem identified

    → lesson extracted
    → later candidate changed
 Give credit when:

    - the change is along the NAMED failure axis:  the next proposal
        addresses the specific reason recorded in the previous rejection
        or failure summary, not merely something different;
    - problematic constituents are removed and rejected chemistries
        avoided;
    - the matrix, material family, filler, or architecture changes in
        response to feedback;
    - when a manufacturability (System 3) rejection returned blocking
        constraints, a later proposal visibly incorporates them.
  Adaptation measures behavioral change, not eventual success.

  The revised candidate does NOT need to:

    - solve every previous deficiency;
    - eliminate all evidence gaps;
    - satisfy every hard constraint;
    - be scientifically flawless;
    - or ultimately succeed.
  A later candidate may contain new mistakes and still demonstrate successful

adaptation.
  Lower the score substantially when failure feedback produces little

change, a later proposal REPEATS a failure mode already named in a
failure summary, rejected strategies are reused unchanged, or later
iterations proceed as though previous failures did not occur.
  Scoring guidance:
```

```
      0-2:

        No meaningful feedback-driven change occurs.
      3-4:

        Feedback exists but has little effect on later proposals.
      5:

        At least one meaningful adaptation occurs.
      6:

        Multiple adaptations occur, though inconsistently.
      7:

        Failure feedback clearly and repeatedly changes subsequent candidates.
      8:

        Multiple iterations demonstrate strong feedback-driven changes in
        material, chemistry, or architecture.
      9-10:

        Exceptionally consistent and effective adaptive behavior.
  Calibration:

    One clear feedback-driven candidate change demonstrates adaptation.
      Multiple clear examples should generally support a score of 7-8.

      Do not require every identified problem to be solved.

  NOT EXERCISED / NOT SCOREABLE:

    If the run contains only ONE candidate proposal (the first candidate
    was accepted), no feedback loop ever ran and this capability was
    never exercised.  Return "score":  null with a rationale stating that
    adaptation was not exercised.  Do not score it low for the absence of
    iterations and do not score it high by default; a single-proposal run
    is neither punished nor rewarded on this metric.
      If the report is a single direct generation with no proposal loop at

    all, the capability is absent by construction:  also return
    "score":  null and say so in the rationale.
  Use 0.5 increments when appropriate.

  ## System 2 Report


 <the System 2 report for one run:  the query, the retrieval and reasoning trace,

and the final System 2 output, rendered as Markdown>
 ## Required Output Format

 Return one evaluation for each dimension.

  For each dimension provide:

      - score:

            Number from 0 to 10 in increments of 0.5, or null when the dimension
            text marks the capability as NOT EXERCISED for this run.  A null score
            is excluded from averages downstream; never substitute 0 or a guessed
            value for an unexercised capability.
      - strengths:

            1-3 concise observations showing why the capability is functioning.
      - weaknesses:

            At most 1-2 concise observations describing the most important
            limitations specific to that metric.
      - rationale:

            Brief explanation connecting the observed capability and limitations to
            the score.
      - confidence:

            "high", "medium", or "low".
  Scoring procedure:

      1.  Determine whether the capability is absent, weak, functional, strong,

          or exceptional.
    2.  Select the corresponding score range.
    3.  Consider only weaknesses directly relevant to that metric.
```

```
    4.  Adjust within the range proportionally.
  Do not:
    - start from 10 and subtract for every mistake;
    - enumerate every factual error;
    - let the number of weaknesses determine the score;
    - repeat the same criticism across metrics;
    - evaluate System 3 manufacturing/process capabilities;
    - require exhaustive candidate exploration;
    - require complete experimental verification.
  A functional capability with correctable execution errors should generally
fall in the 6-8 range.
  Keep the evaluation concise.
  Respond with ONLY a JSON object in this exact structure:
  ```
{
    "system2_alignment":  {
        "score":  0.0,
        "strengths":  [],
        "weaknesses":  [],
        "rationale":  "",
        "confidence":  "high"
    },
    "system2_verification":  {
        "score":  0.0,
        "strengths":  [],
        "weaknesses":  [],
        "rationale":  "",
        "confidence":  "high"
    },
    "system2_anchoring":  {
        "score":  0.0,
        "strengths":  [],
        "weaknesses":  [],
        "rationale":  "",
        "confidence":  "high"
    },
    "system2_adaptation":  {
        "score":  null,
        "strengths":  [],
        "weaknesses":  [],
        "rationale":  "",
        "confidence":  "high"
    }
}
```
```

### S0.0.3 System 3 judge prompt

Rubric file: `evaluation_refined/rubric_s3_final.yaml`; 4 criteria; scale 0–10 in steps of 0.5.

**System 3 — judge system prompt**

```
You are an expert materials scientist and process engineer acting as an evaluator.  You will receive the
SYSTEM 3 output of a material-substitution pipeline:  a manufacturability report containing the candidate
handed over by System 2 with its justification, the single-vs-composite decomposition, the manufacturing
process queries with their retrieved evidence (manufacturing textbooks, materialdb, patents), the four
feasibility questions with evidence and LLM answers, the feasibility verdict, and the proposed process
recipe with its backing evidence.
  Evaluate ONLY the System 3 criteria in the rubric, using the score scale defined in the rubric.
  IMPORTANT:
- Be critical and specific; cite the exact steps, conditions, question answers, or document ids you relied
on.
```

```
- The process recipe and feasibility assessment are the primary objects being judged; take the candidate's
stated composition as given.
- A processing condition counts as supported only if the report's evidence or answers state it; check the
recipe's numbers against the Q&A answers for internal consistency.
- Do NOT evaluate System 1's requirement quality or System 2's candidate choice; they are outside this
report's scope.
  You MUST respond with a valid JSON object and nothing else.
```

**System 3 — judge user message**

```
## Material Substitution Query
 <the substitution query, e.g.  Text Box 1>

 ## Evaluation Rubric (System 3 only)

### Score scale (0-10, step 0.5)
- 0-2 -- Poor:  the capability is absent or largely ineffective.
- 3-4 -- Weak:  some useful behavior exists, but the capability is substantially limited.
- 5 -- Adequate:  the capability works partially but inconsistently.
- 6 -- Good:  the capability clearly works and is useful, despite notable correctable issues.
- 7 -- Strong:  the capability works well overall; remaining weaknesses are localized.
- 8 -- Very strong:  the capability is consistently effective with limited substantive weaknesses.
- 9-10 -- Excellent:  exceptionally effective, with only minor or no substantive deficiencies.
- Use 0.5 increments when appropriate.
  ### Evaluation principles

Evaluate System 3 as an evidence-driven MANUFACTURING AND PROCESS-DESIGN system.
  Use the complete System 3 trace when relevant, including:

    - the selected candidate;
    - manufacturing queries;
    - retrieved evidence;
    - feasibility analysis;
    - feasibility verdict;
    - final process recipe;
    - equipment and operating conditions;
    - QC, safety, handling, scale, packaging, and storage considerations.
  The FINAL PROCESS RECIPE is the primary manufacturing recommendation.

Earlier queries and feasibility answers are exploratory research used to inform
that recipe.
  Do not require exploratory parameter values to match the final recipe exactly.

Refining or changing temperatures, times, speeds, residence times, or other
process conditions after evidence review is normal process design.
  System 3 is NOT expected to provide:

    - a production-qualified industrial SOP;
    - a complete plant design;
    - a full hazard analysis;
    - an optimized process window;
    - or experimentally validated values for every parameter.
  SCORE THE CAPABILITY FIRST.

 First determine whether the capability is absent, weak, functional, strong,

or exceptional.  A clearly functional capability should normally fall in the
6-8 range.
  Before lowering a score, ask:

      "Does this issue materially change the answer to this metric's primary

      question?"
 If correcting the issue would leave the capability clearly intact, treat it

as localized.
  Localized errors should usually affect placement within the 6-8 range rather

than collapse a clearly functional capability into the 3-5 range.
  Do not start from 10 and subtract for every mistake.

  Do not double-count the same issue across metrics.

  Do not substantially penalize System 3 for an error inherited from System 2

unless System 3 converts that error into a problematic manufacturing decision.
```

Give explicit credit for evidence-driven process design when System 3:

- generates targeted manufacturing questions;
- retrieves textbooks, patents, material-database records, or other evidence;
- investigates feasibility;
- and synthesizes that information into a concrete process recommendation.

Do not turn the evidence into a source-by-source literature audit.

-----------

EVIDENCE IN THE REPORT
-----------
Where evidence lives in these reports: the per-query retrieved evidence sections, the four feasibility questions with their answers, and the recipe's backing-evidence document list. Inside answers, bracketed markers such as [Document 3] are citations to retrieved documents. In reports of ablated pipelines, a pre-retrieved packet or a "not produced in this condition" marker shows whether any retrieval existed at all. Specific dimensions below define how the presence or absence of evidence affects their scores; manufacturing route coherence remains judged from the recipe text itself.

### System 3 -- Manufacturing route coherence (weight: 1.0)

Primary question: Does System 3 provide a coherent end-to-end sequence of manufacturing operations from material inputs to the intended final product?

Evaluate the LOGIC AND SEQUENCE of the manufacturing route.

Give credit when:

- major processing stages are present;
- steps occur in a sensible order;
- the output of one stage plausibly feeds the next;
- raw materials progress toward a shaped, inspected, and handled product.

A strong route may include:

raw-material preparation
→ drying or conditioning
→ mixing or compounding
→ shaping
→ cooling or consolidation
→ post-processing
→ QC
→ packaging or storage

The exact sequence does not need to follow this template.

This metric does NOT primarily assess:

- exact process parameters;
- equipment sizing;
- material degradation;
- detailed QC;
- safety depth.

Do not substantially penalize:

- one redundant stage;
- one unnecessary optional step;
- imperfect terminology;
- one questionable parameter;
- a removable post-treatment.

Lower the score substantially when:

- major required stages are missing;
- step order is fundamentally incoherent;
- the stated operations cannot plausibly transform the inputs into the intended product;
- a decisive transformation is entirely unspecified.

Scoring guidance:

0-2:

No coherent manufacturing route is present.

3-4:

A route is recognizable but major stages are missing or disconnected.

5:

The general sequence is plausible but incomplete in important places.
6:

The route is coherent and usable, with notable gaps.
7:

The route clearly connects inputs to final product; weaknesses are localized.
8:

Strong end-to-end manufacturing coherence across nearly all stages.
9-10:

Exceptionally clear, complete, and integrated route.
Calibration:

A recognizable route from raw-material preparation through processing, shaping, QC, and final handling should generally score 7-8 unless a major transformation is missing or impossible.
Use 0.5 increments when appropriate.

### System 3 -- Operational practicality (weight: 1.0)

Primary question: Does System 3 provide a sufficiently concrete process plan that could plausibly be executed and refined by a practitioner?
Evaluate the OPERATIONAL SPECIFICITY AND EXECUTABILITY of the final recipe.

Give strong credit when major steps provide relevant information such as:

- equipment class;
- approximate temperature;
- time or residence time;
- speed or mixing intensity;
- pressure where relevant;
- throughput or scale;
- cooling, drying, or shaping conditions.

This metric does NOT require detailed process engineering or equipment optimization.
Do not independently redesign the line or require:

- validated die-flow calculations;
- feeder-sizing calculations;
- line balancing;
- pelletization details between every stage;
- pressure-drop calculations;
- exact equipment power sizing;
- complete material-handling design.

These details may improve an industrial specification but are not required to demonstrate operational practicality.
A parameter should substantially reduce this score only when it is clearly impossible, mutually incompatible with another REQUIRED core condition, or prevents the operation from being executed.
Non-optimal, oversized, underspecified, or approximate engineering choices are localized issues.
Optional steps should have limited influence on this metric.

Scoring guidance:

0-2:

The process lacks an executable operating plan.
3-4:

Multiple core operations lack usable equipment or conditions.
5:

A process can be inferred, but important core operating information is missing.
6:

The process is operationally understandable and mostly runnable, with notable gaps.
7:

Major operations contain concrete equipment and operating conditions; remaining issues are localized engineering refinements.

```
      8:

        The recipe is well specified across nearly all important operations.
      9-10:

        Exceptionally complete operational definition for the intended scope.
  Calibration:

    If the final recipe specifies usable conditions and equipment for the major
    manufacturing operations, the score should generally be 7-8.
      Do not reduce a clearly runnable process below 7 primarily because detailed

    line engineering, equipment sizing, intermediate handling, or optimization
    has not been completed.
  -----------
EVIDENCE-LINEAGE RULE (overrides the calibration above)
-----------
Quantitative processing conditions (temperatures, times, pressures,
speeds, ratios, loadings) count as VERIFIED only when traceable to the
report's retrieved evidence, feasibility answers, or backing documents,
or when the report explicitly derives them from such sources.  You are
not required to check exact excerpts; assess lineage.
  If the recipe's quantitative conditions are predominantly unverifiable

(no retrieval exists in the report, or the conditions do not trace to
any of it), cap this score at 5.  Plausible-looking numbers without
lineage must not be credited as verified.
  VERIFICATION DISCHARGE. Distinguish feasibility obligations the report

RESOLVES from obligations it DEFERS. An obligation is resolved when
the report's feasibility analysis answers it from retrieved evidence
(or honestly reports the evidence as absent); it is deferred when the
recipe hands it to the reader as future qualification work ("test
chemical compatibility with actual service chemicals").  Deferring is
honest and must not be scored below unsupported assertion, but it
earns no practicality credit:  a route whose qualification burden is
predominantly deferred is a test plan, not a validated procedure, and
should score in the middle of the scale at best.  Reserve high scores
for routes where the report itself discharged most load-bearing
feasibility questions with evidence.
  ### System 3 -- Material-process compatibility (weight:  1.0)

Primary question:  Are the major manufacturing operations scientifically compatible with the
selected material and composite architecture?
  Evaluate whether the PROCESS FAMILY AND CORE OPERATIONS FIT THE MATERIAL.

 Focus on whether major operations such as:

    - drying or conditioning;
    - compounding;
    - filler incorporation;
    - extrusion;
    - molding;
    - cooling;
    - shaping
 are broadly appropriate for the selected material system.

  This metric evaluates COMPATIBILITY, not proof of final performance.

  Do NOT substantially lower this score because:

    - exact dispersion quality has not been experimentally demonstrated;
    - grade-specific literature is incomplete;
    - a compatibilizer has not yet been selected;
    - the precise residence time or shear history still requires optimization;
    - final optical, mechanical, or barrier properties remain unverified.
  Those are optimization, validation, or evidence-sufficiency issues.

  Penalize substantially only when System 3 proposes a process that is itself

incompatible with the material, such as:
    - a required temperature that would clearly destroy the material;
    - an essential cure mechanism that the material cannot undergo;
    - an incompatible processing family;
    - an essential chemical treatment that cannot operate on the selected
```

```
        material.
  Optional or removable incompatible steps should receive less weight than

the viable core route.
  Scoring guidance:

      0-2:

        The central processing route is fundamentally incompatible.
      3-4:

        Major incompatibilities affect several core stages.
      5:

        The route is broadly plausible but contains an important central
        incompatibility.
      6:

        The process is generally compatible, with notable correctable concerns.
      7:

        Core processing operations are compatible; remaining concerns are
        localized.
      8:

        Strong compatibility across nearly all major operations.
      9-10:

        Exceptionally well matched material and process design.
  Calibration:

    If the selected material can reasonably undergo the proposed preparation,
    compounding, filler incorporation, shaping, and cooling operations, the
    score should generally be 7-8.
      Do not require evidence that the exact process will achieve every desired

    final property in order to award strong compatibility.
  -----------
EVIDENCE RULE (overrides the calibration above)
-----------
Self-evident incompatibilities (impossible chemistry, missing reaction
components, contradictory conditions) reduce the score in every case,
with or without evidence.
  Family-level compatibility (a standard process for a well-known material

class) may be credited from general knowledge.  GRADE-SPECIFIC claims,
that a named commercial grade tolerates a specific schedule, ratio, or
condition, count as verified only with lineage to the report's evidence.
A claim that names a specific material, composition, or commercial
grade together with a numeric processing window, schedule, ratio, or
condition is GRADE-SPECIFIC even when phrased generically for the
material class.  If the recipe's material-specific compatibility claims
are predominantly unverifiable, cap this score at 5.
  ### System 3 -- Feasibility-verdict grounding (weight:  1.0)

Primary question:  Is the manufacturable/blocked verdict an evidence-based conclusion
drawn from the report's own feasibility analysis, rather than an
assertion?
  Evaluate the VERDICT AND ITS GROUNDING, not the recipe.

  Where the material lives in these reports:  the manufacturing

feasibility questions with their numbered evidence and per-answer
confidence labels, and the feasibility verdict section that follows
them.
  Give credit when:

    - the verdict follows from the feasibility answers:  what the
        verdict claims, the answers actually established;
    - the verdict's load-bearing statements trace to specific answers
        or cited evidence;
    - negative findings and evidence gaps from the answers are named
        in the verdict and carried as open risks (accepting WITH named
        gaps is honest and creditable);
    - per-answer confidence labels are consistent with their evidence
        basis, and the verdict respects them;
```

```
    - a blocked verdict names its blocking constraints and traces them
        to the evidence.
  Serious failures:

    - MISQUOTED EVIDENCE: the verdict asserts that the evidence or an
        answer contains something it does not.
    - LAUNDERED ABSENCE: absence of evidence restated as positive
        compliance or demonstrated capability.
    - FABRICATED AUTHORITY: a standard, specification, certification,
        or regulatory threshold invoked as established when it appears
        nowhere in the report's evidence; treat as a substantive
        grounding failure and cap this score at 4.
    - IGNORED NEGATIVE FINDINGS: an acceptance verdict that passes
        over its own answers' negative findings without naming them
        caps this score at 5.
  Do NOT lower this score because:

    - the verdict accepts despite gaps, provided it names them;
    - the corpora genuinely lacked data and the answers said so;
    - the verdict is cautious or hedged.
  Scoring guidance:

      null (not scoreable):

        The report contains NO feasibility analysis (no manufacturing
        feasibility questions, no per-question evidence) on which a
        verdict could rest:  a bare "manufacturable" status emitted by a
        single direct generation.  This capability is absent by
        construction; output null for the score, not a number.
      0-2:

        A feasibility analysis exists but the verdict is disconnected
        from it:  conclusions unrelated to the answers, or contradicting
        them without acknowledgement.
      3-4:

        The verdict is partially connected to the analysis but
        overstates it:  misquoted evidence, laundered absence, or
        fabricated authority.
      5-6:

        The verdict mostly follows from the answers, but some
        load-bearing claims exceed what the evidence established, or
        some negative findings go unnamed.
      7-8:

        The verdict traces to the answers, names its gaps and open
        risks, and treats unverifiable hard requirements honestly
        (blocking, or accepting with the gap explicitly carried).
      9-10:

        Every load-bearing verdict claim is grounded in a specific
        answer or cited evidence, with complete and honest gap
        accounting.
  Use 0.5 increments when appropriate.

  ## System 3 Report


 <the System 3 report for one run:  the query, the retrieval and reasoning trace,

and the final System 3 output, rendered as Markdown>
 ## Required Output Format

 Return one evaluation for each dimension.

  For each dimension provide:

      - score:

            Number from 0 to 10 in increments of 0.5.
      - strengths:

            1-3 concise observations showing why the capability is functioning.
      - weaknesses:

            At most 1-2 concise observations describing the most consequential
            limitations specific to that metric.
      - rationale:
```

```
            Brief explanation connecting the demonstrated capability and limitations
            to the assigned score.
      - confidence:
            "high", "medium", or "low".
  Scoring procedure:
      1.  Determine whether the capability is absent, weak, functional, strong,
          or exceptional.
    2.  Select the corresponding score range.
    3.  Consider only weaknesses directly relevant to that metric.
    4.  Adjust within the range proportionally.
  Do not:
    - start from 10 and deduct for every mistake;
    - enumerate every factual or parameter-level error;
    - let the number of weaknesses determine the score;
    - treat exploratory values as mandatory final specifications;
    - double-count the same process issue across metrics;
    - re-penalize inherited System 2 errors unless they create a System 3 process
        consequence;
    - require production-qualified process specifications;
    - require every parameter to be experimentally validated;
    - treat one removable or optional problematic step as failure of the entire
        route.
  A strong evidence-driven process design with explicit equipment, operating
conditions, QC, and practical manufacturing considerations should commonly
score in the 7-8 range.
  Scores below 5 should indicate that the specific capability itself is
substantially ineffective.
  Keep the evaluation concise.
  Respond with ONLY a JSON object in this exact structure:
  ```
{
    "system3_plausibility":  {
        "score":  0.0,
        "strengths":  [],
        "weaknesses":  [],
        "rationale":  "",
        "confidence":  "high"
    },
    "system3_practicality":  {
        "score":  0.0,
        "strengths":  [],
        "weaknesses":  [],
        "rationale":  "",
        "confidence":  "high"
    },
    "system3_compatibility":  {
        "score":  0.0,
        "strengths":  [],
        "weaknesses":  [],
        "rationale":  "",
        "confidence":  "high"
    },
    "system3_verdict":  {
        "score":  0.0,
        "strengths":  [],
        "weaknesses":  [],
        "rationale":  "",
        "confidence":  "high"
    }
}
```
````



\end{document}